\documentclass[fleqn,usenatbib]{mnras}

\usepackage{newtxtext,newtxmath}
\usepackage[T1]{fontenc}
\usepackage{aecompl}
\usepackage{graphicx}	% Including figure files
\usepackage{amsmath}	% Advanced maths commands

\usepackage{booktabs} %Handling tables
\usepackage{multirow} %Handling tables
\usepackage{verbatim}
\usepackage{siunitx} %Handling tables
\usepackage{makecell} %Handling tables
\usepackage[table]{xcolor} %Colouring tables
\usepackage{tikz} %Encircle letters
\usepackage[percent]{overpic} %pics in tables
\usepackage{changepage} %Allows changing page size
\usepackage{bbding} %Fancy checkmarks
\usepackage{pifont}%Fancy symbols: e.g. checkmarks and crosses http://ctan.org/pkg/pifont
\usepackage{scrextend} %ref to the same footnote
\usepackage{enumitem} %Removes indentation in itemized objects
\usepackage{bm} %Bold math symbols

\usepackage{subfig}

\newcommand{\Msun}{M$_\odot$}
\newcommand{\arepo}{\textsc{Arepo}}

\newcommand{\sfdmodels}{\textsc{sf{\scriptsize 3}dmodels}}
\newcommand{\pcafactory}{\textsc{pcafactory}}
\newcommand{\turbustat}{\textsc{TurbuStat}}
\newcommand{\lime}{\textsc{lime}}
\newcommand{\astrodendro}{\textsc{astrodendro}}
\newcommand{\polaris}{\textsc{polaris}}

\newcommand{\treecol}{\textsc{TreeCol}}

\newcommand{\dv}{\delta\upsilon}
\newcommand{\edgeon}{edge-on$_{\phi=0^{\circ}}$}
\newcommand{\edgeonphi}{edge-on$_{\phi=90^{\circ}}$}
\newcommand{\twCOfull}{$^{12}$CO J$=$1$-$0}
\newcommand{\twCO}{$^{12}$CO}
\newcommand{\mean}[1]{\overline{\rm #1}}
\newcommand{\vz}{\upsilon_0}

\newcommand{\xmark}{\ding{56}}% cross for table 1

\definecolor{deepskyblue_mpl}{rgb}{0.0, 0.7490196078431373, 1.0}
\definecolor{magenta_mpl}{rgb}{1.0, 0.0, 1.0}
\definecolor{darkred_mpl}{rgb}{0.5450980392156862, 0.0, 0.0}
\definecolor{dodgerblue_mpl}{rgb}{0.11764705882352941, 0.5647058823529412, 1.0}
\definecolor{green_mpl}{rgb}{0.0, 0.5019607843137255, 0.0}
\definecolor{indigo_mpl}{rgb}{0.29411764705882354, 0.0, 0.5098039215686274}
\definecolor{darkgoldenrod_mpl}{rgb}{0.7215686274509804, 0.5254901960784314, 0.043137254901960784}

\makeatletter
\newcommand\mathcircled[1]{%
  \mathpalette\@mathcircled{#1}%
}
\newcommand\@mathcircled[2]{%
  \tikz[baseline=(math.base)] \node[draw,thick,circle,inner sep=0.8pt] (math) {$\m@th#1#2$};%
}
\makeatother

\newcommand\dashedline[1]{
    \tikz[baseline]{\draw[densely dashed, ultra thick, #1] (0,.5ex)--(.5cm,.5ex);}}
    
\newcommand\solidline[1]{\textcolor{#1}{\rule{0.5cm}{0.7mm}}}

\defcitealias{heyer+1997}{HS97}
\defcitealias{roman-duval+2011}{RD11}
\defcitealias{brunt+2013}{BH13}
\defcitealias{smith+2020}{Paper I}

\title[The Cloud Factory II: Gravoturbulent Kinematics of Clouds]{The Cloud Factory II: Gravoturbulent Kinematics of Resolved Molecular Clouds in a Galactic Potential}

\author[A. F. Izquierdo et al.]{Andr\'es F. Izquierdo,$^{1,2,3}$\thanks{E-mail: andres.izquierdo.c@gmail.com}
Rowan J. Smith,$^{1}$
Simon C. O. Glover,$^{4}$
Ralf S. Klessen,$^{4,5}$
\newauthor
Robin G. Tre{\ss},$^{4}$
Mattia C. Sormani,$^{4}$
Paul C. Clark,$^{6}$
Ana Duarte-Cabral,$^{6}$ and
\newauthor
Catherine Zucker$^{7}$
\\
$^{1}$ Jodrell Bank Centre for Astrophysics, Department of Physics and Astronomy, University of Manchester, Oxford Road, Manchester M13 9PL, UK\\
$^{2}$ European Southern Observatory, Karl-Schwarzschild-Stra{\ss}e 2, 85748 Garching bei M\"unchen, Germany\\
$^{3}$ Leiden Observatory, Leiden University, 2300 RA Leiden, The Netherlands\\
$^{4}$Universit\"{a}t Heidelberg, Zentrum f\"{u}r Astronomie, Institut f\"{u}r theoretische Astrophysik, Albert-Ueberle-Str. 2, 69120 Heidelberg, Germany \\
$^{5}$Universit{\"a}t Heidelberg, Interdisziplin{\"a}res Zentrum f{\"u}r Wissenschaftliches Rechnen, INF 205, 69120 Heidelberg, Germany\\
$^{6}$ School of Physics and Astronomy, Queens Buildings, The Parade, Cardiff University, Cardiff, CF24 3AA\\
$^{7}$ Harvard Astronomy, Harvard-Smithsonian Center for Astrophysics, 60 Garden St., Cambridge, MA 02138, USA
}

\date{Accepted XXX. Received YYY; in original form ZZZ}

\pubyear{2020}

\begin{document}
\label{firstpage}
\pagerange{\pageref{firstpage}--\pageref{lastpage}}
\maketitle

% Abstract of the paper
\begin{abstract}
We present a statistical analysis of the gravoturbulent velocity fluctuations in molecular cloud complexes extracted from our ``Cloud Factory'' galactic-scale ISM simulation suite. For this purpose, we produce non-LTE $^{12}$CO J=1-0 synthetic observations and apply the Principal Component Analysis (PCA) reduction technique on a representative sample of cloud complexes. The velocity fluctuations are self-consistently generated by different physical mechanisms at play in our simulations, which include galactic-scale forces, gas self-gravity, and supernova feedback.
The statistical analysis suggests that, even though purely gravitational effects are necessary to reproduce standard observational laws, they are not sufficient in most cases. We show that the extra injection of energy from supernova explosions plays a key role in establishing the global turbulent field and the local dynamics and morphology of molecular clouds. 
Additionally, we characterise structure function scaling parameters as a result of cloud environmental conditions: some of the complexes are immersed in diffuse (inter-arm) or dense (spiral-arm) environments, and others are influenced by embedded or external supernovae. In quiescent regions, we obtain time-evolving trajectories of scaling parameters driven by gravitational collapse and supersonic turbulent flows.
Our findings suggests that a PCA-based statistical study is a robust method to diagnose the physical mechanisms that drive the gravoturbulent properties of molecular clouds. Also, we present a new open source module, the \pcafactory{}, which smartly performs PCA to extract velocity structure functions from simulated or real data of the ISM in a user-friendly way. Software DOI: \url{10.5281/zenodo.3822718}.
\end{abstract}

% Select between one and six entries from the list of approved keywords.
% Don't make up new ones.

%%%%%%%%%%%%%%%%%%%%%%%%%%%%%%%%%%%%%%%%%%%%%%%%%%

%%%%%%%%%%%%%%%%% BODY OF PAPER %%%%%%%%%%%%%%%%%%

%\input{section-progression.tex}
\section{Introduction} \label{sec:introduction}

The relative importance of the physical mechanisms involved in star formation has been subject to intense debate over the last decades. Undoubtedly, gravitational effects govern the concluding stages of individual star-forming systems \citep{maclow+2004, krumholz+2007, keto+2010, ballesteros-paredes+2011, traficante+2018a, traficante+2018b}, but additional factors may play a role on the larger scales where gas is assembled into molecular clouds and successive fragmentation takes place \citep{bergin+2007, klessen+2016}. Observational data from the latest generation of telescopes have confirmed that, far from being isolated systems, stars are formed within large-scale molecular cloud complexes (10$-$60\,pc) that form in the cold interstellar medium \citep{blitz+1993, williams+2000}. These cloud complexes consist of interconnected molecular clouds (2$-$20\,pc) which, at the same time, exhibit high degrees of sub-structuring over subsequent scales \citep{falgarone+1992} and filamentary signatures \citep{andre+2010, ragan+2014, smith+2014b, arzoumanian+2019}. Thus, studying the dynamics of molecular structures in different spatial regimes becomes highly relevant to uncover the nature and evolution of star formation properties. 

\cite{larson+1979, larson+1981} discovered a systematic increase of the global velocity dispersion $\Delta\upsilon$ (km s$^{-1}$) with the projected size $L$ (pc) of diverse molecular associations ($\Delta\upsilon \propto L^{0.38}$) using mostly optically thin tracers ($^{13}$CO, H$_2$CO, NH$_3$). Larson interpreted this hierarchical behaviour to be a consequence of energy transport across successive spatial scales as it is reminiscent of the Kolmogorov structure law ($\dv \propto l^{1/3}$, where lower case $\dv$ and $l$ indicate internal velocity and spatial scales), derived from the statistical framework developed by \cite{kolmogorov+1941} and \cite{onsager+1949} for viscous incompressible (subsonic) turbulent fluids\footnote{However, \cite{larson+1981} also hinted at the possibility of supersonic turbulence in molecular clouds given the steeper scaling exponent compared to that of Kolmogorov's law. Larson interpreted this as a lack of velocity fluctuations at small scales caused by energy dissipation at larger scales via supersonic shocks.}. This linewidth-size relationship, often termed the (first) Larson law, would lay the groundwork for subsequent literature on the role of turbulence in setting dynamical signatures of the ISM.

\citet{solomon+1987} focused on a more homogeneous sample of clouds and reported a similar but slightly steeper linewidth-size relationship ($\dv \propto l^{1/2}$) using \twCO{} data. They interpreted this result as a consequence of virial equilibrium under the premise that the mean surface density of clouds is independent of size. However, this idea would be contradicted a few decades later by \cite{heyer+2009} who re-examined the same objects using a lower opacity tracer ($^{13}$CO) and higher spectral and angular resolution. In any case, classical statistical-hydrodynamic theories \citep[e.g.][]{kraichnan+1974, fournier+1983} derived the same velocity scaling index ($\gamma_2 = 1/2$) for fluids in a compressible (supersonic) turbulent field, which thereby suggests that energy dissipation in molecular clouds not only occurs at small scales (where viscosity dominates) but can be driven by supersonic shocks at larger scales as well \citep{mckee+2007}.

Turbulence is essential not only for triggering primordial density enhancements and seeding star formation, but also for regulating the onset of new stellar systems. Compressible (supersonic) turbulent velocity fields generate large-scale converging flows and strong density fluctuations, which, by the action of gravity, may end up collapsing and forming new stars in the most massive regions \citep{maclow+2004}. At the same time, turbulence is a key mechanism for controlling star formation rates as it acts against gravity, which alongside support from magnetic forces, prevents runaway gravitational collapse \citep{falgarone+1992, federrath+2018}. Further details about a gravoturbulent scenario for fragmentation in molecular clouds and its implications for star formation properties can be found in \cite{klessen+2004}. 

Additionally, the interplay between supersonic turbulence and local gravitational forces produces particular gas density distributions. High column densities associated with massive regions, dominated by self-gravity, exhibit power-law probability density functions (PDFs) \citep{ballesteros-paredes+2011, schneider+2015b}. Conversely, low column densities dominated by turbulent supersonic motions yield log-normal PDFs \citep{vazquez-semadeni+1994, kainulainen+2009} which can also exhibit non-Gaussian wings due to intermittency effects \citep{federrath+2010}. The range of densities in cloud complexes is typically wide \citep[$10^2-10^5$ cm$^{-3}$,][]{maclow+2004}, which implies that density distributions from realistic scenarios \citep[see e.g.][]{schneider+2002} are in general a combination of both profiles \citep{hennebelle+2008, kainulainen+2009, burkhart+2018}. Gravoturbulent mechanisms are hence crucial to establish stellar and core initial mass functions which may be closely related to the mass distribution of parental clouds \citep{padoan+2002, hennebelle+2009}.

A great deal of effort has also been expended on understanding the origin of non-thermal motions in the cold ISM. \cite{heyer+1997} adapted the principal component analysis (PCA) reduction technique to investigate the turbulent behaviour of indivial cloud complexes using spectroscopic data. The method consists in finding non-redundant representative components of (molecular) line emission data to extract velocity fluctuations $\dv$ (km s$^{-1}$) associated with characteristic spatial scales $l$ (pc) of the analysis region. They applied the algorithm to synthetic and real objects and found power-law dependencies analogous to the Larson linewidth-size relationship. Several studies were then carried out to connect the scaling parameters retrieved from this method to their intrinsic hydrodynamic structure function \citep{brunt+2002, brunt+2003c, heyer+2004, roman-duval+2011, brunt+2013}, responsible for describing the three-dimensional velocity fluctuations field as a function of the spatial separation of particles in the fluid. Other works tested the sensitivity of the technique to different feedback conditions. \citet{heyer+2006} found distinct relationships for clouds inside ($\dv=(1.00\pm0.04) l ^{0.79\pm0.06}$) and outside ($\dv=(0.70\pm0.03) l ^{0.66\pm0.06}$) an ionization front driven by a cluster of massive stars in the Rosette cloud complex. \cite{bertram+2014} used numerical simulations of molecular clouds with imposed turbulent fields and noticed variations in PCA-derived exponents when changing mean densities and optical depths. Using \twCO{} intensity, they infer a steeper relationship $\dv \propto l ^{0.82\pm0.03}$ for clouds with gas mean density $\overline{n}=300$ cm$^{-3}$, compared to the $\dv \propto l ^{0.59\pm0.02}$ for $\overline{n}=100$ cm$^{-3}$. Also, they suggest that using $^{13}$CO, which is an optically thinner tracer, can lead to slightly different relationships ($\dv \propto l ^{0.74\pm0.02}$ for $\overline{n}=300$ cm$^{-3}$). These findings make the technique an interesting tool to investigate the nature of non-thermal motions in the ISM.

MHD simulations carried out by \cite{deavillez+2005} and \citet{joung+2009} included a global galactic context to consistently investigate the ISM evolution. They found that several observational properties of the ISM turbulence can be reproduced in supernova feedback-dominated scenarios. However, due to the achievable spatial resolution ($\sim$1.5\,pc) and the lack of local gravitational effects, they could not study the internal structure and dynamics of molecular clouds in detail.

Later, in order to uncover the structure of turbulent motions in molecular clouds, \cite{federrath+2010} simulated synthetic turbulent fields made up of two different forcing components, solenoidal and compressive, within periodic uniform grids assuming isothermal gas. They suggest that molecular clouds have generally different mixtures of forcing, in which the solenoidal component is associated with quiescent regions with low star formation activity, and the compressive component to regions dominated by sources of strong energy feedback. This is supported by observations of quiescent and active star-forming regions or a combination of both scenarios \citep{heyer+2006, hacar+2016}.

More sophisticated high-resolution simulations were then developed to try to explain the origin and nature of these turbulent motions. \cite{klessen+2010} provided analytic and numerical calculations including magnetic fields, self-gravity and a standard ISM cooling function to show that accretion processes can drive the observed turbulence on several scales, from galaxies to protostellar disks. They used converging flows of accretion, incoming from the computational boundary with superimposed mean velocities and fluctuations, and obtained a linewidth-size relation $\Delta\upsilon=0.8(L/\rm{pc})^{0.5}$ km s$^{-1}$ compatible with Larson's law. This suggests that the turbulent scenario is similar to the classical energy cascade process from large to small scales, driven by outside-cloud phenomena. However, further work on energy injections from supernovae explosions \citep{gatto+2015, walch+2015, girichidis+2016, pan+2016}, stellar outflows \citep{nakamura+2007, cunningham+2011, federrath+2014} and HII regions \citep{peters+2017, haid+2018}, would demonstrate that the role of stellar feedback is also essential in configuring the turbulent field of molecular clouds \citep[for a summary see also][]{klessen+2016}. In particular, ISM simulations presented by \cite{padoan+2016a, padoan+2016b, padoan+2017} claimed that the structure and dynamics of molecular clouds are a natural consequence of a supernovae-driven scenario, and suggest that supernovae energy injection is necessary to set and maintain the turbulent cascade observed in molecular clouds. They generated random supernovae over a periodic cubic box of 250\,pc, with high (sub-parsec) spatial resolution, but at the cost of considering neither the large-scale gravitational potential nor differential rotation.

Our Cloud Factory simulations seek to address this limitation by including both supernova feedback and the large-scale galactic environment with high enough resolution to study the internal turbulence within clouds. We take into account the global galactic context using a multi-component gravitational potential and galactic differential rotation, while, at the same time, resolving selected molecular clouds with cell masses as small as 0.25\,M$_\odot$. We include stellar feedback in the form of supernovae, both randomly distributed across the Galaxy and tied to sites of star formation, as well as local gravitational forces and molecular chemistry. In this work, we use full non-LTE radiative transfer calculations and the PCA technique on our cloud complexes to investigate the detailed signatures of non-thermal motions over a wide range of spatial scales provided by our simulations. Full radiative transfer modelling is necessary to produce realistic synthetic observations that can be readily compared to observational data with analogous methods. Our analysis aims at investigating the role of clustered supernova feedback and local and large-scale gravitational forces in configuring the velocity fluctuations field of the cold ISM. We conclude that our simulations are able to self-consistently generate cloud complexes, with realistic turbulent fields, that can be used in future for studies of clustered star formation in a galactic context.

We briefly present the main aspects of our Cloud Factory simulation suite and the selected cloud complexes in Section \ref{sec:cloudfactory}. Sections \ref{sec:radiativetransfer} and \ref{sec:non-thermal_motions} are dedicated to the radiative transfer setup and statistical description of velocity fluctuations in fluids.
In Section \ref{sec:methodology} we outline the general workflow and explain the three PCA extraction methods explored in the paper. We then present in Section \ref{sec:results} the results split by physical scenario (\ref{subsec:results_physicalscenario}), line-of-sight projection (\ref{subsec:results_orientation}), time snapshot (\ref{subsec:results_time}) and analysis scale (\ref{subsec:results_inner}). We provide a discussion on the resemblance of our self-consistently generated clouds to observational data and the role of supernovae feedback in Section \ref{sec:discussion}, and wrap up with the conclusions of the work in Section \ref{sec:conclusions}. In Appendix \ref{sec:appendixfigures} we add supporting figures including edge-on cloud column densities, line emission and optical depth profiles, and also show variations in PCA-derived parameters when assuming LTE and LVG level populations for the radiative transfer.

\section{The Cloud Factory Simulation Suite} \label{sec:cloudfactory}

\subsection{The Hydrodynamic Code and Physical Ingredients} \label{subsec:hydrocloud}

The cloud complexes that we examine in this work are extracted from our Cloud Factory simulation suite \citep[][hereafter Paper I]{smith+2020}, which is built on a version of the \arepo{} code \citep{springel+2010, pakmor+2015} customised with a set of physical/chemical modules that account for various mechanisms taking place in the cold molecular ISM such as: 

\begin{itemize}
    \item The galactic gravitational potential. 
    \item Time evolution of CO and hydrogen chemistry.
    \item Ultraviolet extinction considering H$_2$ and CO shielding properties, and dust absorption.
    \item Star formation via sink particles. 
    \item Injection of (energy/momentum) feedback from supernova explosions. 
\end{itemize}

We use an analytic description of the large-scale gravitational potential of the Galaxy to efficiently determine and control its influence on the dynamics of mesh cells in each time step of the simulation. The potential is a combination of a dark-matter halo, a bulge, and a gas disc with thin and thick components. We use the best-fitting model of \cite{mcmillan+2017}, which is constrained by observations of the Milky Way. Additionally, we include a four-armed spiral component from \cite{cox+2002} and a consistent spiral perturbation to the potential, already implemented in \cite{smith+2014a}. The density profiles spawning the large-scale potential are fully described in \citetalias{smith+2020}. 

Our gas chemistry description adopts the approach of \cite{nelson+langer1997}, where the CO evolution is a simplified treatment that assumes a direct conversion between the C$^+$ and CO abundances (intermediate species are neglected). The CO formation is triggered by a radiative association between C$^+$ and H$_2$ to form hydrides that react afterwards with atomic oxygen. The CO destruction depends on the ultraviolet (UV) photo-dissociation rate from \cite{dejong+1977} and \cite{falgarone+1985}, which is a function of the gas number density and the strength of the UV portion of the interstellar radiation field \citep[assumed in our simulations to be that of the solar neighbourhood derived by][]{draine+1978}.  
We use the \treecol{} algorithm \citep{clark+2012} to compute the UV extinction of the medium by considering H$_2$ and CO self-shielding, the shielding of CO by H$_2$, and the shielding of both by dust absorption. 

Regarding the convergence of CO abundance in our simulations, at our highest resolution points for a number density of 10$^4$\,cm$^{-3}$ we have a resolution of 0.05\,pc or smaller \citepalias[see Fig. 4 in][]{smith+2020}, which compares well with the converging flow spatial resolution requirement from \cite{joshi+2019}. It is unclear that we meet their convergence criterion everywhere as our resolution is spatially variable. However, when we plug a conservative estimate of the internal velocity dispersion in 0.1\,pc scales in their criterion, we meet the resolution requirement at these densities. 

The non-equilibrium hydrogen chemistry from \cite{glover+2007a, glover+2007b}, implemented in our Cloud Factory, involves reactions between molecular (H$_2$), atomic (H) and ionised (H$^+$) hydrogen, electrons, cosmic rays, dust grains and the UV radiation field. This encompasses H$_2$ formation on grains, collisional and photo-dissociation of H$_2$, cosmic rays and collisional ionization of H, and H$^+$ recombination in the gas phase or on dust grains. The net energy exchange due to radiative and chemical gas heating or cooling is computed using an atomic and molecular cooling function as outlined in \cite{clark+2019}. %See the reaction formulas in Table 1 and their analytical treatment in Section 2.2 of \cite{glover+2007a}

Our model of star formation uses a hybrid approach based on sink particles that can represent either individual stellar systems or clusters of stars depending on the target mass resolution of the region where they form. In order to become a sink particle, following \cite{bate+1995} and \cite{federrath+2010b}, a cell and its neighbours have to be above a critical density $\rho_c$ and satisfy energy checks to confirm that they are bound and the internal collapse is runaway: the cells must be located on a local minimum of the gravitational potential, outside the accretion radius of any other sink particle and have inwardly directed velocities and accelerations. Sink particles behave like non-gaseous bodies that interact gravitationally with the surrounding medium and can accrete material from neighbouring bound cells that are within a given sink accretion radius. Sinks are especially helpful to set up a natural halt threshold in the code and prevent excessive mesh refinements \citep{hubber+2013} but at the same time to keep track of the sites where stellar feedback will be injected in form of thermal energy and/or momentum from supernovae explosions \citep{gatto+2015, walch+2015, girichidis+2016, padoan+2016a, padoan+2016b, pan+2016}. In this paper, we only consider supernovae as they are the most energetic source of stellar feedback and generally accepted to be the dominant driving mechanism of turbulence in star-forming galaxies \citep{maclow+2004, padoan+2016a}. Either way, other mechanisms such as outflows/jets \citep{nakamura+2007, cunningham+2011, federrath+2014}, stellar winds \citep{dale+2008, peters+2017, gatto+2017} and local photoionising radiation \citep{peters+2017, haid+2018} are also present in real scenarios.

We consider two ways of injecting stellar feedback: (i) purely random supernova explosions and (ii) a mixture of random supernovae and supernovae tied to star formation sites. For the first approach, we randomly distribute the supernovae according to the gas density profile of the galactic disc. We adopt a rate of 1 event every 50 years as estimated from Milky way observations of Gamma-ray emission in massive stars \citep{diehl+2006}. Our second approach produces bursts of strong feedback from the spiral arms. It assumes a star formation efficiency and a realistic stellar initial mass function in order to compute the number of massive stars ($> 8$\,M$_\odot$) that will undergo supernovae explosions at the end of their lifetime. We use the stellar mass function from \cite{kroupa+2002} and calculate the number of massive stars out of the stellar content of sink particles following \cite{sormani+2017}. 
When the target mass is large (100 M$_\odot$; see Section \ref{subsec:simulationsetup} below), the sink particles introduced in the simulation correspond to portions of clouds with size scales larger than individual star-forming cores, and hence a large fraction of gas in the sinks should not actually form stars. We account for this by adopting a low star formation efficiency for these sinks of 1$-$2\% based on the work of \cite{krumholz+2007}.
When the simulation target mass is small (< 10 M$_\odot$), the sink particles more closely correspond to individual star-forming cores and so in this case we adopt a higher star formation efficiency of 33\%, based on \cite{matzner+2000}. In this approach we also use random supernovae but at a lower rate of 1 event every 300 years to account for Type Ia supernovae. 
The way in which energy from supernovae is released into the gas depends on whether the Sedov-Taylor phase of the supernovae expansion is resolved, similar to the approach of \cite{hopkins+2014} and \cite{gatto+2015}, and introduced analytically by \cite{blondin+1998}. In our case, if the expansion phase is resolved by 32 cells we inject thermal energy directly into the surrounding gas, otherwise we inject terminal momentum pointing radially outwards. Further details of our supernovae model can be found in \citetalias{smith+2020} and in \cite{tress+2020}. 

\subsection{Simulation Setup and Refinement Scheme} \label{subsec:simulationsetup}

The initial gas distribution of our simulation is based on observational constraints and theoretical modelling of the Milky Way presented in \cite{mcmillan+2017}. They suggest an exponential profile for the H and H$_2$ densities as a function of the radius of the galaxy disc. We rather take the mass contribution from both profiles and start with a single gas distribution consisting only of H, from which H$_2$ will form self-consistently according to our chemical treatment as the Galaxy evolves. %The domain of our simulation comprises radii between 4 kpc and 12 kpc as we are interested in analysing properties of clouds outside the Milky Way's central bar.  

In the first stage of the simulation we let our Galaxy evolve for 150\,Myr under the effects of the large-scale potential and the energy/momentum feedback from random supernovae to naturally form spiral arms and reach a steady state. In this stage, the cell target mass resolution is set to 1000 \Msun{} and the mesh refinement operates accordingly.

Next, we start the middle phase of the simulation by turning on a co-rotating 3 kpc high resolution box centred at a galactic radius of 8 kpc. This phase lasts for around 70\,Myr, or two spiral arm passages. In this phase we launch 3 runs undergoing different physical mechanisms as follows: (a) a potential dominated scenario in which the ISM dynamics respond only to the large-scale gravitational potential and the random supernova feedback as it was set up during the first stage of the simulation, (b) same as the previous case but this time gas self-gravity between cells is included, and (c) a feedback dominated scenario in which both the large-scale potential and self-gravity effects operate, and the supernova feedback is mixed. By mixed feedback we mean that both the random and supernovae tied to sink particles are turned on (see our supernova implementation in Sec. \ref{subsec:hydrocloud}). The target mass resolution of this phase is initially set to 100 \Msun{} but is further lowered down to 10 \Msun{}  for the final 10\,Myr of the middle phase.  

In the final stage of our simulation, and in order to resolve substructures within the processed cold ISM, we further increase the resolution on individual cloud complexes of $\sim$100\,pc radius within the high resolution box by injecting Monte Carlo tracer particles \citep{genel+2013} everywhere the gas density is above 100 cm$^{-3}$. The target mass is lowered down to 0.25 \Msun{}  where tracer particles are present, which allows us to achieve high spatial resolutions (e.g. cell diameters of $\rm d_{cell}=1$\,pc at $n=5$ cm$^{-3}$, or $\rm d_{cell}=0.03$\,pc at $n=10^5$ cm$^{-3}$). For this target mass, we set a sink creation density of $\rho_c=10^4$ cm$^{-3}$, which according to \cite{maclow+2004} (and based on the size scales reachable in this phase) corresponds either to individual star-forming clumps or protostellar cores. Unlike Paper I, in this phase we split the potential-dominated scenario in two cases, with and without self-gravity, in order to explore the effects of local gravitational forces on the dynamical signatures retrieved from our cloud complexes. 

Regardless of the stage of the simulation, we require that the Jeans length is resolved by at least 4 cells everywhere in the mesh to adequately check energy and bounding conditions and avoid artificial fragmentation \citep{truelove+1997, federrath+2011}. If sink creation densities are achieved but the gas fails to pass the energy checks (see Sec. \ref{subsec:hydrocloud}), we continue to resolve the gas until it is unambiguously bound as long as it remains above the sink creation density.   

\subsection{The Selected Cloud Complexes}

In order to comprehensively investigate non-thermal motions in our synthetic clouds we use the same regions as in Paper I: A, B, C and D, but include two more complexes, A$_0$ and B$_0$, in which self-gravity is switched off. 

These cloud complexes arise from contrasting environments in the Galaxy. Complexes A$_0$ and A are (at the same location) in a dense spiral arm, whereas B$_0$ and B are in an inter-arm, more diffuse region. Cloud complexes C and D, representative of the feedback-dominated scenario, are the densest regions in the high resolution box and were born after a burst of clustered supernova feedback. However, complex D, the denser of the two, gets to form massive stars that undergo supernovae explosions over time, whereas complex C is only influenced by external feedback as it does not produce embedded supernovae during the analysis time. This information is briefly summarised in Table \ref{table:cloudcomplexes}. A figure showing the exact location of our cloud complexes can be found in \citetalias{smith+2020}.

\setlength{\tabcolsep}{1.5pt} %pad between columns

\begin{table}
\centering
{\renewcommand{\arraystretch}{1.5}%pad between rows
\begin{tabular}{ |c|cccc|l| } 

\toprule
\hline
\multirow{2}{*}{\makecell[c]{Cloud \\ Complex}} & \multirow{2}{*}{\makecell[c]{Galactic \\ potential}} & \multirow{2}{*}{Self-gravity} & \multicolumn{2}{c|}{Supernova feedback} & \makecell[c]{\multirow{2}{*}{Description}} \\
\cline{4-5}
& & & Random & On sinks & \\
\hline
A$_0$ & \Checkmark & \xmark & \Checkmark & \xmark & Inside arm \\
B$_0$ & \Checkmark & \xmark & \Checkmark & \xmark & Inter-arm region \\
\hline
A & \Checkmark & \Checkmark & \Checkmark & \xmark & Inside arm \\
B & \Checkmark & \Checkmark & \Checkmark & \xmark & Inter-arm region\\
\hline
C & \Checkmark & \Checkmark & \Checkmark & \Checkmark & No embedded SNe \\
D & \Checkmark & \Checkmark & \Checkmark & \Checkmark & Embedded SNe \\

\hline
\bottomrule

\end{tabular}
 \caption{Cloud complexes analysed in this work and physical mechanisms operating in each, with a short description of their surrounding environment.}
  \label{table:cloudcomplexes}
  }
\end{table}

We consider two different time snapshots for complexes A$_0$, B$_0$, C and D, and (for time evolution analysis) three snapshots for complexes A and B. In all cases, the first snapshot was extracted at a time when no massive sink particles had formed yet. Also, we explore three different cloud orientations, which we refer to as face-on, \edgeon{} and \edgeonphi{} views. In cylindrical coordinates, the face-on line-of-sight points towards the $\{-\hat{e}_z\}$ direction, meaning that the cloud complex is viewed from above the Galaxy. The \edgeon{} and \edgeonphi{} lines-of-sight point, respectively, towards the $\{+\hat{e}_\phi\}$ and $\{-\hat{e}_{R}\}$ directions.

Figure \ref{fig:column_densities} shows face-on projections of H, H$_2$ and \twCO{} column densities from the selected set of cloud complexes, 2\,Myr after tracer refinement has commenced (\edgeonphi{} views can be found in Fig. \ref{fig:appendix_column_densities}). As a short comment, note that the CO density does not necessarily trace the density of hydrogen species, which is a consequence of collisional and photo-dissociation processes induced by supernova explosions and the interstellar radiation field in our simulations. This is particularly apparent in feedback-dominated complexes C and D, hinting at high amounts of CO-dark molecular gas \citep[see, e.g.,][]{smith+2014a}, which is sensitive to variation of the feedback conditions in clouds.

\begin{figure*}
\begin{center}
\begin{tabular}{cc}

\begin{overpic}[width=0.77\textwidth]{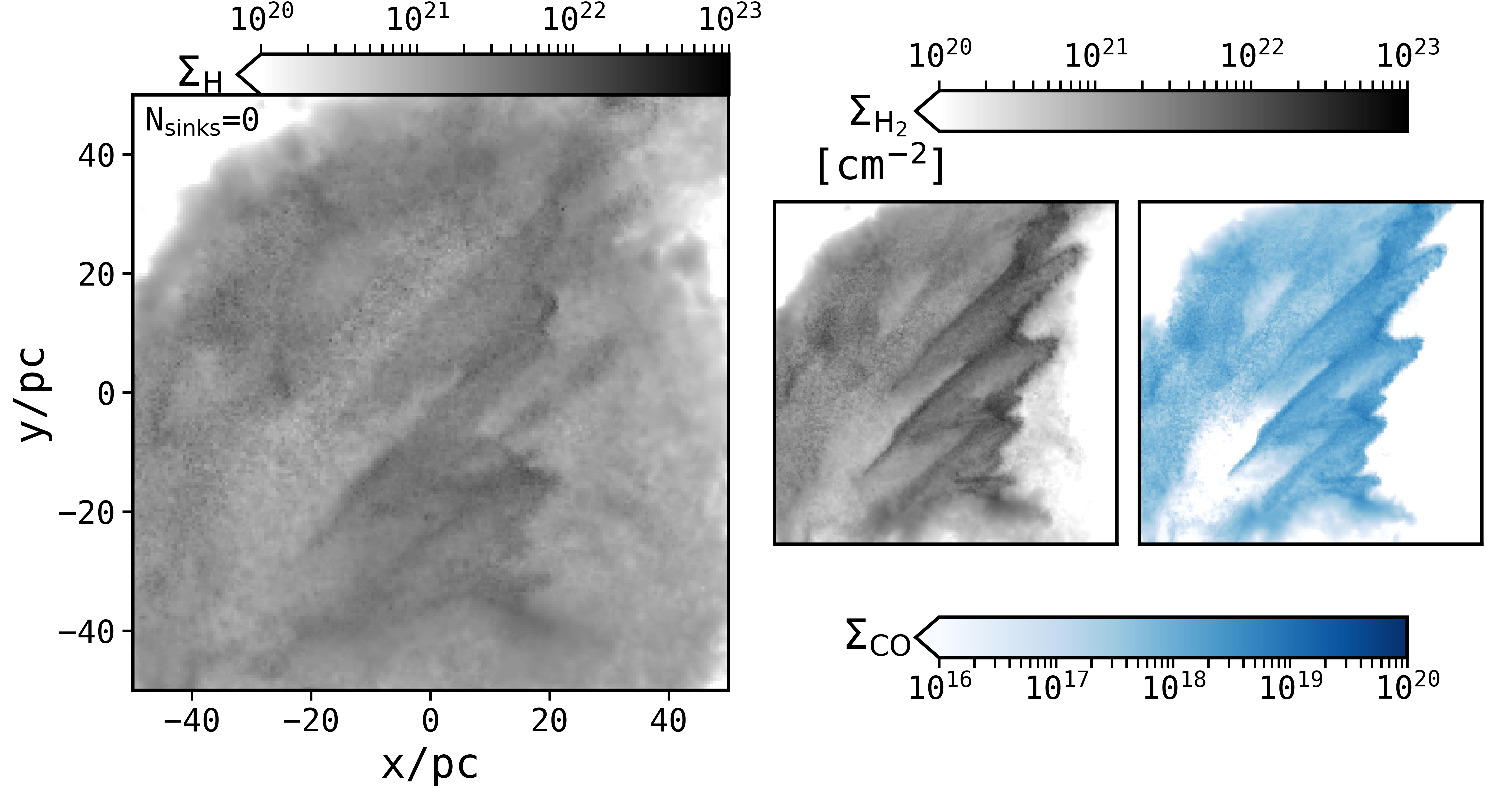}
\end{overpic} \hspace{0.5cm}
\put (40,120) {\makebox(0,0){{\huge A$_0$}}} \hspace{0.5cm}
\\

\begin{overpic}[width=0.77\textwidth]{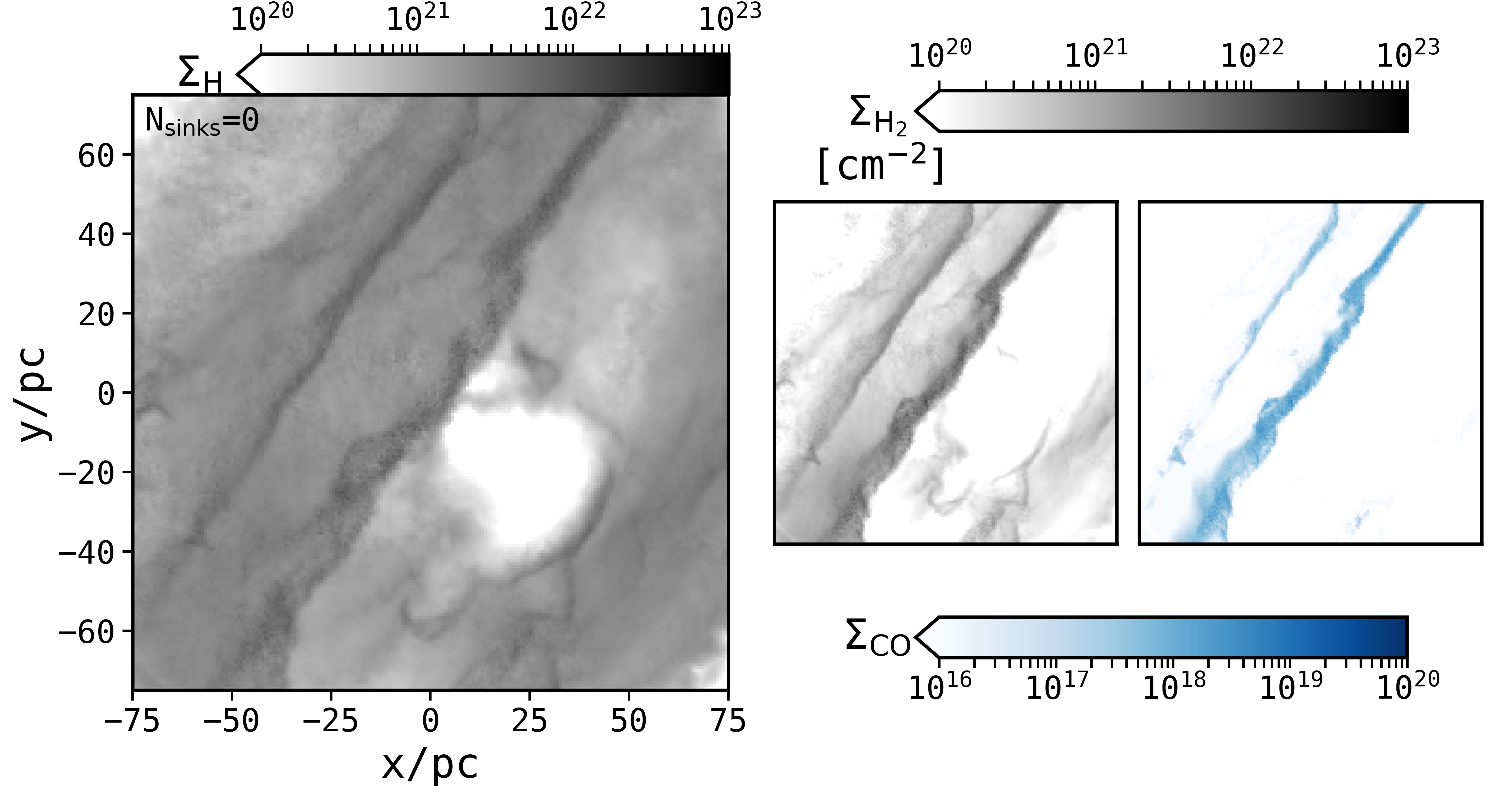}
\end{overpic} \hspace{0.5cm}
\put (40,120) {\makebox(0,0){{\huge B$_0$}}} \hspace{0.5cm}
\\

\begin{overpic}[width=0.77\textwidth]{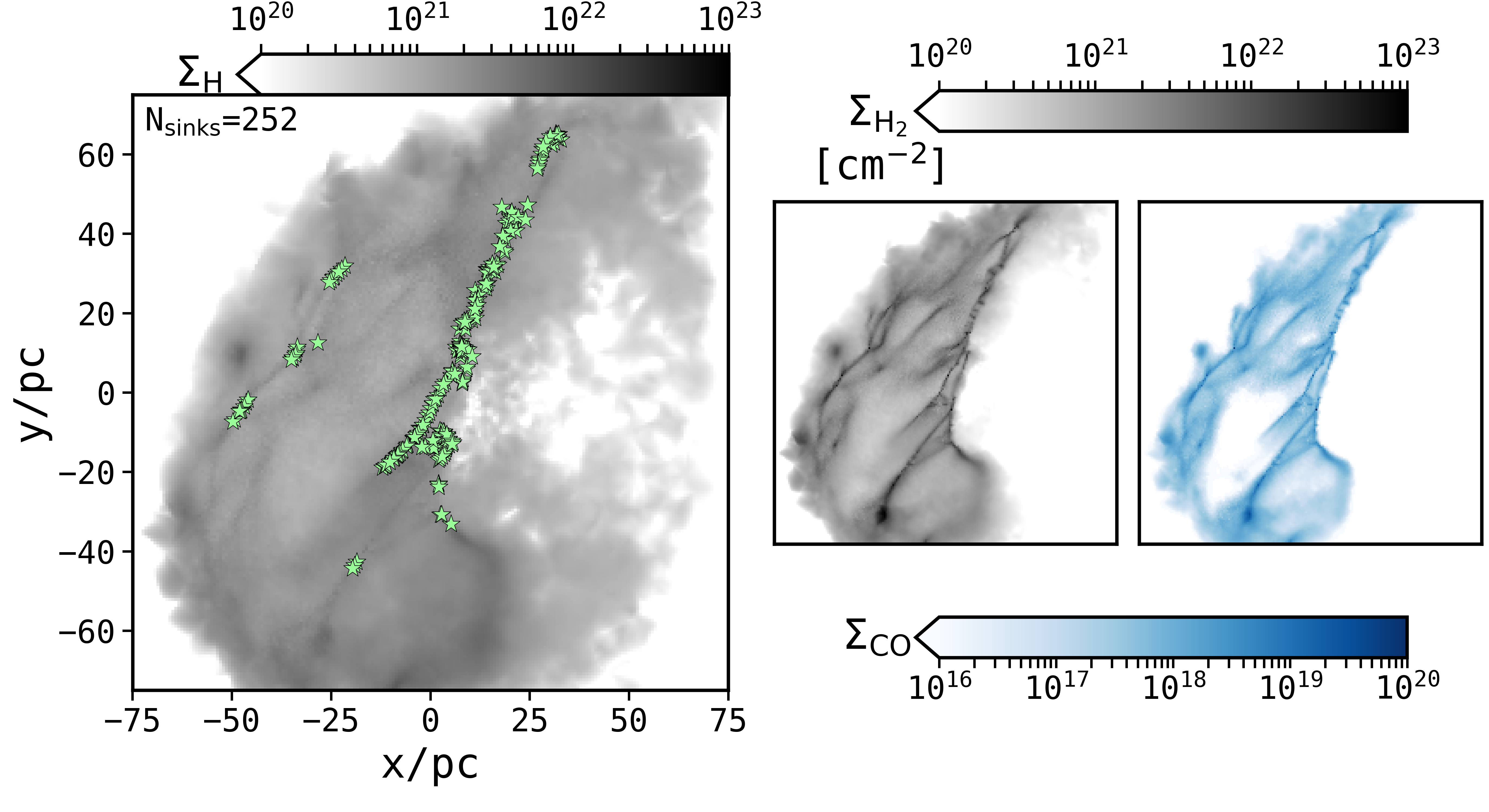}
\end{overpic} \hspace{0.5cm}
\put (40,120) {\makebox(0,0){{\huge A}}} \hspace{0.5cm}

\end{tabular}
\caption[]{Face-on projections of H, H$_2$ and $^{12}$CO column densities ($\Sigma$) from cloud complexes (labeled on the right) extracted 2\,Myr after injecting tracer particles in the simulations. If any, sink particles are overlaid on H maps as star markers.}
\label{fig:column_densities}
\end{center}
\end{figure*}

\begin{figure*}
\ContinuedFloat
\begin{center}
\begin{tabular}{cc}

\begin{overpic}[width=0.77\textwidth]{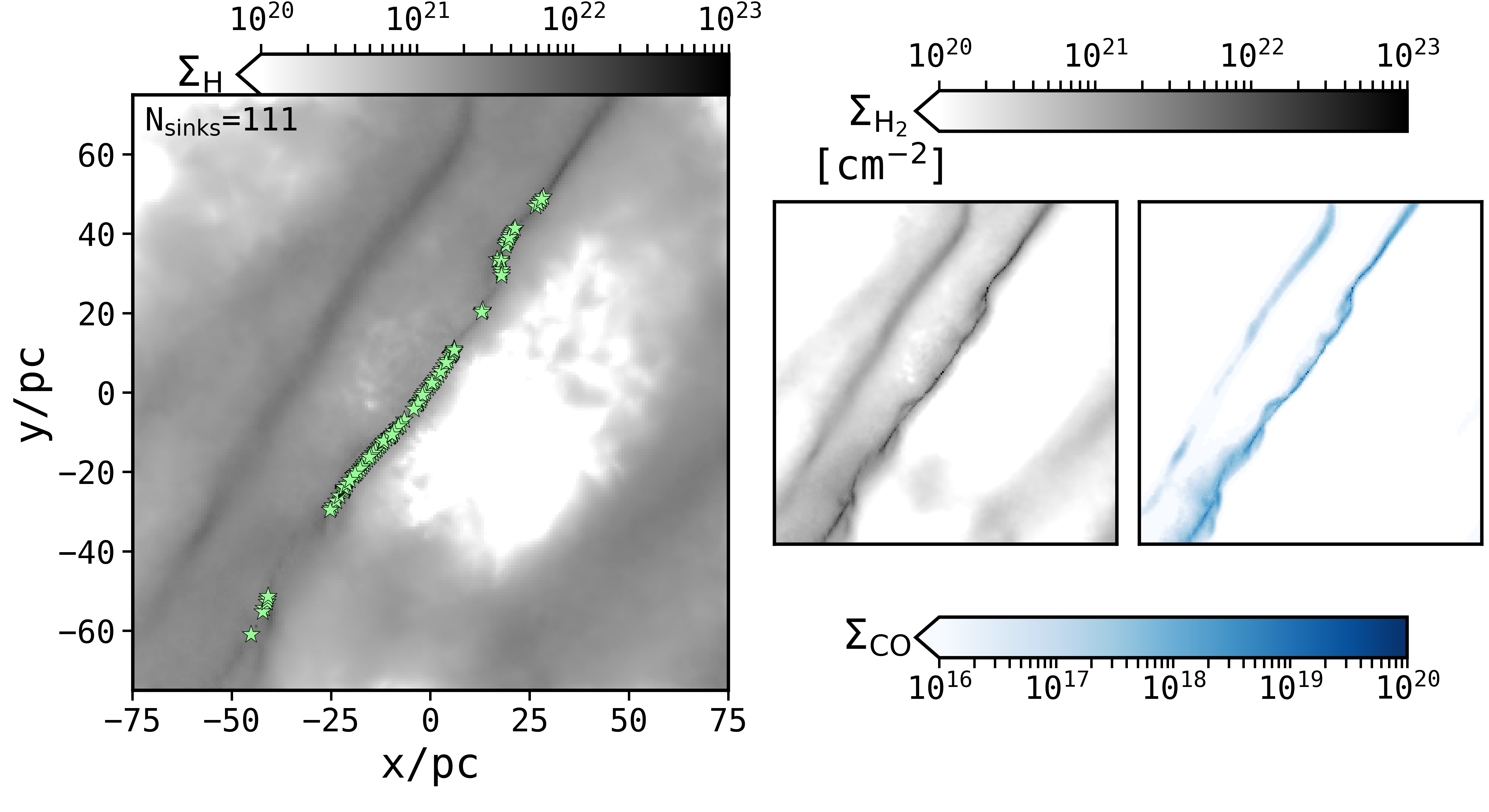}
\end{overpic} \hspace{0.5cm}
\put (40,120) {\makebox(0,0){{\huge B}}} \hspace{0.5cm}
\\

\begin{overpic}[width=0.77\textwidth]{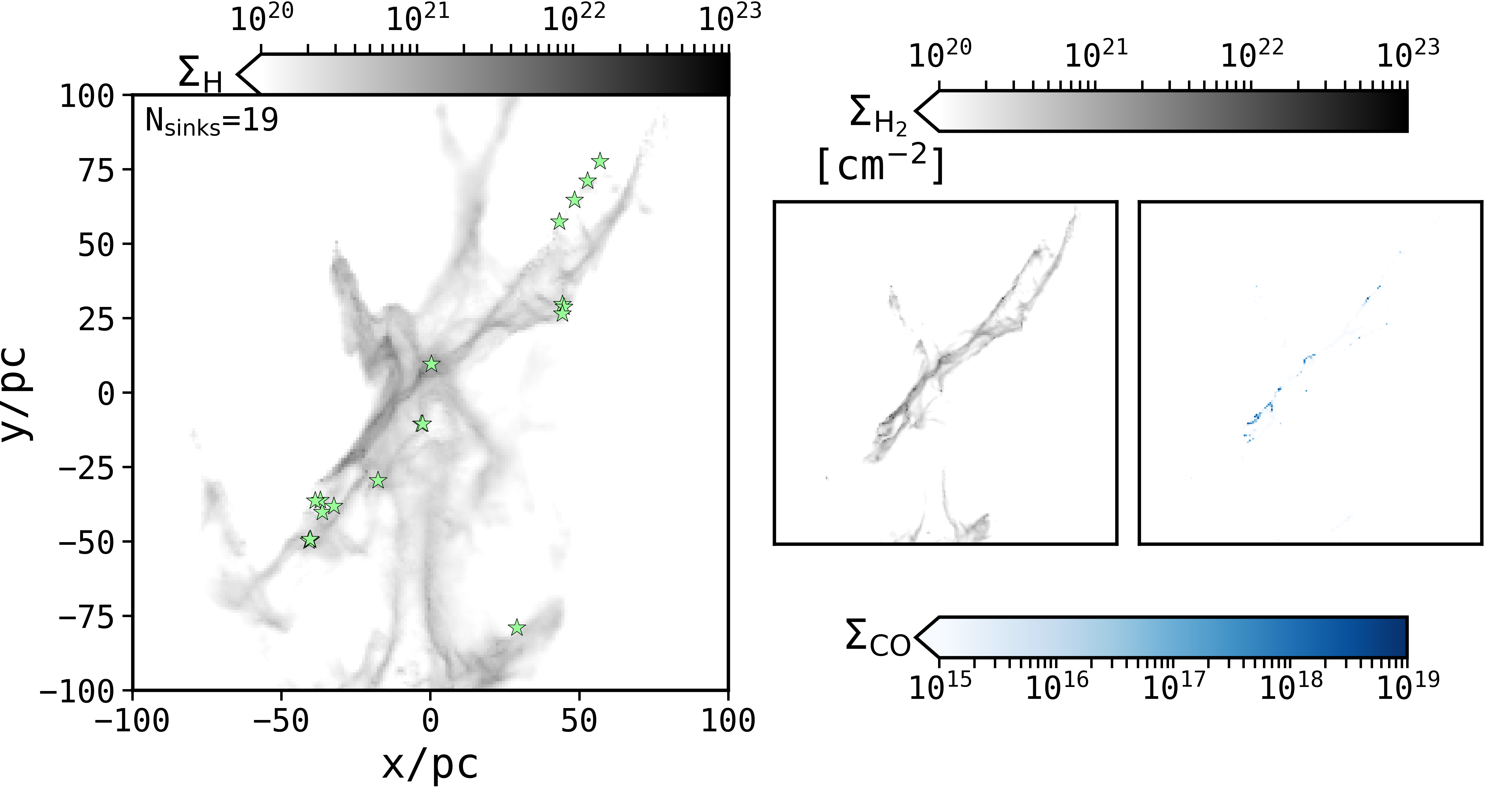}
\end{overpic} \hspace{0.5cm}
\put (40,120) {\makebox(0,0){{\huge C}}} \hspace{0.5cm}
\\

\begin{overpic}[width=0.77\textwidth]{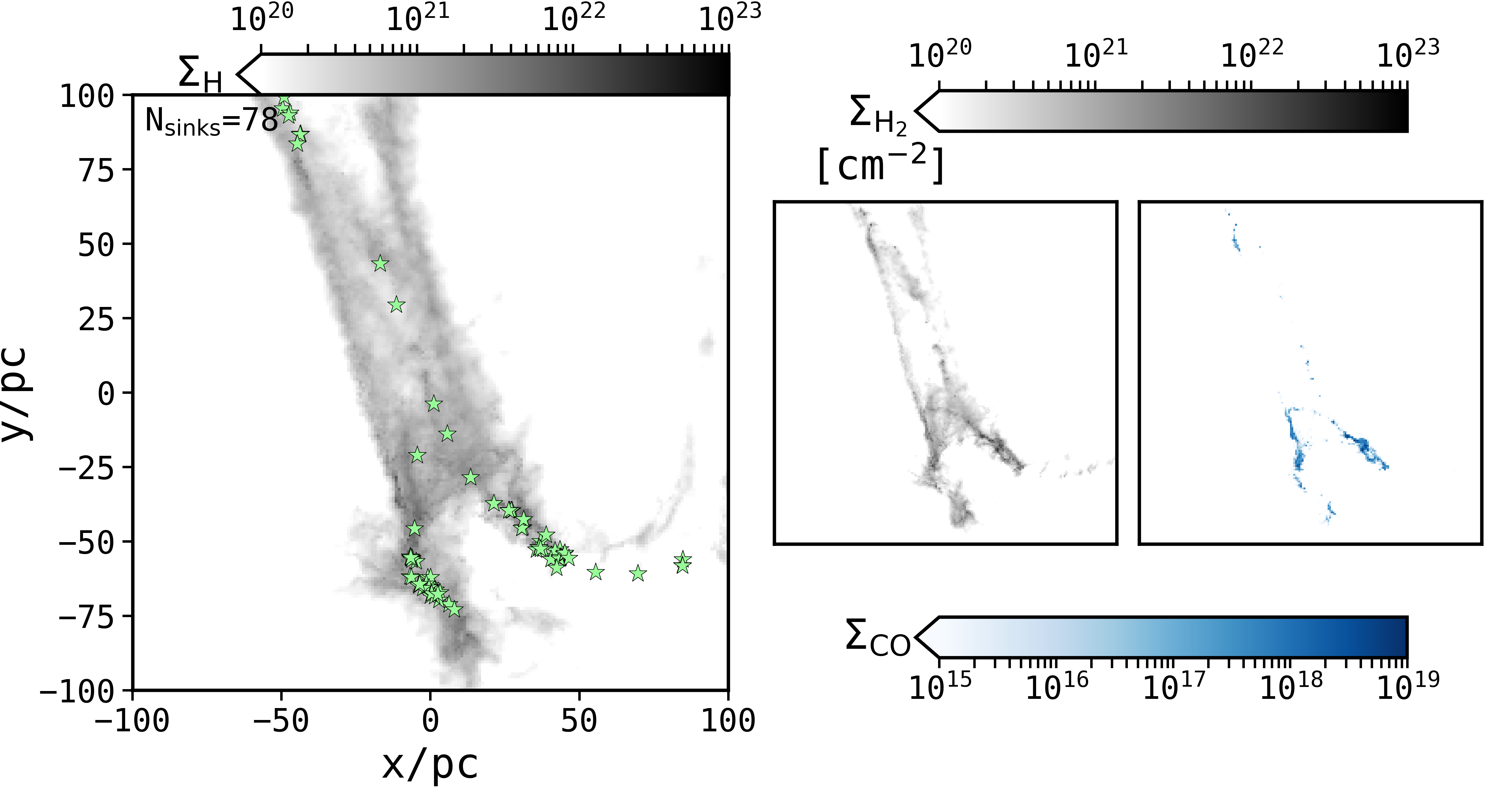}
\end{overpic} \hspace{0.5cm}
\put (40,120) {\makebox(0,0){{\huge D}}} \hspace{0.5cm}

\end{tabular}
\caption[]{(continued)}
\end{center}
\end{figure*}

\section{Radiative Transfer Modelling} \label{sec:radiativetransfer}

We perform radiative transfer simulations of our synthetic cloud complexes using the Line Modelling Engine code \citep[\lime{}\footnote{\url{https://lime.readthedocs.io}},][]{brinch+2010} and the Polarized Radiation Simulator \citep[\polaris{}\footnote{\url{http://www1.astrophysik.uni-kiel.de/~polaris}},][]{reissl+2016}. Both are flexible codes that predict molecular line and dust continuum emission from arbitrary 3D geometries in (sub-)millimetre and infrared wavelengths. 

For a given grid point distribution, in our case with information on position, gas density, temperature, and velocity, the codes construct a Delaunay triangulation and its corresponding Voronoi mesh where they iteratively propagate photons and integrate the radiative transfer equation. In particular, \lime{} comprises two approaches to solve the line excitation problem, suitable for matter in different equilibrium states: 

(i) A Local Thermodynamic Equilibrium (LTE) approximation in which the radiative features of the gas are fully and uniquely determined by the local kinetic temperature and its internal properties, namely, the atomic/molecular level populations are dominated by particle collisions which obey a Maxwell-Boltzmann distribution law. The code uses this to calculate the level populations and the Kirchoff's law for thermal radiation to solve the transfer equation \citep[see e.g.][]{rybicki+1986}. 

(ii) A non-LTE mode for media in which the level populations are not only ruled by collisions but also have a non-negligible contribution from the local radiation field. This problem needs to be addressed iteratively over the physical domain, taking into account the outgoing radiation from all the grid cells with each other. \lime{} solves this by propagating randomly oriented packages of photons from every cell of the grid, along lines of the Delaunay triangulation. In each cell, the algorithm computes provisional level populations using the incoming local radiation and collisional rates and releases a number of photons proportional to the number of neighbouring cells. The calculation stops when the propagating photons escape the physical domain, and the whole process is repeated, ideally, until convergence (i.e. populations in equilibrium) is reached. As convergence depends on the input physical distributions, the number of iterations is not fixed by \lime{} but left as a free parameter. 

On the other hand, besides LTE, \polaris{} supports a Large Velocity Gradient (LVG) approximation that we also explore in this work. This approach assumes that velocity variations over a given size scale are larger than microturbulent and thermal velocities, which simplifies the computation of optical depths and level populations \citep{sobolev+1960}. 

Once the level populations are determined in either of the approaches, the codes integrate the radiative transfer equation along isotropic tracer rays that cross the Voronoi grid
until they hit the border of the physical domain. The resulting specific intensity is then used to compute the observed intensity at the distance, spatial and spectral resolutions established by the user. The output FITS file is a 3-dimensional $n_{\rm xpix}\times n_{\rm ypix}\times n_{\rm chan}$ position-position-velocity (PPV) cube containing intensity (in units of either Jy pix$^{-1}$, Kelvin or L$_\odot$ pix$^{-1}$) or optical depth ($\tau$) information as a function of the spectral channel (in m s$^{-1}$).

Full radiative transfer modelling is necessary to produce synthetic emission maps that can be readily compared to observational data with analogous methods, especially at the present time with the advent of new telescopes and techniques that allow resolving non-ideal and highly coupled regimes. Evidently, the resulting cubes also inherit observational limitations such as spatial and spectral finite resolution and natural constraints from projection and optical depth effects. 

In the Appendix, we show variations in line emission profiles (Fig. \ref{fig:line_profiles}) and optical depth maps (Fig. \ref{fig:line_profiles_tau}) after considering the radiative transfer approaches (LTE, LVG and non-LTE) outlined in this Section, as well as their impact on PCA-derived structure function parameters (Fig. \ref{fig:pca_summary_distribution_LTE}). 

\subsection{Non-LTE \twCOfull{} Line Excitation}
\label{subsec:nonlte_12co}

Our analysis focuses on the emission of carbon monoxide in its ground state rotational transition \twCOfull{}, which, along with other isotopologues, have been used extensively to trace molecular, relatively dense, cold gas \citep{vandishoeck+1998, vandishoeck+2004} as well as in previous statistical studies \citep{larson+1981, heyer+2004, roman-duval+2011, bertram+2014} of both simulated and observed molecular associations. 

We use the non-LTE implementation of \lime{} because, in our simulations, a considerable fraction of the H$_2$ density (the main collisional partner of \twCO{}) is below the critical density ($n_{\rm crit} \sim 2\times10^3$ cm$^{-3}$) to collisionally populate \twCO{} at the upper level of its ground transition, which is valid within a wide range of temperatures \citep[$2-3000$ K,][]{yang+2010}. This implies that the first rotational level of \twCO{} is populated by different mechanisms in our cloud complexes; it is dominated by radiation in diffuse regions and thermally controlled by collisions in dense regions \citep[see Sec. 12.4 of][]{wilson+2013}. 
We note that we do not consider collisions between CO and H in our modelling, which may be important for diffuse CO gas near the edges of clouds due to turbulent mixing.

\subsection{New Tools for Handling \textsc{Arepo}$-$Like Meshes in Radiative Transfer Codes}
\label{subsec:tools_arepomeshes}

\subsubsection{Input Unstructured Meshes and Implementation of \textsc{kdtree} Algorithm in \lime{}}

Our previous customisation of \lime{} allowed the user to compute the radiative transfer solutions on analytic 3D models of star-forming regions generated with the \sfdmodels{} package \citep{izquierdo+2018}. These models were based on a uniform Cartesian grid and then mapped by \lime{} via random grid points weighted by the density distribution of the model. Cartesian grids are computationally efficient and a good approach to problems with low-dynamic range. Our current simulations, however, are based on (highly) non-uniform Voronoi grids specially adapted to track the effects of physical mechanisms governing the gas dynamics at several time and length scales. On top of that, our statistical analysis requires well-determined spatial scales to properly and unambiguously uncover the gravoturbulent nature of molecular clouds via velocity structure functions (see Sec. \ref{subsec:structure_function}). 
For these reasons, we modified \lime{} to also handle unstructured meshes pre-processed with \sfdmodels{}. To this end, we halt the default randomly-weighted generation of grid points in \lime{} and force it to rather take the simulation points to reconstruct the Voronoi mesh where the radiative transfer is solved.

Another addition to the code is motivated by the fact that, unlike Cartesian grids, finding neighbouring cells in unstructured grids is non-trivial. This is necessary during the ray-tracing algorithm in which the radiative transfer equations are computed iteratively over subsequent cells. Clearly, a `brute force' search that minimises the distance from the test cell to the whole set of cells is rather slow (of order $O(N^2)$) with increasing number of cells $N$. To solve this, we implement a k-d tree algorithm that splits the search domain in representative areas to discard unnecessary distance checks to remote cells. We use the third-party, open source, C library \textsc{kdtree}\footnote{\url{https://github.com/jtsiomb/kdtree}}, which is pre-built in our customised version of \lime{} and does not need any particular installation.

Additionally, the \lime{} domain is spherical and surrounded on its surface by randomly distributed points called `sink-points' (not to be confused with the sink particles that we use to represent star formation in the \arepo{} simulations), through which tracer rays emerge from the physical model to make up the synthetic image. Since the input unstructured grid can have any shape, large voids between sink-points and inner physical grid points are likely to exist, leading to artefacts at the borders of the image as the reconstructed Voronoi cells are larger there than in the original mesh. We soften this effect by including empty `dummy' grid points close to the border of the radiative transfer domain using our new \textsc{grid.fillgrid} module incorporated in \sfdmodels{}. We distribute $N$/10 dummy points randomly between a radius enclosing 90 per cent of the total gas mass and the maximum radial extent of the grid. However, this needs to be done with caution. For instance, including dummy points starting at small radii could induce multiple artificial holes in actual regions of the simulation and may also lead to an underestimation of gas masses as the volume of the original grid cells would have to decrease to make room for the newly inserted dummies.

\subsubsection{Removal of Twin \arepo{}  Cells after a Refinement Step}

According to \cite{springel+2010}, a cell meeting user-defined refinement criteria is split along its centroid into two cells. Initially, the position locators of the newborn cells coexist with the original cell centre. During subsequent time steps, the new locators are separated via mesh-regularisation techniques until they reach the actual centroids of the split cells. This bears the possibility of finding two or more cell locators at the same position in a single time snapshot of the simulation. Such subtlety can lead either to errors during the grid reconstruction by the triangulation algorithms of the radiative transfer codes, or it can slow down intermediate grid smoothing steps.

We have written an efficient iterative algorithm included in the \textsc{arepo.UniqueCells} module of \sfdmodels{} to get rid of these `twin' cells. Broadly speaking, the algorithm constructs an array of cell ids based on unique radially-sorted cells. These unique cells are then compared only to their nearest neighbours to check if they share the exact same location in 3D. The algorithm recycles the mass of the twin cells into the surviving cell using a direct summation, whereas the other physical properties remain unchanged as they are approximately equal in all the twin cells. The algorithm returns a clean dictionary with unique cells and their new physical properties. 

The \sfdmodels{} package and the latest customised version of \lime{} are open source and documented online\footnote{\url{https://star-forming-regions.readthedocs.io}}.

\subsection{Radiative Transfer Setup}
\label{subsec:rt_setup}

The front domain of the cloud complexes is deliberately set 2.4 kpc away from the observer, a typical distance to large nearby star-forming regions \citep[e.g. W33,][]{immer+2013}. The pixel size of the PPV cubes is 26$''$, which translates into a projected physical resolution of 0.3\,pc. This pixel size is a good compromise between resolving most of the cells from our \arepo{} meshes (which contain cells as small as $\sim0.03$\,pc, see Figure 3 in \citetalias{smith+2020}) and the processing time of the radiative transfer. We assume an optimal scenario where no beam smearing or noise are considered. The spectral resolution, or channel width, for each cloud complex is determined using the relation $\Delta _\textrm{chan}=(\upsilon_{\rm max} - \upsilon_{\rm min})/(n_{\rm chan}-1)$, where $\upsilon_{\rm max}$ and $\upsilon_{\rm min}$ are the maximum and minimum projected velocities along each of the three lines-of-sight explored in this work (face-on, \edgeon{} and \edgeonphi{}). The number of channels is constant ($n_{\rm chan}=101$) to ensure the same dimensionality of the principal component analysis. Typical channel-widths range from $\sim$\,0.2\,km\,s$^{-1}$ in potential-dominated complexes to $\sim$\,0.4\,km\,s$^{-1}$ in feedback-dominated complexes\footnote{We also analysed cubes with constant $\Delta_\textrm{chan}=0.1$\,km\,s$^{-1}$ for each physical scenario, at one time step, and in all cases the PCA-derived parameters remained unchanged except for the associated errors which decreased systematically by a factor $\leq3$.}. For the calculation of level populations in the non-LTE setup we assume 50 iterations as the populations converge after approximately 30 cycles. We ran tests with 100 iterations for selected regions and found differences of $<5\%$ in mean intensities, which is reasonable within the expected random uncertainty. 

For consistency, we include micro-turbulence as an additional source of line broadening to account for non-thermal motions on scales smaller than the cell size. This contribution is assumed to be equal to the sound speed of each cell of the mesh and added in quadrature to the standard thermal broadening during the line transfer. A typical mass-weighted average temperature in our cloud complexes is $11$\,K, which yields a sound speed of $0.25$\,km\,s$^{-1}$ for H$_2$. In cold dense cells, where the \twCO{} mass is non-negligible, micro-turbulence is generally much smaller than the bulk speed of the cells ($\times 10-100$ lower). Note that there may be additional unresolved small scale turbulence or organised motions (e.g. driven by gravity) within the cell so our assumed micro-turbulence is a lower limit.

\lime{} takes the number density of hydrogen nucleons and CO abundance available from our simulations to compute the CO mass consistently. We assume that the CO collisional partners, the spin isomers of molecular hydrogen (para-H$_2$ and ortho-H$_2$), have a 1:1 ratio corresponding to the expected value for molecular gas with an age of a few Myr \citep[see e.g.][]{flower+2006}. However, we note that the CO excitation rate is only weakly sensitive to the ortho-to-para ratio and hence our results should not be affected by this choice.

\section{Statistical Description of Velocity Fluctuations in Molecular Associations} \label{sec:non-thermal_motions}

\subsection{Velocity Structure Function of a Fluid} \label{subsec:structure_function}

The concept of a generalised function to describe non-thermal velocity fluctuations in a 3D fluid originated with the work of \cite{kolmogorov+1941}. The author considered an incompressible viscous fluid with very large Reynolds numbers ($\rightarrow \infty$), namely, locally dominated by isotropic turbulent motions, and assumed all the components of the turbulent velocity to be homogeneous and statistically random variables. Using similarity hypotheses on time and length-scale energy dissipation rates, the author found that the (averaged) velocity dispersion is a function of the spatial separation between test points of the fluid. This work envisioned the idea of rapid successive transport of turbulent kinetic energy, from large (low order) to small (high order) scales of the fluid, as a cascading process. Higher order scales end up dispersing energy in the form of heat as the effect of viscosity exceeds the magnitude of velocity fluctuations on small scales. 
Later, \cite{onsager+1949} used an analogous theoretical basis to demonstrate that 3D vorticities can accelerate the turbulent cascade and explain the rapid viscous dissipation of energy with increasing wave number. Based on this, he found the characteristic energy spectrum of Kolmogorov-like fluids
\begin{equation} \label{eq:1}
E(k)dk = A Q^{2/3} k^{-5/3} dk,    
\end{equation}

\noindent where $A$ is a dimensionless constant, $k=2\pi/l$ is the wave number associated with a given size scale $l$ in a Fourier expansion of the velocity field, and $E(k)dk$ stands for the kinetic energy distribution within an interval $dk$, which is being dissipated as heat at a rate $Q$. 

However, incompressible flows of the kind considered in Kolmogorov's scenario are rare in molecular clouds, where non-thermal motions are not negligible but rather transonic or supersonic \citep{maclow+2004, mckee+2007}. In particular, supersonic motions lead to shock-dominated turbulence \citep[Burgers-like turbulence,][]{burgers+1939, kraichnan+1974, fournier+1983, passot+1988, frisch+2001}, which serves as a mechanism of energy diffusion at large scales apart from just heat dissipation at small scales. This makes the energy spectrum decay faster at high order wave-numbers,
\begin{equation} \label{eq:2}
    E(k)\propto k^{-2}.
\end{equation}

Moreover, it is well known from early studies that the gradient of energy dissipation in non-ideal turbulent fluids depends on the properties of the medium \citep{kraichnan+1974} and the turbulent cascade must reflect this on different size scales. Hence, a power-law generalisation of the energy spectrum is a reasonable approach to account for intrinsic velocity fluctuations as a function of the input environment \citep[see, e.g.,][]{federrath+2010},
\begin{equation} \label{eq:3}
E(k) \propto k^{-\beta}.
\end{equation}

The exponent $\beta$ is known as the spectral index of the energy spectrum in a three-dimensional turbulent fluid and is an intrinsic property of the velocity distribution in the medium. In addition, this power-law dependence is supported in \cite{onsager+1949} by the fact that the total vorticity of a fluid is in general a linear combination of the wave-number scale vector $\vec{k}$. 

Using Eq. \ref{eq:3}, it is possible to compute the mean square velocity fluctuations at a given size scale $l$ by summing up the energy contributions from higher order (smaller-sized) scales as follows
\begin{equation} \label{eq:4}
 \langle | \dv(l)|^2 \rangle \propto \int_{2\pi/l}^{\infty} E(k)dk \propto \int_{2\pi/l}^{\infty} k^{-\beta}dk \propto l^{(\beta-1)}; \;\; (\beta>1),
\end{equation}

\noindent from which is straightforward to derive the root-mean-square (rms) velocity, 
\begin{equation} \label{eq:5}
\begin{aligned}
    \langle |\dv_l|^2 \rangle ^{1/2} &= \vz l^{\gamma_2}; \;\; \textrm{with} \;\; \gamma_2 = (\beta-1)/2,
\end{aligned}
\end{equation}

\noindent where $\vz$ and $\gamma_2$ are the intrinsic scaling parameters of the rms velocity fluctuations field.

A generalised description of velocity fluctuations, the so-called velocity structure function, was introduced by \cite{kolmogorov+1941} and further developed by \cite{anselmet+1984} and \cite{frisch+1995} in order to (statistically) explain multi-component turbulent motions, which the scaling parameters of the rms-velocity field are unable to model comprehensively \cite[see][]{brunt+2003c}. In molecular clouds, multi-component turbulence arise from several dissipation mechanisms such as shocks, magnetic fields, radiative cooling, and heat diffusion \citep{boldyrev+2002}, but also from energy-injection mechanisms like stellar feedback, both affecting different time and length scales of the region. The velocity structure function is written as 
\begin{equation} \label{eq:6}
\begin{aligned}
S_p(l) &= \langle | \dv(l)|^p \rangle;
\;\; \textrm{with} \;\; \dv = \upsilon(r) - \upsilon(r+l),
\end{aligned}
\end{equation}

\noindent where the exponent $p$ accounts for the order of the velocity fluctuations $\dv$ and provides information about the degree of coherence of the velocity field when subject to spatial variations $l$. This relationship can be modelled as a power-law of the size scale $S_p(l) \propto l^{\zeta_p}$ analogous to the approach followed in Eqs. \ref{eq:4} and \ref{eq:5} for the rms velocity, then
\begin{equation} \label{eq:6b}
\begin{aligned}
\langle | \dv(l)|^p \rangle ^ {1/p} &= \vz l^{\gamma_p};
\;\; \textrm{with} \; \; \gamma_p = \zeta_p / p,
\end{aligned}
\end{equation}

\noindent being $\gamma_p$ (or the equivalent $\zeta_p$) the intrinsic scaling exponent of order $p$ and $\vz$ the magnitude of velocity fluctuations known as scaling coefficient or normalisation of the velocity structure function.\footnote{Note that this expression satisfies the upper limit of the H\"older condition for the ($l$, $\dv$) metrics, which implies that the velocity distribution is uniformly continuous within the spatial domain of the fluid.} In Section \ref{subsec:pca_calibration}, we expand the discussion on structure function variations with changing $p$-orders. 

As a final comment, note from Eq. \ref{eq:5} that Kolmogorov's law ($\beta=5/3$) and the Burgers-like turbulence ($\beta=2$) yield rms scaling exponents, $\gamma_2=1/3$ and $\gamma_2=1/2$, respectively.

\subsection{Principal Component Analysis} \label{subsec:pca}

In this subsection we briefly summarise the main aspects of the principal component analysis (PCA) technique applied to the study of the structure and dynamics of molecular associations in the ISM.
%, either clouds, clumps, cores or a combination of them.

PCA is a statistical multivariate method that transforms an input dataset of $n$, possibly correlated, variables into a new object spanned by a set of $m$ orthogonal uncorrelated components (called principal components) in such a way that the variance of the initial dataset, which can be seen as the amount of information as a function of the original variables, is (non-redundantly) maximized along subsequent principal components. This property allows the dimensionality of the analysis to be reduced to only the components that hold most of the variance of the data, i.e. $m \leq n$. 

%ACF are calculated...
The method's theoretical framework was originally presented by \cite{pearson+1901} and \cite{hotelling+1936} but adapted for the first time for the study of ISM dynamics in the work of \cite{heyer+1997}, who described the formalism of the technique when considering position-position-velocity (PPV) data cubes and demonstrated its ability to retrieve velocity fluctuations within characteristic spatial scales from synthetic models with well-known line profiles and noise level. However, it was not until the study of \citet[][hereafter BH13]{brunt+2013} that the method, applied to spectroscopic images, acquired a formal theoretical foundation. 

Making use of the PPV intensity cube $T_{ij} = T(\bm{r}_i, \upsilon_j)$, composed of $n = n_x \times n_y$ pixels, and $n_c$ velocity channels, the method first finds the associated covariance matrix defined as,
\begin{equation} \label{eq:7a}
S_{jk} = \frac{1}{n}\sum_{i=1}^{n} T_{ij}T_{ik}.
\end{equation}
%It also determines its corresponding eigenvalues
The method then computes the corresponding eigenvalues $\lambda_m$, and eigenvectors $\bm{u}_m$, of the covariance matrix, by solving the eigenequation $Su = \lambda u$. The subscript $m$ stands for the $m$th principal component of the covariance matrix. Next, the intensity cube is projected on to the eigenvectors to construct the associated eigenimages,
\begin{equation} \label{eq:7b}
I_m(\bm{r}_i) = \frac{1}{n}\sum_{j=1}^{n_c} T_{ij}u_{mj}.
\end{equation}
The characteristic width of the three-dimensional autocorrelation functions of the eigenvectors $\bm{u}_m$ and eigenimages $I_m$, i.e. the velocity and spatial lags for which the autocorrelation functions have decreased by a factor of 1/e from their peak values, determine, respectively, the $m$th order velocity fluctuations $\dv_m$ and spatial scales $\delta l_m$. Further details can be found in \citetalias{brunt+2013} and references therein.

\cite{heyer+1997} applied this PCA technique to a sample of four cloud complexes and found, for each, a correlation of the form 
\begin{equation} \label{eq:7}
\dv = \vz l^\alpha,
\end{equation}
\noindent linking velocity fluctuations $\dv$ to characteristic size scales $l$, where the normalisation $\vz$ is the magnitude of velocity fluctuations in a given cloud complex. Note that we use $\alpha$ to refer to the PCA-derived scaling exponent, but in general it differs from the intrinsic exponent $\gamma_p$ of the velocity structure function as we briefly discuss in Section \ref{subsec:pca_calibration}. In any case, this pseudo-structure function seems analogous to the empirical linewidth-size relationship,
\begin{equation} \label{eq:8}
\Delta\upsilon=CL^\Gamma,
\end{equation}

\noindent found by \cite{larson+1979,larson+1981} after computing global velocity dispersions $\Delta\upsilon$ from an ensemble of varied molecular associations as a function of their projected sizes $L$. We use capitals $C$ and $\Gamma$ to indicate that the scaling coefficient and exponent, respectively, were derived from a global linewidth-size analysis which combined data from independent clouds rather than from a local-scale study of velocity fluctuations within individual clouds.

Both relationships were systematically investigated in subsequent works, some of which are listed in Table \ref{table:references} for further comparison with our results. Also, both of them resemble a structure function-like dependence (see Eq. \ref{eq:6}), but actually do not represent the natural behaviour of non-thermal motions mainly due to projection and radiative transfer effects. It is possible, however, to connect these pseudo-structure functions to intrinsic structure functions via calibration relations and universality principles of turbulence as summarised in Sections \ref{subsec:pca_calibration} and \ref{subsec:pca_universality}.

We use the \textsc{pca} module included in the \turbustat{}\footnote{\url{https://turbustat.readthedocs.io}} package \citep{koch+2019} to retrieve characteristic size and spectral scales ($l$, $\dv$) from our cloud complexes. We constrain the PCA algorithm to keep most of the data variance in principal components (95$-$99 per cent of the total variance), but not too much (i.e. $\gg$99 per cent) in order to avoid artificial clustering of points at the minimum recoverable scales of intensity cubes. The PCA pseudo-structure functions and their corresponding scaling parameters ($\vz$, $\alpha$) are computed separately by our \pcafactory{} package according to the three extraction methods summarised in Section \ref{sec:methodology}.

\setlength{\tabcolsep}{4.5pt} %pad between columns

\begin{table*}
\centering
{\renewcommand{\arraystretch}{2.0}%pad between rows

\resizebox{\textwidth}{!}{
\begin{tabular}{ |l|c|c|c|cc|l| }  %alignment in the cells: centre (c), align decimal points (S), left (l), right (r) 
\toprule
\hline
\multirow{2}{*}{Reference} & \multirow{2}{*}{Marker} & \makecell[c]{Larson-like \\ exponent} & \makecell[c]{PCA-derived \\exponent} & \multicolumn{2}{c|}{Structure function} &
\makecell[r]{\multirow{2}{*}{Short description}} \\ %the 'r' (right) cancels out the 'l'. 'c' is not enough to centre the header. 
%\multicolumn{1}{c|}{Short description}\\ 
& & {$\Gamma$} & {$\alpha$} & {$\upsilon_0$  (km s$^{-1}$)} & {$\gamma$} & \\
\hline

\cite{kolmogorov+1941} & \dashedline{blue}  & -- & -- & -- & $1/3$ & \makecell[l]{$\bullet$ Kolmogorov's law for ideal incompressible turbulence.} \\

\cite{kraichnan+1974} & \dashedline{red} & -- & -- & -- & $1/2$ & \makecell[l]{$\bullet$ Compressible supersonic shock-dominated turbulence \\ (Burger's turbulence).} \\

\midrule

\cite{larson+1979, larson+1981} & $\mathcircled{\textrm{\bf L}}$ & $0.38$ & $-$ & $1.1$ & $0.38$ & \makecell[l]{$\bullet$ Molecular clouds, clumps, cores, H \textsc{ii} regions, mapped by \\ different tracers.} \\

\cite{solomon+1987} & $\mathcircled{\textrm{\bf S}}$ & $0.50\pm0.05$ & $-$ & $1.0\pm0.1$ & $0.50\pm0.05$ & \makecell[l]{$\bullet$ \twCO{} emission from homogeneous sample of 273 clouds.} \\

\cite{heyer+1997} & $\mathcircled{\textrm{\bf \scriptsize 1a}}$ & $-$ & $0.43\pm0.04$ & $1.23\pm0.08$ & $0.24\pm0.10$ & \makecell[l]{$\bullet$ \twCO{} emission of cloud complex Sh 155.} \\

\cite{heyer+1997} & $\mathcircled{\textrm{\bf \scriptsize 1b}}$ & $-$ & $0.55\pm0.03$ & $0.78\pm0.05$ & $0.42\pm0.09$ & \makecell[l]{$\bullet$ \twCO{} emission of cloud complex Sh 235.} \\

%\cite{brunt+2003c} & $\mathcircled{\textrm{\bf 2}}$ & --  & -- & -- & -- & $\bullet$ Simulated intermittent and non-intermittent velocity fields. \\

\cite{heyer+2004} & $\mathcircled{\textrm{\bf H}}$ & --  & $0.65\pm0.01$ & $0.87\pm0.02$ & $0.57\pm0.07$ & $\bullet$ \twCO{} emission from 27 molecular clouds. \\

\cite{heyer+2006} & $\mathcircled{\textrm{\bf \scriptsize 3a}}$ & -- & $0.74\pm0.04$ & $0.73\pm0.03$ & $0.71\pm0.12$ & $\bullet$ Rosette cloud complex as a whole. \\

\cite{heyer+2006} & $\mathcircled{\textrm{\bf \scriptsize 3b}}$ & -- & $0.79\pm0.06$ & $1.00\pm0.04$ & $0.79\pm0.04$ & \makecell[l]{$\bullet$ Rosette cloud complex, zone I: inside ionization front,\\feedback from H \textsc{ii} region, using \twCO{} (similar for $^{13}$CO).} \\

\cite{heyer+2006} & $\mathcircled{\textrm{\bf \scriptsize 3c}}$ & -- & $0.66\pm0.06$ & $0.70\pm0.03$ & $0.59\pm0.14$ & $\bullet$ Rosette cloud complex, zone II: outside ionization front. \\

\cite{bolatto+2008} & $\mathcircled{\textrm{\bf 4}}$ & $0.60\pm0.10$ & $-$ & $0.76\pm0.27$$^\star$ & $0.60\pm0.10$ & $\bullet$ Molecular clouds in extragalactic systems, using \twCO{}. \\

\cite{roman-duval+2011} & \solidline{green_mpl} & -- & $0.62\pm0.20$ & $-$ & $0.53\pm0.35$ & \makecell[l]{$\bullet$ Average from 368 molecular clouds, with a resolution limit \\ of 1 FHWM (48''), using $^{13}$CO emission.} \\

\cite{hacar+2016} & $\mathcircled{\textrm{\bf M}}$ & -- & -- & $0.66$$^\star$ & $0.58$ & \makecell[l]{$\bullet$ Musca cloud as a whole, isolated from stellar sources \\ of feedback, using $^{13}$CO and C$^{18}$O(2-1) emission.}  \\

\cite{rice+2016} &  & $0.49\pm0.04$ & $-$ & $0.66\pm0.09$$^\star$ & $0.49\pm0.04$ & $\bullet$ Dendrogram-based catalog of 611 outer Galaxy clouds. \\

\cite{rice+2016} &  & $0.52\pm0.03$ & $-$ & $0.87\pm0.09$$^\star$ & $0.52\pm0.03$ & $\bullet$ Dendrogram-based catalog of 428 inner Galaxy clouds. \\

\cite{miville+2017} &  & $0.63\pm0.30$ & $-$ & $0.83$$^\star$ & $0.63\pm0.30$ & $\bullet$ \twCO{} catalog of 8107 clouds in the Galactic plane. \\

\cite{traficante+2018b} & \solidline{darkgoldenrod_mpl} & $0.09\pm0.04$ & $-$ & $-$ & $0.09\pm0.04$ & $\bullet$ Gravity-dominated regions, core and clump scales. \\

\midrule

\cite{federrath+2010} & \solidline{darkred_mpl} & -- & $0.66\pm0.05$ & -- & $0.59\pm0.13$ & $\bullet$ Simulations with purely solenoidal forcing ($\nabla \cdot f = 0$). \\

\cite{federrath+2010} & \solidline{dodgerblue_mpl} & -- & $0.76\pm0.09$ & -- & $0.74\pm0.19$ & $\bullet$ Simulations with purely compressive forcing ($\nabla \times f = 0$). \\

\cite{klessen+2010} & $\mathcircled{\textrm{\bf 5}}$ & -- & -- & 0.8 & 0.5 & \makecell[l]{$\bullet$ Simulations with continuous accretion of diffuse material.} \\

\cite{bertram+2014} & \solidline{indigo_mpl} & -- & $0.82\pm0.03$ & $-$ & $0.83\pm0.11$ & \makecell[l]{$\bullet$ Non-isothermal simulations  of chemically evolving clouds. \\ Values for \twCO{} emission and mean density $n=300$ cm$^{-3}$.}  \\

\cite{padoan+2017} & $\mathcircled{\textrm{\bf 6}}$ & -- & -- & 0.82$^{\star,1}$ & $0.5\pm0.1$ & $\bullet$ Simulated clouds with supernovae-driven turbulence. \\

\hline
\bottomrule

\end{tabular}
}
 \caption{Scaling parameters from Larson-like and PCA-derived structure functions reported in selected literature. For comparison, calibrations indicated in Section \ref{subsec:pca_calibration} are used to translate parameters to rms structure function parameters when not provided. From top to bottom, the horizontal rules separate classical, observational and theoretical parameters. $^\star$Converted from 1D to 3D velocity dispersion assuming isotropy. $^1$Extracted manually from their Fig. 4. We provide a graphical version of this summary in Fig. \ref{fig:literature_summary}.}
  \label{table:references}
  }
\end{table*}

\subsubsection{Calibration from PCA to Structure Function Scaling Parameters} \label{subsec:pca_calibration}

\cite{brunt+2002} tested the sensitivity of the technique to different energy spectra $E(k)\propto k^{-\beta}$ from simulated data and found an empirical calibration that relates the PCA-derived scaling exponent $\alpha$ with the spectral index $\beta$, which is intrinsic to the 3D velocity distribution. Equivalently, \citet[][hereafter RD11]{roman-duval+2011} obtained almost the same calibration using a wider range of synthetic scenarios with different intermittency (both spatial and temporal sporadic fluctuations in the turbulent flow) and power spectra for density and velocity. It is valid below a relatively high level of density variability and for spectral indices $\beta$ between $1.2-2.6$. Later, \citetalias{brunt+2013} derived a slightly steeper but very similar calibration from an analytic point of view. However, in this work we use the \citetalias{roman-duval+2011} calibration because, as \citetalias{brunt+2013} suggest, their analytic result should be seen as a supportive element rather than a replacement of the previous empirical estimations given the approximations used in their derivation. The \citetalias{roman-duval+2011} calibration can be written as:
\begin{equation} \label{eq:9}
    \beta = (0.20 \pm 0.05) + (2.99 \pm 0.09)\alpha,
\end{equation}
\noindent which is useful for connecting PCA scaling parameters derived from genuine observables to the intrinsic energy distribution within a given 3D fluid volume. Observables in molecular clouds result from line-of-sight projected averages of velocity fluctuations in the best case, but there might also be optical depth effects involved. 

The retrieved spectral indices $\beta$ are related to the 2$^{\rm nd}$ order structure function scaling exponent $\gamma_2$ via Eq. \ref{eq:5}. However, as mentioned in Section \ref{subsec:structure_function}, the rms velocity is generally not a full representation of the fluctuations field in a fluid due to its multi-component nature triggered by several energy dissipation and injection mechanisms. 

\cite{brunt+2003c} demonstrated that Eq. \ref{eq:9} is a good link between $\alpha$ and $\gamma_2$, which holds even for orders $p \neq 2$ as long as the velocity field of the region is non-intermittent and therefore reproducible by a single $\gamma$. For completeness of details, it is straightforward to show from Eq. \ref{eq:6b} that $\gamma_p$ is constant ($ \gamma_p = \gamma_2 \equiv \gamma$) if and only if $\zeta_p$ depends linearly on $p$ ($\zeta_p = p\gamma$). This is not the case for intermittent velocity fluctuations, in which the exponent $\gamma_p$ is no longer constant but depends on the order $p$ of the function. Furthermore, they concluded on intermittent fields that the PCA-derived exponent $\alpha$, after calibration, is better correlated with structure function exponents of orders $p=1/2,1$, which means that the translation given by the combination of Eqs. \ref{eq:5} and \ref{eq:9} is preferentially a measure of the intrinsic scaling exponent $\gamma_p$ of low-order rather than rms velocity fluctuations. Qualitatively, fields that are intermittent exhibit extreme differences in the magnitude of velocity fluctuations and/or density distribution over a given spatial scale \citep{brunt+2003c}, which is why multiple structure functions would be needed to fully describe such a scenario.

\subsubsection{Universality of Turbulence in the Molecular ISM} \label{subsec:pca_universality}

Based on earlier works, \cite{larson+1981} collated three-dimensional rms velocity dispersions along with projected sizes $L$ (in pc) of varied molecular associations and derived the relation $\Delta\upsilon=1.10L^{0.38}$ for 0.1$\lesssim L \lesssim$100\,pc. This can even be extended up to larger-scale $\sim$1000\,pc interstellar motions \citep{larson+1979}. He found complementary relationships that connect mass and density to the dynamics of molecular clouds. Similarly, but from a more homogeneous sample, \cite{solomon+1987} found the relation $\Delta\upsilon=(1.0\pm0.1)L^{0.5\pm0.05}$ \citep[see also][for a compilation of data covering five orders of magnitude in spatial scale]{falgarone+2009}. For the velocity dispersion calculation, they extracted centroid and line-width based velocity differences and added them in quadrature. 

These relationships describe multiple evaluations of the structure function in particular cases $S_2(l=L)$, which, when combined, turn out to follow the same functional form as if they were part of a single fluid, despite the fact that most of the objects analysed were born far from each other and do not interact. 

\cite{heyer+2004} applied PCA on individual molecular clouds and performed Monte Carlo simulations to prove that this resemblance is a consequence of the turbulence universality and self-similarity over different scales of the molecular ISM in the Galaxy, which hints at common formation mechanisms for molecular clouds. For this reason, let us adopt the following 1-to-1 translations from Larson-like to structure function parameters,
\begin{equation} \label{eq:10}
\begin{aligned}
    &C \approx \upsilon_0 \\
    &\Gamma \approx \gamma_2. 
\end{aligned}
\end{equation}

For standardisation purposes, we use this turbulence universality condition and the 2$^{\rm nd}$ order calibrations (Eqs. \ref{eq:5} and \ref{eq:9}) to translate, respectively, Larson-like exponents $\Gamma$ and PCA-derived exponents $\alpha$ to rms scaling exponents $\gamma_2$, which we will simply call $\gamma$ from now on.

\subsubsection{Optical Depth Effects on PCA Pseudo-Structure Functions}
\label{subsec:pca_opticaldepth}

\cite{larson+1981} considered optically thin regions, mostly traced by $^{13}$CO, H$_2$CO and NH$_3$, and only two \twCO{} optically thick regions, where the large-scale velocity variations were dominant (compared to the smaller-scale fluctuations derived from linewidths) in order to avoid any effects from line saturation. However, numerical and observational studies have concluded that PCA-derived scales are nearly insensitive to the opacity regime of the emission \citep{heyer+1997, brunt+2003b, brunt+2004, heyer+2006, brunt+2009, roman-duval+2011}. This is a strong point of the method as it allows us to study a broad range of objects/scales traced by opaque emission. \cite{brunt+2013} suggest that this is a consequence of the centroid velocity not being affected by saturation in the optically thick regime as long as it is symmetric to the line central frequency \citep[see also the discussion in][]{bertram+2014}. 

In Section \ref{subsec:results_orientation}, we discuss optical depth effects by comparing the retrieved scaling parameters from cloud complexes at different orientations.
\section{General Methodology} \label{sec:methodology}
 
We implement three PCA extraction methods to investigate the response of the retrieved pseudo-structure functions to different analysis scales:

\begin{itemize}
    \item The Mixed method: This method consists of deriving PCA scales ($l$, $\dv$) from small (30\,pc) cloud portions\footnote{This size is typical of giant molecular clouds, used as individual objects for PCA in, e.g., \cite{heyer+2004}.}, which are semi-automatically extracted from the simulated cloud complexes ($\sim$200\,pc). The resulting scales are then combined to construct a single fit equation (often called ``Mixed fit'') which represents the pseudo-structure function of the complex. This is the method that we use the most throughout the work. 
    \item The Complex method: In this method, we analyse the cloud complexes as they are, without any sub-portioning. 
    \item The Individual Cloud method: For this method, we do not combine the PCA-derived scales from the individual (30\,pc) portions but rather use them to compute their pseudo-structure functions separately, as individual objects. This will allow us to do further statistics, especially to analyse cloud environmental effects on the retrieved scaling parameters. Henceforth, we refer to individual portions interchangeably as ``individual clouds'' or just ``clouds''.
\end{itemize}

The Complex method is straightforward to implement. We take the intensity cubes obtained from the radiative transfer calculations and pass them directly through the \textsc{turbustat.pca} module to compute spatial ($l$) and velocity scales ($\dv$) from our cloud complexes, as single objects. The Mixed and Individual Cloud methods require, however, further steps that we explain below in detail.

First, we determine zeroth moment maps using the cloud complex intensity cubes. The zeroth moment is defined as the integrated intensity along the velocity (or frequency) axis $\psi$ as follows,
\begin{equation} 
\mathrm{M_0}(x,y)=\int_\psi T(x,y; \upsilon) \mathrm{d\upsilon}, \label{eq:11}
\end{equation}

\noindent where $x$ and $y$ refer to the pixel location in the synthetic cube. 

Then, we use the \astrodendro{}\footnote{\url{http://www.dendrograms.org}} package with each zeroth moment map as input to compute a hierarchical tree, called a dendrogram, which divides the map into closed subregions denoted by equal-valued pixels.

\begin{figure}
	\includegraphics[width=1.0\columnwidth]{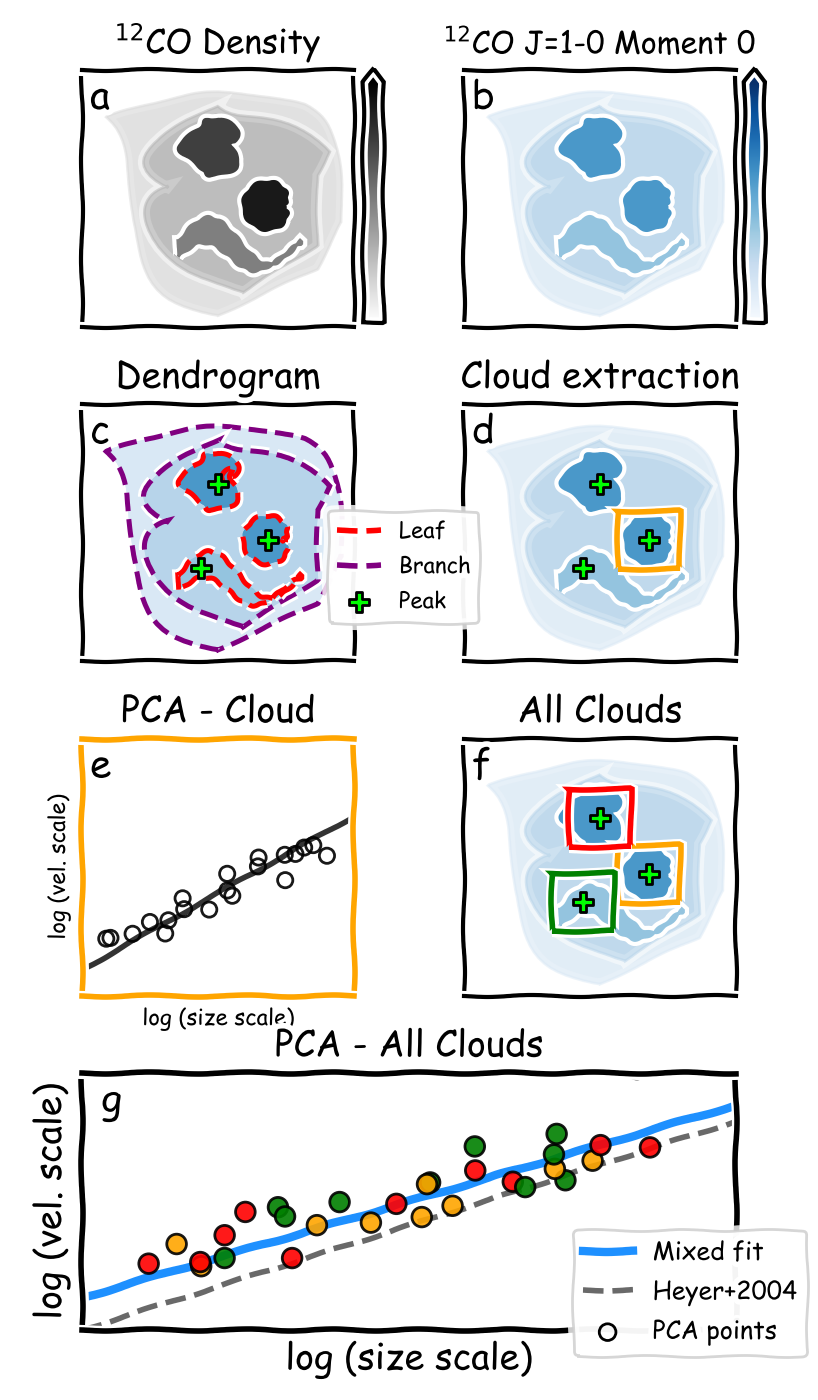}
    \caption{Basic workflow to retrieve PCA pseudo-structure functions of cloud complexes formed in our Cloud Factory simulation suite. This figure illustrates the Mixed PCA extraction method.}
    \label{fig:sketch}
\end{figure}

From the computed dendrograms, we focus on extracting the innermost, irreducible, substructures of the region, called ``leaves'', which are usually within larger-scale ``branches'' that may also be part of more extended structures (see panel c in Fig. \ref{fig:sketch}). To construct the dendrograms we set the three main parameters, $\rm min\_value=0.1\,\overline{\rm M}_0$, $\rm min\_delta=\overline{\rm M}_0$ and $\rm min\_npix$ between 100 and 400, where $\overline{\rm M}_0$ is the mean value of the zeroth moment. For reference, typical values for complex D are $\rm min\_value=9.3$, $\rm min\_delta=92.9$ and $\rm min\_npix=300$. Next, we find peak pixels in 
``leaves'' and centre (30\,pc)$^2$ portions on them which we use afterwards to slice the original \twCO{} cubes. Ideally, these portions (or individual clouds) should cover the cloud complex moment map as much as possible so that the least intensity is left out of the analysis, regardless of the velocity channel. In order to achieve this, we set the dendrogram parameters in such a way that the ``leaves'' are enough in number and sufficiently separated from each other. Finally, PCA is computed on individual clouds to construct the Mixed and Individual Cloud fits. Figure \ref{fig:sketch} is a schematic representation of this procedure.  

For the fitting process we randomise the PCA-derived scales ($l$, $\dv$) $N$ times using their uncertainties ($\sigma_l, \sigma_{\dv}$) as Gaussian standard deviations. We run 1000 random realisations for the Mixed and Complex fits and 200 for each Individual Cloud fit. The reported fit consists of the mean scaling parameters ($\vz$, $\alpha$) obtained from the $N$ random fits, and the errors correspond to the standard deviations ($\sigma_{\vz}, \sigma_{\alpha}$) from the mean values. We also report the associated coefficient of determination, $R^2$, and the reduced chi-square, $\chi^2$, to evaluate the goodness of the fits. This is all recorded in the supplementary database provided with the manuscript.

Despite the fact that we assign a single power-law structure function to each cloud complex, all of them can be seen as a composition of different power-laws, each of which is associated with a different portion of the complex (one per colour in Fig. \ref{fig:pca_fits_faceon}). The structure functions extracted from individual clouds (e.g. Fig. \ref{fig:time_complexB}, bottom row) are all collected in the Supplementary file. The dispersion of individual cloud scaling parameters is related to the level of intermittency of the hosting complex (see Sec. \ref{subsec:pca_calibration}).

This pipeline and analysis tools are all built-in and executed by our \pcafactory{} package.
\section{results} \label{sec:results}

In this section we present the Principal Component Analysis (PCA) of a set of cloud complexes extracted from our Cloud Factory simulation suite using \twCOfull{} intensity PPV cubes. We report the impact on velocity structure functions after adding physical mechanisms such as gas self-gravity or supernova feedback on star-forming sites (Section \ref{subsec:results_physicalscenario}). Likewise, we present variations driven by different line-of-sight projections (Section \ref{subsec:results_orientation}) and time snapshots (Section \ref{subsec:results_time}). Lastly, we investigate cloud environmental effects and explore variations in velocity fluctuations when applying PCA locally on individual molecular clouds or globally on full cloud complexes (Section \ref{subsec:results_inner}). We study six cloud complexes (two per physical scenario) as summarised in Table \ref{table:cloudcomplexes}, but, since we consider (three) different lines-of-sight and (two/three) time snapshots, there are in practice 42 different objects: 12 for the potential-dominated scenario with no self-gravity; 18 for the potential-dominated scenario with self-gravity, for which we extracted an extra time snapshot to study the time evolution of clouds under local gravitational effects (see Sec. \ref{subsec:results_time}); and 12 for the feedback-dominated scenario.
 
For reasons of space, the corresponding PCA figures and line emission profiles from all the cloud complex configurations can be found in the supplementary file provided along with this manuscript\footnote{\label{noteweb}\url{https://github.com/andizq/andizq.github.io/tree/master/pcafactory-data}}. Additionally, the whole catalog of ($l$, $\dv$) and ($\upsilon_0$, $\alpha$) pairs extracted from individual clouds and cloud complexes is also available online. The software developed to carry out this work, the \textsc{pcafactory}, is open source and available on GitHub\footnote{\url{https://github.com/andizq/pcafactory}}.

\subsection{Structure Function Dependence on the Physical Scenario} \label{subsec:results_physicalscenario}

\begin{figure}
	\includegraphics[width=\columnwidth]{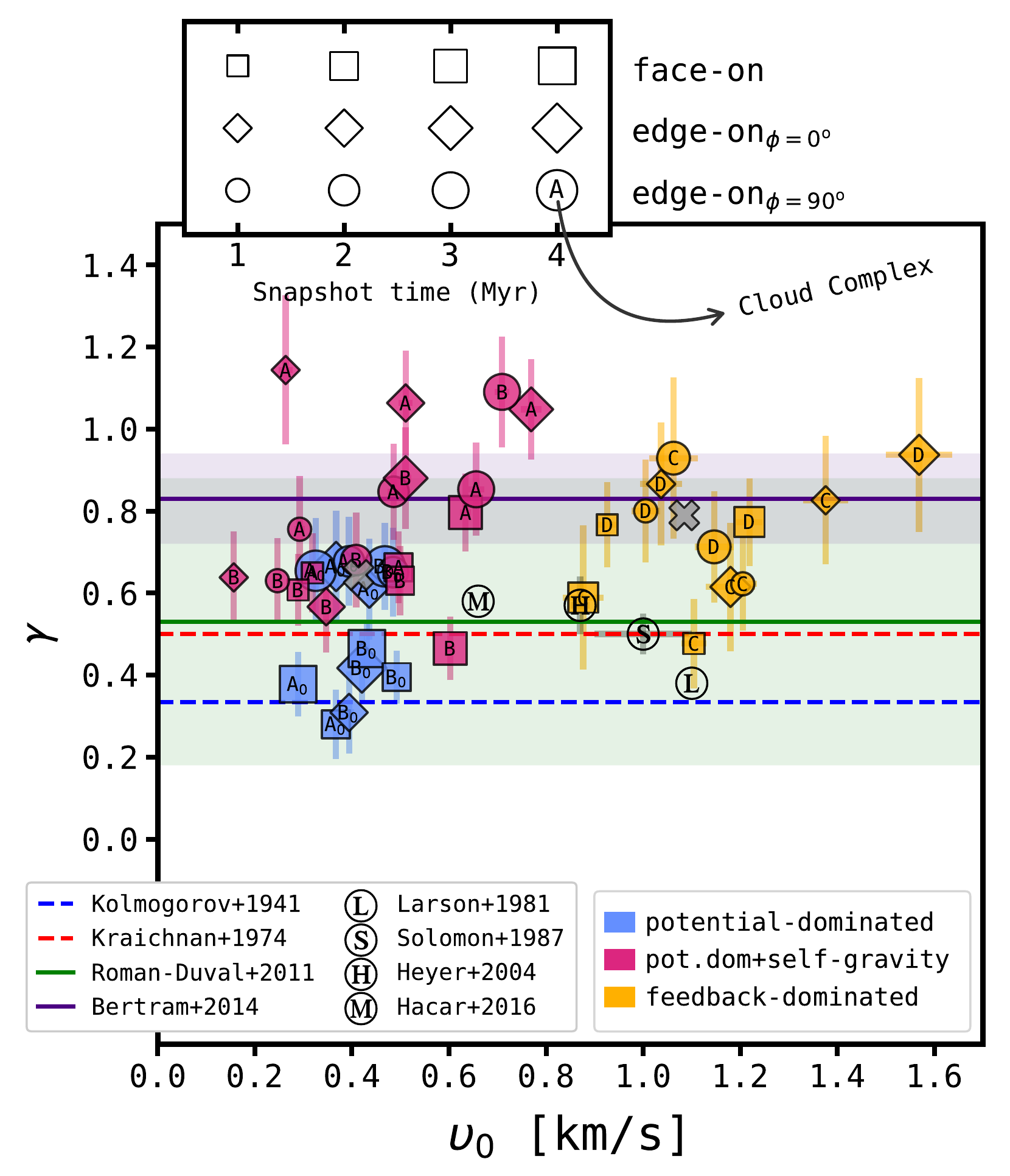}
    \caption{Structure function parameters ($\vz$, $\gamma$) of cloud complexes extracted from different physical scenarios (marker colour), orientations (marker style) and time snapshot (marker size) after tracer refinement has commenced. The parameters are derived from PCA pseudo-structure functions using the Mixed method (see Sec. \ref{sec:methodology} and Fig. \ref{fig:sketch}). Grey crosses are the best cluster centres obtained from a K-means clustering analysis. Empty circles and horizontal lines, with errors as shades, are selected parameters from previous literature (see Table \ref{table:references}).}
    \label{fig:pca_summary}
\end{figure}

In Figure \ref{fig:pca_summary} we present the structure function scaling parameters ($\vz, \gamma$) derived for all our cloud complexes. The scaling parameters were retrieved using the Mixed PCA extraction method (see Sec. \ref{sec:methodology}), which combines PCA-derived scales ($l$, $\dv$) of smaller cloud portions to find the associated pseudo-structure function of each cloud complex. 
The translation from PCA-derived scaling exponents ($\alpha$) into the corresponding structure function exponents ($\gamma$) assumes the 2$^{\rm nd}$ order calibrations given by Eqs. \ref{eq:5} and \ref{eq:9}. The ($\vz$, $\gamma$) pairs are computed for different line-of-sight projections and time snapshots for each cloud complex. For comparison, we include representative observational and synthetic values reported in previous literature as well as theoretical regimes for subsonic and supersonic turbulence (see Table \ref{table:references}). 

The most obvious aspect is that the scaling parameters agglomerate in different zones according to the physical mechanisms governing the cloud complexes. To assess this, we perform a K-means clustering analysis along with a Silhouette Coefficient score test which suggests that the parameter distribution is better represented by two clusters (score\,$=$\,0.56), followed by three clusters (score\,$=$\,0.51). The cluster centres are also shown in Fig. \ref{fig:pca_summary}. Overall, the structure function scaling exponents ($\gamma$) are similar regardless of the physical scenario and lie mostly around the range of values obtained by \citetalias{roman-duval+2011} and \cite{bertram+2014}, suggesting that the structure exponent alone is not sufficient to fully determine the physical nature of cloud complexes. Scaling coefficients ($\vz$) must also be taken into account to effectively distinguish between clouds.   

We summarise in Table \ref{table:mean_pars_pdf} the mean and standard deviations of scaling parameters derived from all the cloud complex configurations, using three PCA extraction methods detailed in Section \ref{sec:methodology}. For the potential-dominated case with no self-gravity, the mean scaling exponent $\mean{\gamma}=0.51\pm0.15$ corresponds to that of the classical Burger-like turbulence ($\gamma=1/2$). Turning on self-gravity increases scaling exponents to $\mean{\gamma}=0.78\pm0.19$ while maintaining low scaling coefficients $\mean{\vz}=0.46\pm0.17$\,km\,s$^{-1}$. On the other hand, feedback-dominated complexes show similar exponents but suffer a substantial increase in scaling coefficients to $\mean{\vz}=1.14\pm0.18$\,km\,s$^{-1}$. \footnote{We warn the reader that the reported $\gamma$ should be used with caution as it more reliably traces low-order fluctuations (rather than rms velocities) when the region is intermittent (see Sec. \ref{subsec:pca_calibration}).} Additionally, we show in Table \ref{table:12co_vs_13co} a comparison between PCA parameters derived from $^{12}$CO and $^{13}$CO J$=$1$-$0 cubes for some cloud complexes and orientations. We assume that the $^{13}$CO abundance is a factor of 69 lower than that of \twCO{} \citep[see][]{visser+2009}. Despite the contrast in opacity between the two tracers, no significant differences are found for scaling exponents. This indicates that both tracers are valid descriptors of cloud kinematics as long as velocity fluctuations are supersonic (generally satisfied), which softens opacity effects. We find that even in edge-on orientations, line emission profiles from small cloud portions can span over a wide range of velocities and exhibit substructure suggesting that the emission can still arise from different depths in the cloud (see e.g. Fig. \ref{fig:line_profiles}). Scaling coefficients, however, decrease systematically in the case of $^{13}$CO, which we attribute to the fact that $^{13}$CO lines are less extended in velocity channels and hence the retrieved fluctuations decrease for all size scales. 

\begin{figure*}
\begin{overpic}[width=0.94\textwidth]{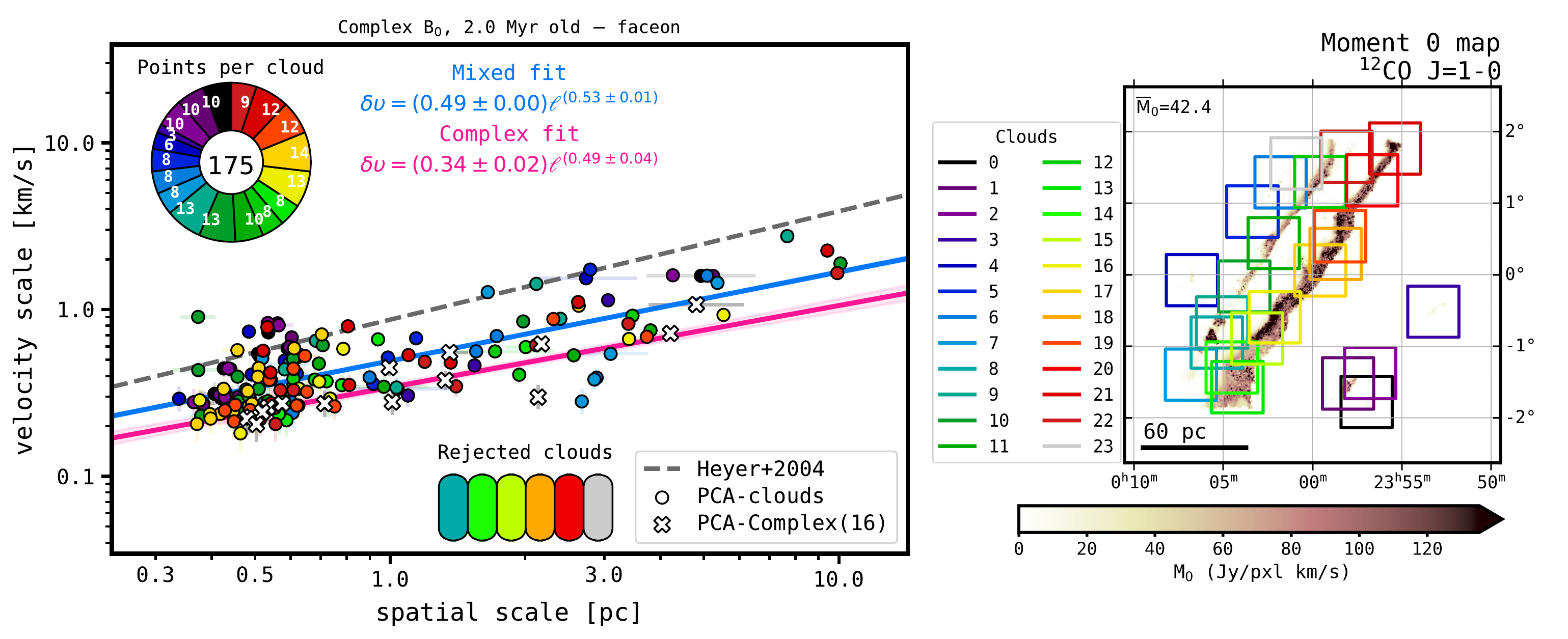}
\end{overpic} \\
\begin{overpic}[width=0.94\textwidth]{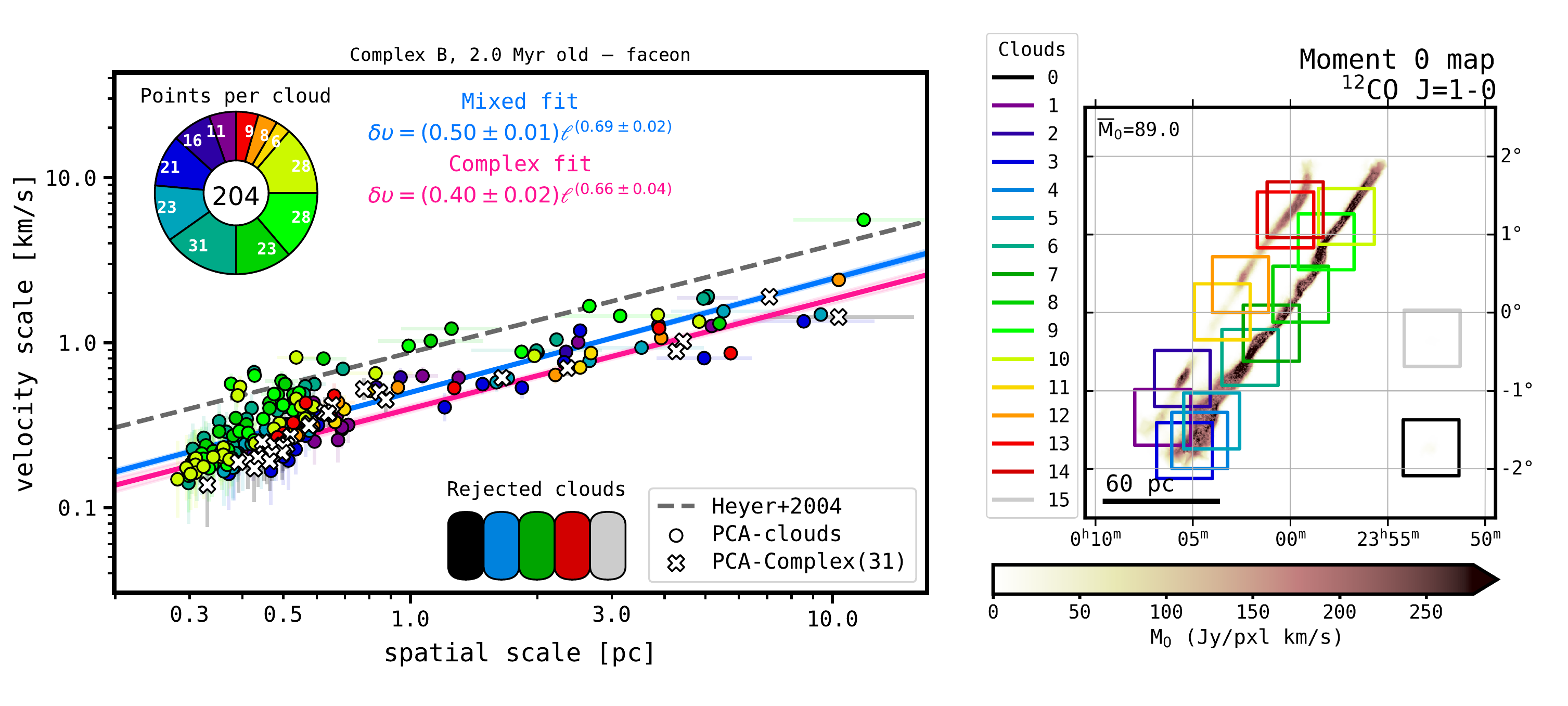}
\end{overpic} \\
\begin{overpic}[width=0.94\textwidth]{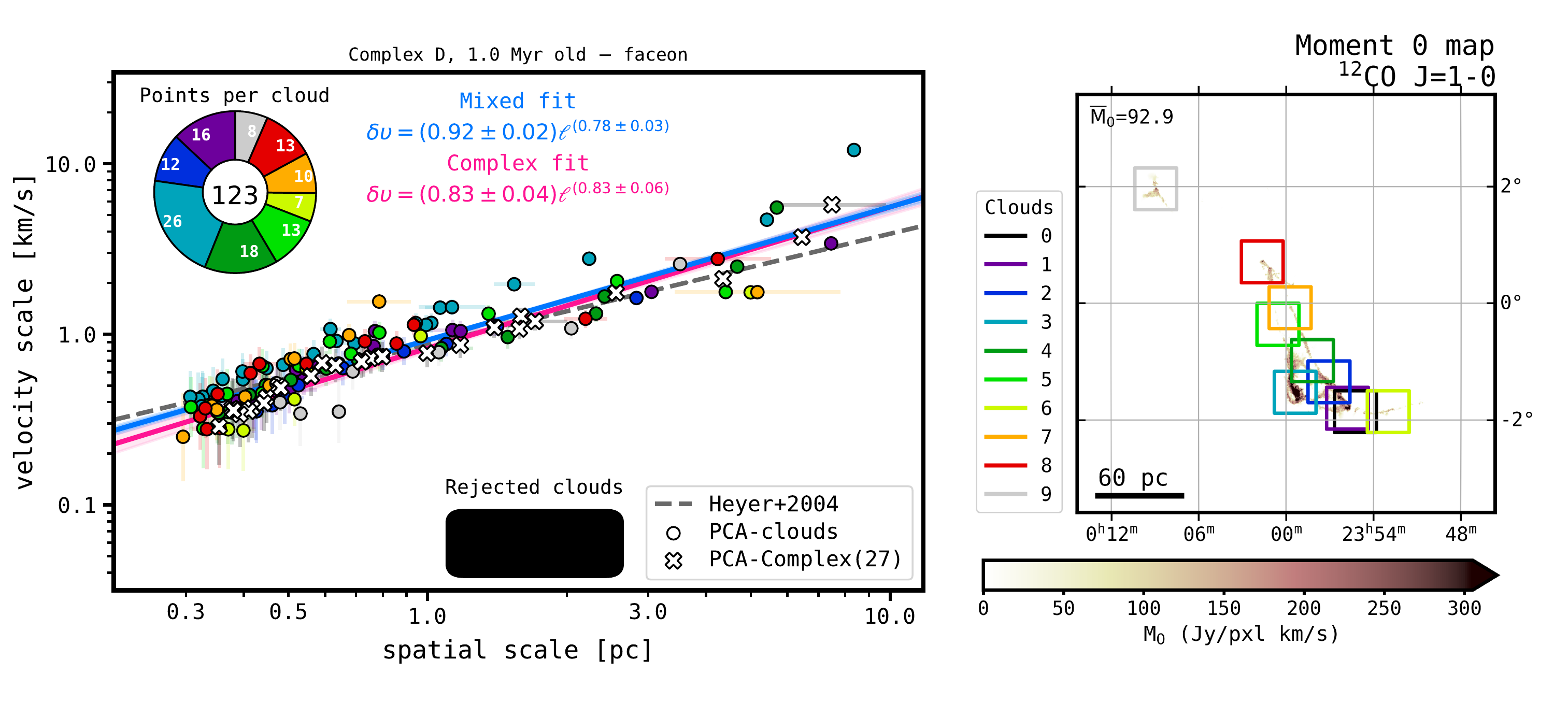}
\end{overpic}

\caption{\textit{Left column}: Velocity structure function from the face-on view of cloud complexes B$_0$ (top), B (middle) and D (bottom) at selected time snapshots. The coloured circles are the PCA-derived scales ($l,\dv$) from the same-coloured (30\,pc)$^2$ clouds on the right panel. The white crosses are the PCA-derived scales from the cloud complex as a whole. The pie chart indicates the number of scales extracted from each cloud and the total number of scales in the middle. Overlapping cloud portions are rejected and shown in colours at the bottom of the left panels. For reference, the dashed line is the PCA-derived best fit found by Heyer \& Brunt (2004) from an ensemble of 27 molecular clouds. The blue and fuchsia lines are, respectively, the fit from considering all the points from individual clouds simultaneously (Mixed method) and the fit to the points from the cloud complex as a whole (Complex method). \textit{Right column}: Cloud portions extracted for PCA, overlaid on the synthetic \twCOfull{} zeroth moment maps of the cloud complexes.} 
\label{fig:pca_fits_faceon}

\end{figure*}

For illustration, in Figure \ref{fig:pca_fits_faceon} we show the PCA relationships derived for three cloud complexes, B$_0$, B and D, each from a different physical scenario. All of them are seen face-on and extracted at similar time steps after starting tracer refinement so that the effects of varying physical scenario prevail. The PCA already hints at differences in pseudo-structure functions ($\dv$) from case to case, increasing both the scaling exponent ($\gamma$) and coefficient ($\vz$) as we add more physical processes in the simulation. However, if we look at the full set of results for $\gamma$ shown in Figure \ref{fig:pca_summary}, which includes values for the other cloud complexes and other projections, we see that while there is still a clear sign that including self-gravity increases $\gamma$, the difference between the feedback- and (both) potential-dominated cases is less clear.

The coloured circles in the left panel of Figure \ref{fig:pca_fits_faceon} are the PCA-derived scales ($l$, $\dv$) from selected cloud portions correspondingly illustrated as coloured squares in the right panel, overlaid on the \twCOfull{} zeroth moment map of the cloud complex. The crosses in the left panel are the PCA-derived scales from the cloud complex as a whole. The pie chart shows the number of PCA scales extracted from each cloud portion with the total net scales in the middle. Yet, a cloud portion is subject to rejection so long as (a) the centre of its associated window is enclosed by another window or (b) it has no scales retrievable by the PCA. This is shown by the coloured round boxes at the bottom of the left panel. The blue line is the associated fit to the combined points from cloud portions (Mixed method) and the pink line is the fit to the points derived from the cloud complex as a whole (Complex method). The blue and pink shaded regions are the uncertainties of these fits (see details in Sec. \ref{sec:methodology}). For reference, the dashed gray line is the best fit found by \cite{heyer+2004} using PCA-derived scales from 27 molecular clouds, which, in this case, most closely resembles the feedback-dominated cloud complex D. 
 
Also, note that the number of PCA points is more sensitive to higher zeroth moment values than to larger \twCO{} flux extents, as one might think at first. This is because the zeroth moment is related to intensity variance along the velocity axis. If density fluctuations and velocity intermittence in the region are low, a high-valued pixel in zeroth moment spans into a wide range of values in the covariance matrix of the line cube. Thus, it is equivalent to the amount of information that principal components can keep and hence the PCA algorithm will raise more ($l$, $\dv$) pairs. Furthermore, the number of pairs usually depends on the characteristic scaling exponent of the region. There are fewer pairs (per number of cloud portions) for cloud complexes with lower exponents \citep[see also][]{brunt+2002}, which is in turn related to the diagonality of the covariance matrix, driven by a typically higher degree of line-centroid variation in higher exponent velocity fields \citep[see e.g. Fig. 1 of][]{brunt+2013}.

\setlength{\tabcolsep}{10.5pt} %pad between columns

\begin{table*}
\centering
{\renewcommand{\arraystretch}{1.5}%pad between rows
\begin{tabular}{ lcccc } 

\toprule
\toprule
PCA Extraction Method & Physical Scenario & Cloud complex & $\mean{\upsilon}_0$ (km\,s$^{-1}$) & $\mean{\gamma}$ \\
\midrule

\multirow{3}{*}{Mixed} & a & A$_0$ \& B$_0$ & $0.41\pm0.06$ & $0.51\pm0.15$ \\ 
& b & A \& B & $0.46\pm0.17$ & $0.78\pm0.19$ \\
& c & C \& D & $1.14\pm0.18$ & $0.74\pm0.14$ \\
\midrule

\multirow{3}{*}{Complex} & a & A$_0$ \& B$_0$ & $0.31\pm0.05$ & $0.48\pm0.15$ \\ 
& b & A \& B & $0.36\pm0.14$ & $0.74\pm0.19$ \\
& c & C \& D & $1.13\pm0.20$ & $1.02\pm0.18$ \\
\midrule

\multirow{6}{*}{Individual Clouds} & a & A$_0$ & $0.36\pm0.09$ & $0.64\pm0.26$ \\ 
& a & B$_0$ & $0.48\pm0.11$ & $0.64\pm0.31$ \\ 

& b & A & $0.49\pm0.20$ & $1.01\pm0.26$ \\
& b & B & $0.45\pm0.22$ & $0.82\pm0.35$ \\

& c & C & $1.37\pm0.55$ & $0.84\pm0.48$ \\
& c & D & $1.20\pm0.46$ & $0.85\pm0.29$ \\

\bottomrule
\bottomrule

\end{tabular}
 \caption{Mean and standard deviation of structure function parameters derived from different PCA extraction methods. The mean is taken over all time snapshots and lines-of-sight. Conventions: a. Potential-dominated; b. Potential-dominated with self-gravity; c. Feedback-dominated.}
  \label{table:mean_pars_pdf}
  }
\end{table*}

\setlength{\tabcolsep}{12.5pt} %pad between columns

\begin{table*}
\centering
{\renewcommand{\arraystretch}{1.5}%pad between rows
\begin{tabular}{ |ccc|cc|cc| } 

\toprule
\toprule
Complex & Time step (Myr) & Orientation & \multicolumn{2}{c|}{$^{12}$CO} & \multicolumn{2}{c|}{$^{13}$CO} \\
& & & {$\upsilon_0$  (km s$^{-1}$)} & {$\alpha_{\rm PCA}$} & {$\upsilon_0$  (km s$^{-1}$)} & {$\alpha_{\rm PCA}$} \\
\midrule

A$_0$ & 2 & face-on & $0.30\pm0.01$ & $0.66\pm0.03$ & $0.16\pm0.01$ & $0.57\pm0.05$ \\ 
A$_0$ & 2 & \edgeonphi{} & $0.49\pm0.01$ & $0.69\pm0.03$ & $0.27\pm0.01$ & $0.79\pm0.05$ \\
\midrule
B & 3 & face-on & $0.77\pm0.01$ & $0.61\pm0.02$ & $0.60\pm0.02$ & $0.73\pm0.03$ \\ 
B & 3 & \edgeonphi{} & $0.88\pm0.03$ & $0.87\pm0.05$ & $0.49\pm0.02$ & $0.93\pm0.05$ \\
\midrule
C & 1 & face-on & $0.83\pm0.03$ & $0.75\pm0.07$ & $0.59\pm0.03$ & $0.74\pm0.09$ \\ 
C & 1 & \edgeonphi{} & $0.91\pm0.04$ & $0.94\pm0.09$ & $0.79\pm0.04$ & $0.88\pm0.10$ \\
D & 1 & face-on & $0.79\pm0.02$ & $0.88\pm0.04$ & $0.79\pm0.03$ & $0.99\pm0.05$ \\ 
D & 1 & \edgeonphi{} & $1.04\pm0.04$ & $0.92\pm0.06$ & $0.87\pm0.04$ & $0.90\pm0.07$ \\

\bottomrule
\bottomrule

\end{tabular}
 \caption{Comparison between PCA-derived scaling parameters from LVG $^{12}$CO and $^{13}$CO J$=$1$-$0 cubes for different cloud complexes and orientations, using the Mixed extraction method. We also show the distribution of parameters for both isotopologues in the LVG panel of Fig. \ref{fig:pca_summary_distribution_LTE}. 
 }
  \label{table:12co_vs_13co}
  }
\end{table*}

On the other hand, the fact that the velocity scaling coefficients are generally low for complexes A and B, at all times, reflects a high level of velocity coherence due to weak stellar feedback in these regions. This favours the development of long filamentary structures present in both complexes (stretched out by differential rotation of the Galaxy), their preservation over time, and a sustained emergence of stellar systems represented by sink particles. This agrees with the weak feedback provided by randomly distributed supernovae, and also with the low velocity gradients in complex B and in long filaments of complex A reported in \citetalias{smith+2020}.

Conversely, clustered supernova feedback plays a significant role in taking cloud complexes out of that coherent state. As previously found in \citetalias{smith+2020}, supernovae tied to star formation sites make complexes C and D have shorter and less massive filaments. This type of feedback also induces stronger shear that reduces filament lifetimes and consequently also their star formation efficiencies. In this work this manifests as high scaling coefficients, which indicates a larger degree of velocity fluctuations in complexes C and D. Again, this is in good agreement with the physical mechanisms we know are governing the regions and suggests that the velocity scaling coefficient of the structure function can provide valuable information about the cold dense molecular ISM.
 
In summary,
\begin{itemize}
    \item Cloud complexes governed by different physical mechanisms produce separate clusters of points in the ($\vz$, $\gamma$) parameter space (see e.g. Fig. \ref{fig:pca_summary}). 
    \item The scaling coefficient $\vz$ is an excellent reference to distinguish between potential-dominated and feedback-dominated cloud complexes, whereas the role of $\gamma$ is only discernible when self-gravity is turned off (see Fig. \ref{fig:pca_summary} and Table \ref{table:mean_pars_pdf}).
    \item Low velocity scaling coefficients are associated with (quiescent) regions with coherent velocity fields, which favours the development and preservation of long filamentary structures. Conversely, strong stellar feedback takes cloud complexes out of their coherent state, producing smaller and short-lived structures.
\end{itemize}

\subsection{Structure Function: Cloud Orientation Effects} \label{subsec:results_orientation}

\begin{figure}
	\includegraphics[width=\columnwidth]{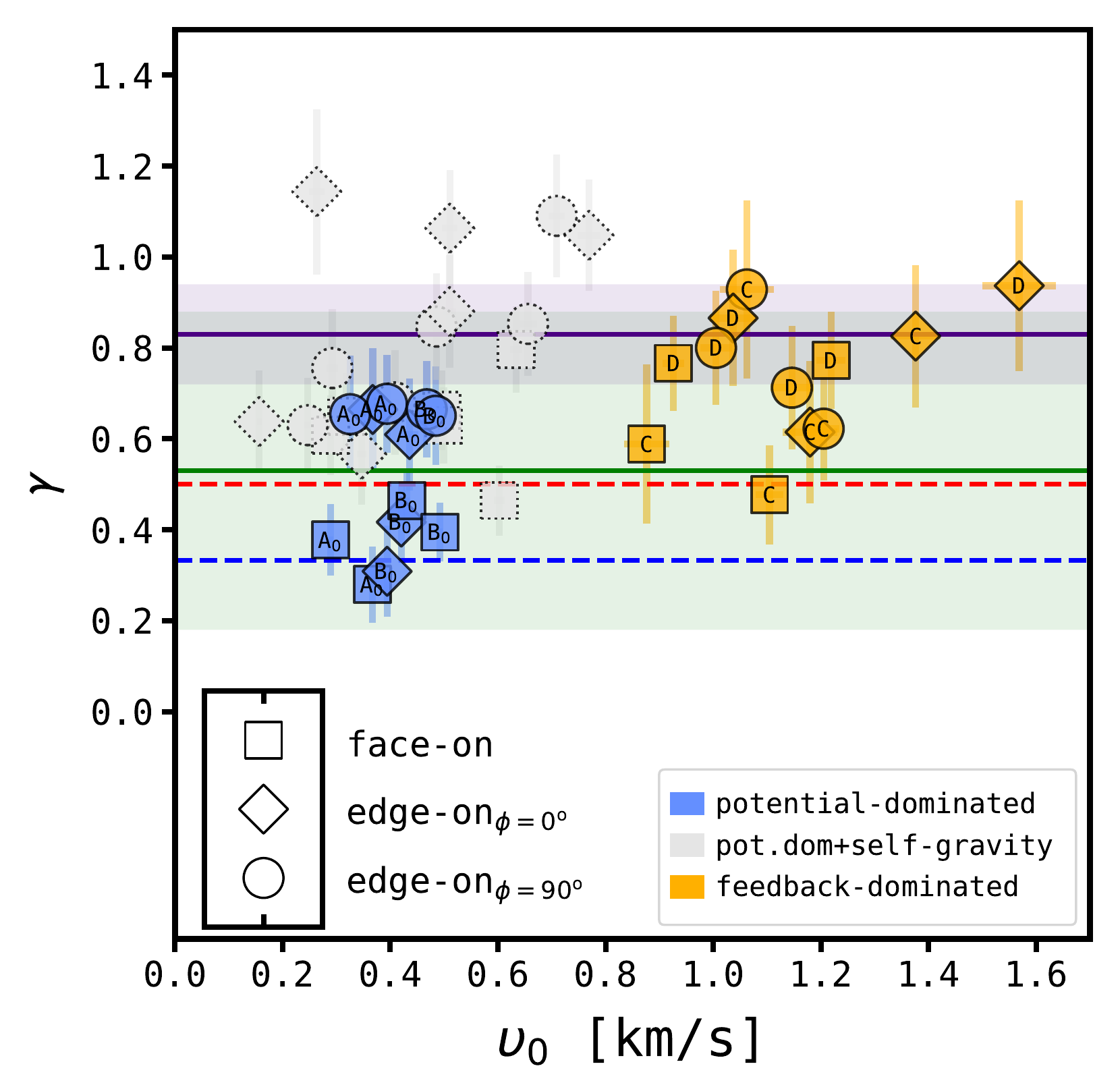}
    \caption{Structure function parameters ($\vz$, $\gamma$) of cloud complexes extracted from different physical scenarios (marker colour) and orientations (marker style) using the Mixed method. Solid lines (and their errors as shades) are reference literature values whose legends can be found in Table \ref{table:references}. Markers in gray are potential-dominated self-gravitating complexes studied separately in Sec. \ref{subsec:results_time} as a function of time.}
    \label{fig:pca_summary_los}
\end{figure}

We also vary the line-of-sight projection of cloud complexes during the radiative transfer calculations to explore the influence of density/velocity anisotropies and optical depth effects on the retrieved structure functions. As shown by the different markers in Figure \ref{fig:pca_summary_los}, there are no systematic variations in scaling parameters when changing the orientation of cloud complexes; they rather depend on the particular geometry of each complex and hence on their evolution over time.

Edge-on projections of cloud complexes with low scale-heights (A$_0$, B$_0$, A and B) exhibit flat and continuous CO distributions but actually consist of both nearby and distant structures that comprise each cloud complex as a whole (see Figs. \ref{fig:column_densities} and \ref{fig:appendix_column_densities}). Background emission is hence susceptible to being blocked by foreground gas with similar line-of-sight velocities if optically thick tracers are used as in our case with \twCO{}. However, opacity effects are softened by the fact that velocity fluctuations are generally supersonic and, in consequence, part of the background emission can still reach the observer.

Projection effects for A$_0$ and B$_0$, which are our least realistic cases, seem to be driven by column density variations. As shown in Figure \ref{fig:pca_summary_los}, complex A$_0$ yields comparable scaling exponents both for the \edgeon{} and \edgeonphi{} orientations ($\gamma \approx 0.6$ at any time) but lower values for the face-on views ($\gamma \approx 0.3$). The column densities are very high when this complex is viewed edge-on (either $\phi=0^{\circ}$ or $\phi=180^{\circ}$) compared to the face-on view ($\approx 3\times$ higher). 

Cloud complex B$_0$ does not follow the same pattern. In this case, variations induced by line-of-sight projections are smaller. This is in good agreement with our previous interpretation because gas column densities are this time nearly the same for all projections, with differences smaller than 30 per cent between face-on and edge-on column densities. 

When compared to PCA applied over cloud complexes as a whole, the variability of complex B$_0$ decreases even more (see Figure \ref{fig:pca_summary_cloud}). For B$_0$, all the scaling parameters lie around $\vz=0.35$ km s$^{-1}$ and $\gamma=0.4$ regardless of orientation. This suggests that variations in scaling parameters in this complex are due to individual cloud velocity fluctuations driven either by local collisions or random supernova feedback. On the other hand, cloud complex A$_0$ does maintain the same behaviour when studied as a whole, namely, the face-on orientations still yield the lowest scaling exponents at all times and the gap to edge-on values is roughly the same. Both results strengthen our argument that larger column densities yield steeper scaling exponents, and equivalently, isotropic column densities result in similar scaling parameters. We emphasise, however, that these are small variations that in no way resemble the variations induced by changing feedback conditions.

Complexes A and B show similar signatures but their parameters are strongly affected by local gravitational forces, displaying rather systematic variations over time. Hence, we prefer to leave this for Section \ref{subsec:results_time} where cloud complexes are analysed from their evolutionary context.

For feedback-dominated complexes, C and D, the \edgeon{} orientation (parallel to the $\phi-$axis) produces high scaling exponents and the highest scaling coefficients in most of the cases. This is because for both complexes the axes of filamentary structures are preferentially stretched out along the $\phi-$direction due to differential rotation of the Galaxy. This makes the \edgeon{} projection contain more gas mass within smaller projected scales than the other orientations. Additionally, the scale-height of these regions is naturally high due to nearby (in C) and internal (in D) energy feedback from supernova explosions, which enables more gas to contribute to the retrieved velocity fluctuations as optical depth effects are softened (see Figure \ref{fig:line_profiles_tau}).

In summary,
\begin{itemize}
    \item In potential-dominated complexes without self-gravity, higher column densities lead to steeper scaling exponents. For comparable column densities, variations in scaling parameters are smaller and driven by localised cloud-scale disturbances (e.g. cloud collisions, isolated supernovae). 
    \item In feedback-dominated complexes, where optical-depth effects are less prominent, projections with larger column densities yield higher scaling parameters. Especially for \edgeon{} lines-of-sight, which are parallel to the long axis of filamentary structures.
\end{itemize}

\subsection{Structure Function: Time Evolution} \label{subsec:results_time}

\begin{figure*}
\includegraphics[width=1\textwidth]{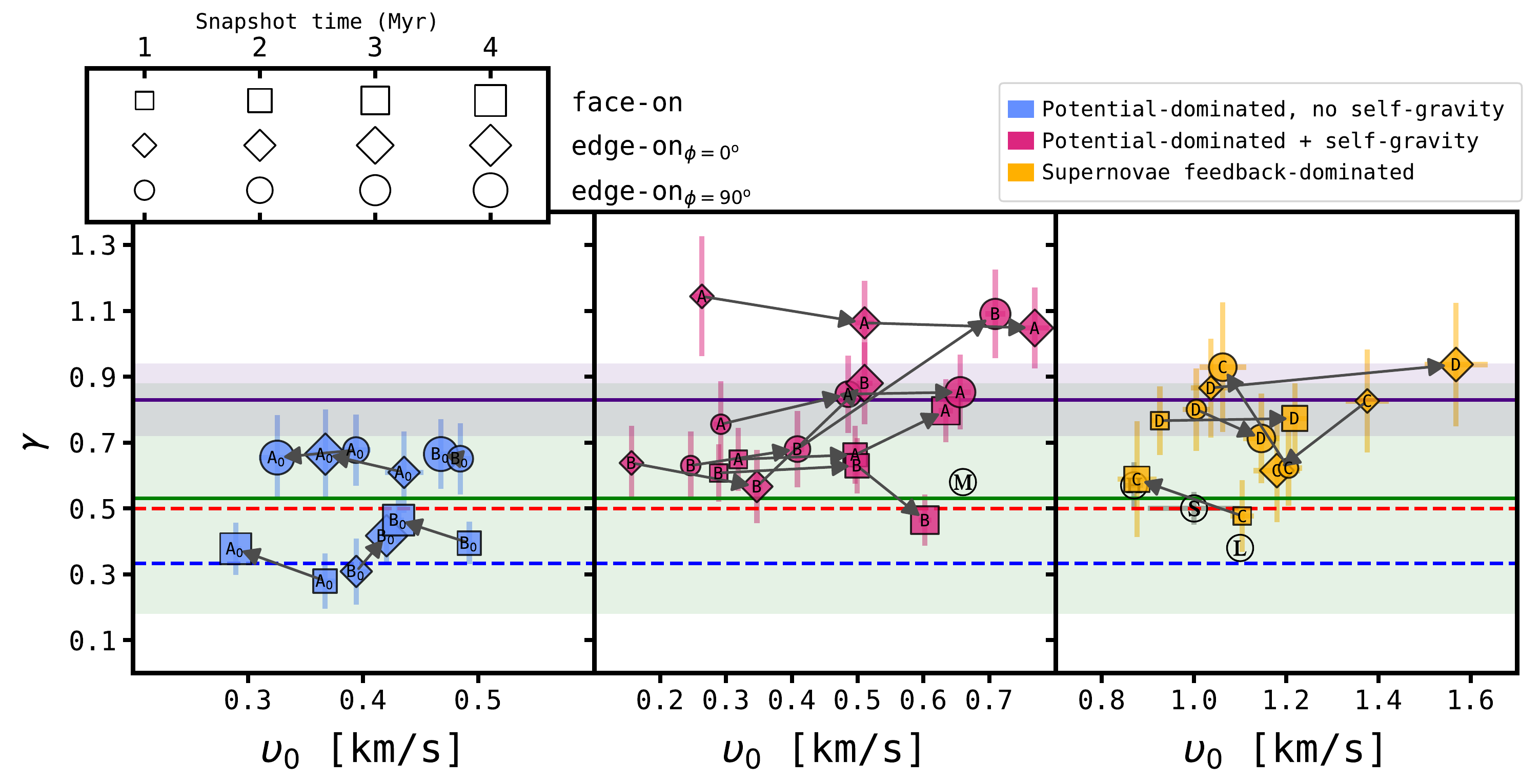}\\
\includegraphics[width=1\textwidth]{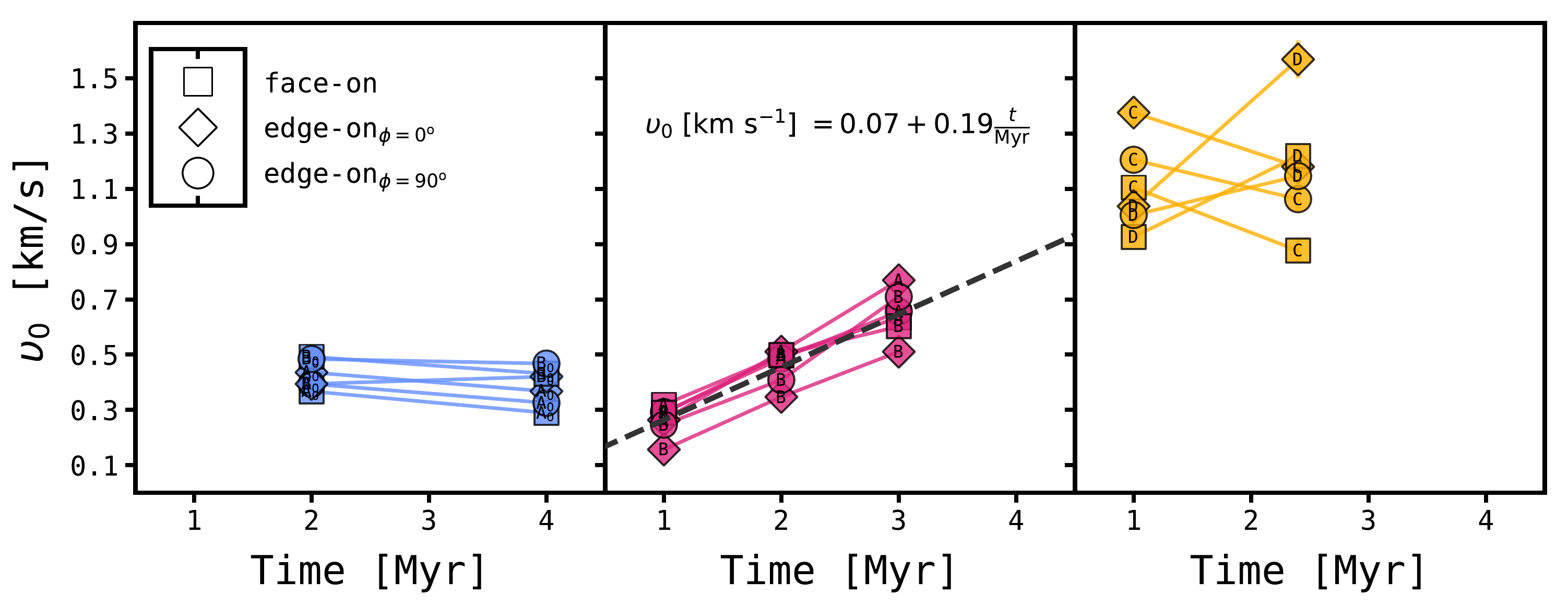}

\caption{Top row: structure function parameters split by physical scenario to illustrate variations over time. Arrows show the evolution of cloud complex parameters for a given line-of-sight. Literature values are shown for reference; marker codes and descriptions are summarised in Table \ref{table:references}. Bottom row: time evolution of scaling coefficients $\vz$ for each cloud complex and orientation. The dashed line in the middle panel is the linear best-fit obtained for complexes in the potential-dominated scenario with self-gravity.}
\label{fig:pca_time_evolution}
\end{figure*}

Figure \ref{fig:pca_time_evolution} follows the evolution of structure function parameters of cloud complexes for each physical scenario and orientation. 
Potential-dominated complexes with no self-gravity yield stable scaling exponents ($\gamma$) over time but present a systematic reduction of velocity scaling coefficients ($\vz$). This suggests a sustained lack of turbulence fueling sources to compensate the natural accelerated energy decay over length scales characteristic of Kolmogorov-like fluids \citep[see e.g.][]{onsager+1949}, which, on small scales, rapidly lose energy via viscous dissipation. Also, this indicates that in these complexes, driving of large-scale velocity fluctuations by random supernovae is unable to maintain the level of turbulence within the clouds (see also \citealp{ibanez-mejia+2017} and \citealp{seifried+2018} who find similar results).

Turning on self-gravity dramatically changes this behaviour. In this case, the scaling coefficient (or magnitude of velocity fluctuations, $\vz$) increases with time for complexes A and B. The gradient of this increment appears linear and is similar for both complexes, which allows us to capture the time evolution of velocity scaling coefficients with the following relationship:
\begin{align}
{\upsilon}_0(t) &= 0.07 + 0.19 \left(\frac{t}{\rm Myr}\right) \label{eq:12} \; [{\rm km \; s^{-1}}].
%\\ {\upsilon}_0(t) &= 0.05 + 0.16 \left(\frac{t}{\rm Myr}\right) \label{eq:13} \; [{\rm km \; s^{-1}}],
\end{align}
As expected, we also find a (nearly linear) sustained increase in mean surface densities over time. For instance, cloud complex B yields mean surface number densities $\Sigma_{\rm H} (t=1\,\textrm{Myr})= 7.65\times10^{19}$\,cm$^{-2}$, $\Sigma_{\rm H} (t=2\,\textrm{Myr})= 1.07\times10^{20}$\,cm$^{-2}$ and $\Sigma_{\rm H} (t=3\,\textrm{Myr})= 1.44\times10^{20}$\,cm$^{-2}$. This is in excellent agreement with the dependence of $\vz$ on the cloud surface density found by \cite{heyer+2009}, and suggests that the physics behind such a dependence is dominated by local gravitational interactions. 

\begin{figure*}
\includegraphics[width=\textwidth]{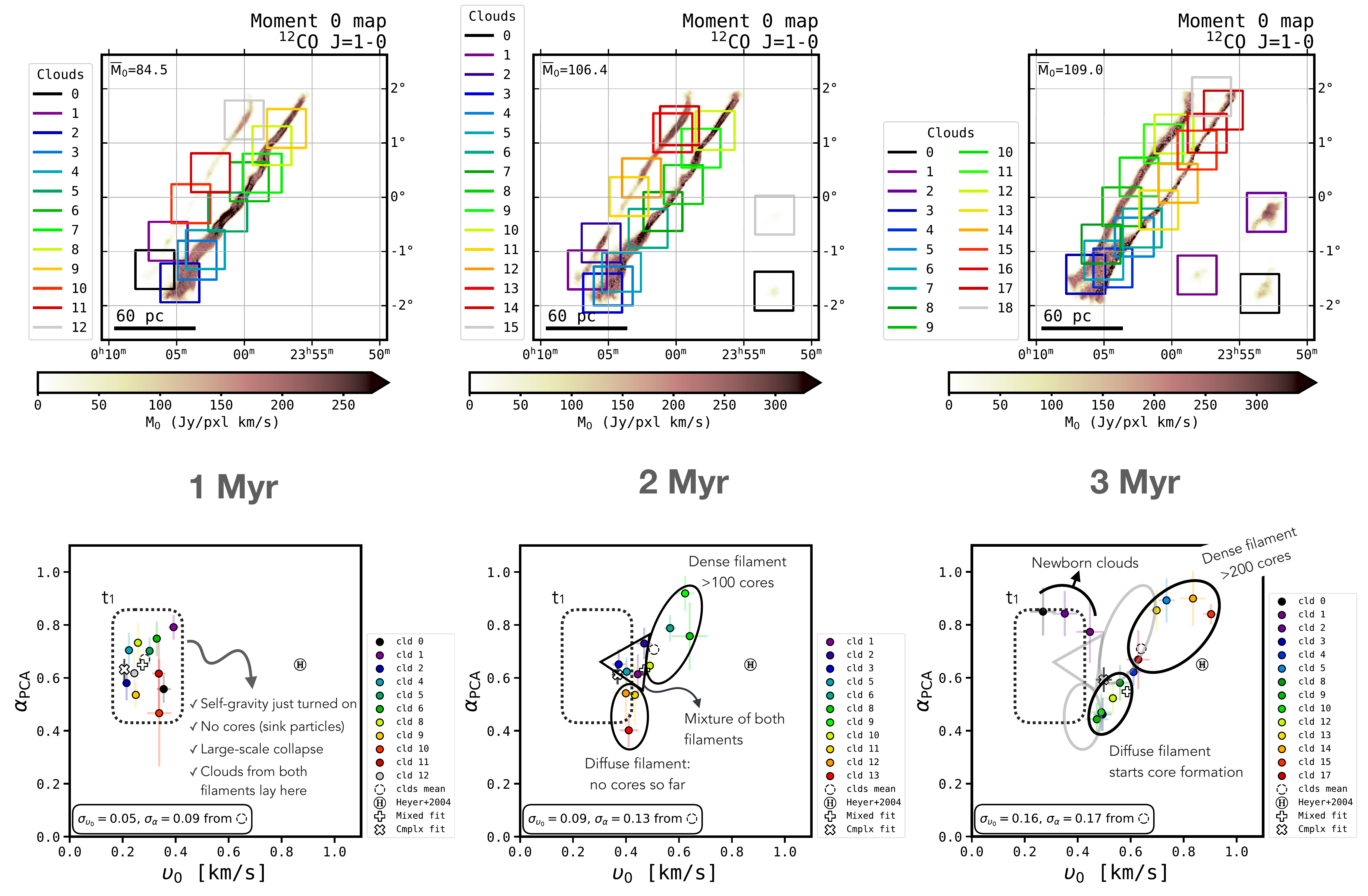}

\caption{Time evolution of scaling parameters in cloud complex B. The top row shows zeroth moment maps at different time steps after commencing tracer refinement. The bottom row shows the PCA-derived pseudo-structure function parameters ($\vz$, $\alpha$) from individual clouds in complex B. The standard deviation of cloud parameters is shown in the lower left corner of the panels. The colour code of individual portions in top panels matches the marker colours in bottom panels. Individual clouds migrate over time in the ($\vz$, $\alpha$) space in response to different gravitational processes taking place in the complex.} 
\label{fig:time_complexB}
\end{figure*}

Conversely, the evolution of scaling exponents $\gamma$ is less predictable, which might indicate that local gravitational forces trigger velocity fluctuations across a wide range of spatial scales. To assess this, we zoomed into cloud complex B to track the evolution of individual filament scaling parameters as a response to local gravitational effects (see Figure \ref{fig:time_complexB}). At $t=1$\,Myr, when there are still no sink particles, individual cloud parameters cluster around a common zone in the ($\vz$, $\gamma$) space. Overall, scaling exponents are steeper ($\mean{\gamma}=0.75$) and velocity fluctuations weaker ($\mean{\upsilon}=0.28$\,km\,s$^{-1}$) compared to those of the potential-dominated case with no self-gravity ($\mean{\gamma}=0.51$, $\mean{\upsilon}=0.42$\,km\,s$^{-1}$). We propose that this is a consequence of global (large-scale) collapse processes that reduce the degree of turbulence in the region by increasing coherence. 

Later, at $t=2$\,Myr, there is a burst of star formation in the long dense filament (N$_{\rm sinks}>100$) but none in the diffuse filament yet. This produces a clear separation of individual cloud parameters coming from each filament as shown in the middle panel of Fig. \ref{fig:time_complexB}. The PCA scaling exponents of the short diffuse filament are this time shallower because gravitational fragmentation has commenced. This favours small-scale interactions in the region and is likely to be a signature of pre-core stages in molecular clouds. The long dense filament skipped this phase as it formed several individual clumps/cores much faster. Such a rapid emergence of stellar systems drastically lowers velocity fluctuations on small scales and, hence, increases the scaling exponent ($\gamma$). This `burst' of new cores also establishes multiple point-like centres of collapse affecting inter-core gas predominantly on intermediate and large scales. For reference, the region in the long filament where sink particles are more numerous has a mean sink separation of $\approx$\,0.5\,pc (see Fig. \ref{fig:column_densities}). As a side note, a consequence of our sink particle implementation is that there is some missing information from gas velocities on small scales because it reduces the bound gas mass around the cores to point-mass gravitating particles. Hence, in real observations, shallower (but still high) scaling exponents should be found in core stages because gas from cores/clumps do still contribute to small-scale velocity fluctuations in the region via line broadening. 

Finally, at $t=3$\,Myr, the short filament commences the formation of cores at its upper tip. For that particular cloud (cld 13 at $t=2$\,Myr, cld 12 at $t=3$\,Myr), both scaling parameters increase and favour the zone occupied formerly by clouds from the long filament, but not as high as them, because of the much lower core formation rates. As expected, cloud portions with no cores retain the same $\gamma$, but with higher $\vz$, because fragmentation is still carrying on. Both parameters continue to increase for the long filament as the number of cores grows over time. Interestingly, new molecular associations appear on the right side of the complex and yield scaling parameters close to the same zone where clouds from both of the filaments started at $t=1$\,Myr, which strengthen the idea that quiescent clouds have gravity-driven time-dependent trajectories in the ($\vz$, $\alpha$) space. This behaviour also holds for cloud complex A, namely individual cloud scaling parameters also respond differently depending on local core formation stages. Note that both fragmentation processes in pre-core stages, as well as in subsequent formation of cores, continuously lead to an increase in the magnitude of velocity fluctuations ($\vz$) across the cloud complex. 

Feedback-dominated cloud complexes evolve according to their particular environment. As seen in Figure \ref{fig:pca_time_evolution}, the common pattern is that complex C migrates toward lower scaling coefficients $\vz$ whereas complex D moves toward higher $\vz$. Local gravitational effects do not play the same role in these complexes as in quiescent regions because mass instabilities are this time more difficult to reach. In complex C, external feedback from supernova explosions set large turbulence driving scales that decay over time through smaller scales as there are no internal fueling sources of turbulence. Other mechanisms of internal feedback (beyond the scope of this work) such as winds or photoionising radiation could potentially help inject fresh energy into the cloud \citep[see e.g.][]{peters+2017}. In complex D, however, energy re-injection from internal supernovae seems to sustain and increase the level of turbulence through expanding supernova bubbles.    

In summary,

\begin{itemize}
    \item Random supernovae alone are unable to sustain turbulence in molecular clouds at a level consistent with observations.
    \item Quiescent self-gravitating molecular clouds have time-dependent trajectories in the ($\vz$, $\gamma$) parameter space: 
    \item[$-$] There is a common zone in the parameter space where molecular clouds are born.
    \item[$-$] The scaling coefficient $\vz$ increases steadily over time. This is associated with gravitational fragmentation processes (both in pre-core and core formation stages) that increase the magnitude of velocity fluctuations over the cloud life time.
    \item[$-$] The evolution of scaling exponents $\gamma$ is less predictable. It is associated with variations in the characteristic driving size-scale of gravitational evolutionary stages in cloud portions of a complex.
    Thereby, large-scale collapse, pre-core and core formation stages, which may all take place at the same time in a cloud complex, respond differently in the ($\vz$, $\gamma$) space.
    \item Mean surface densities in quiescent self-gravitating cloud complexes also increase steadily over time, suggesting a connection between the evolutionary stage of local gravitational processes and the dependence of $\vz$ on surface density reported in previous literature.
    \item Pre-core stages favour small-scale fluctuations ruled by turbulence and local fragmentation, yielding shallower exponents. 
    \item Core stages increase scaling exponents as cores act as multiple point-like centres of collapse affecting inter-core gas at medium and large scales predominantly.
    \item Local gravitational effects do not seem to play the same role in feedback-dominated cases because mass instability is more difficult to achieve and supernovae effects prevail.
    \item In feedback-dominated complexes, driving of large-scale velocity fluctuations by clustered supernovae can sustain the level of turbulence within the clouds. 
    \item Unlike external feedback, internal clustered supernovae substantially increase the magnitude of velocity fluctuations $\vz$ over time, which in turn destroys the cloud faster.
\end{itemize}

\subsection{Structure Function: Variability within Cloud Complexes and Environmental Effects} \label{subsec:results_inner}

We explore three PCA extraction methods to address any variations in scaling parameters that might arise when different analysis scales are used to retrieve structure functions. As a reminder, the three approaches are the Mixed, the Complex and the Individual Cloud methods. The Mixed method consists of combining PCA-derived scales from cloud portions to construct the structure function representative of their hosting complex; the Complex method applies the PCA algorithm on whole complexes, without sub-portioning; and the Individual Cloud method extracts PCA-derived structure functions from cloud portions as if they were individual objects, without mixing them. See further details in Section \ref{sec:methodology}. 

\begin{figure*}
\begin{center}

\begin{overpic}[width=0.33\textwidth]{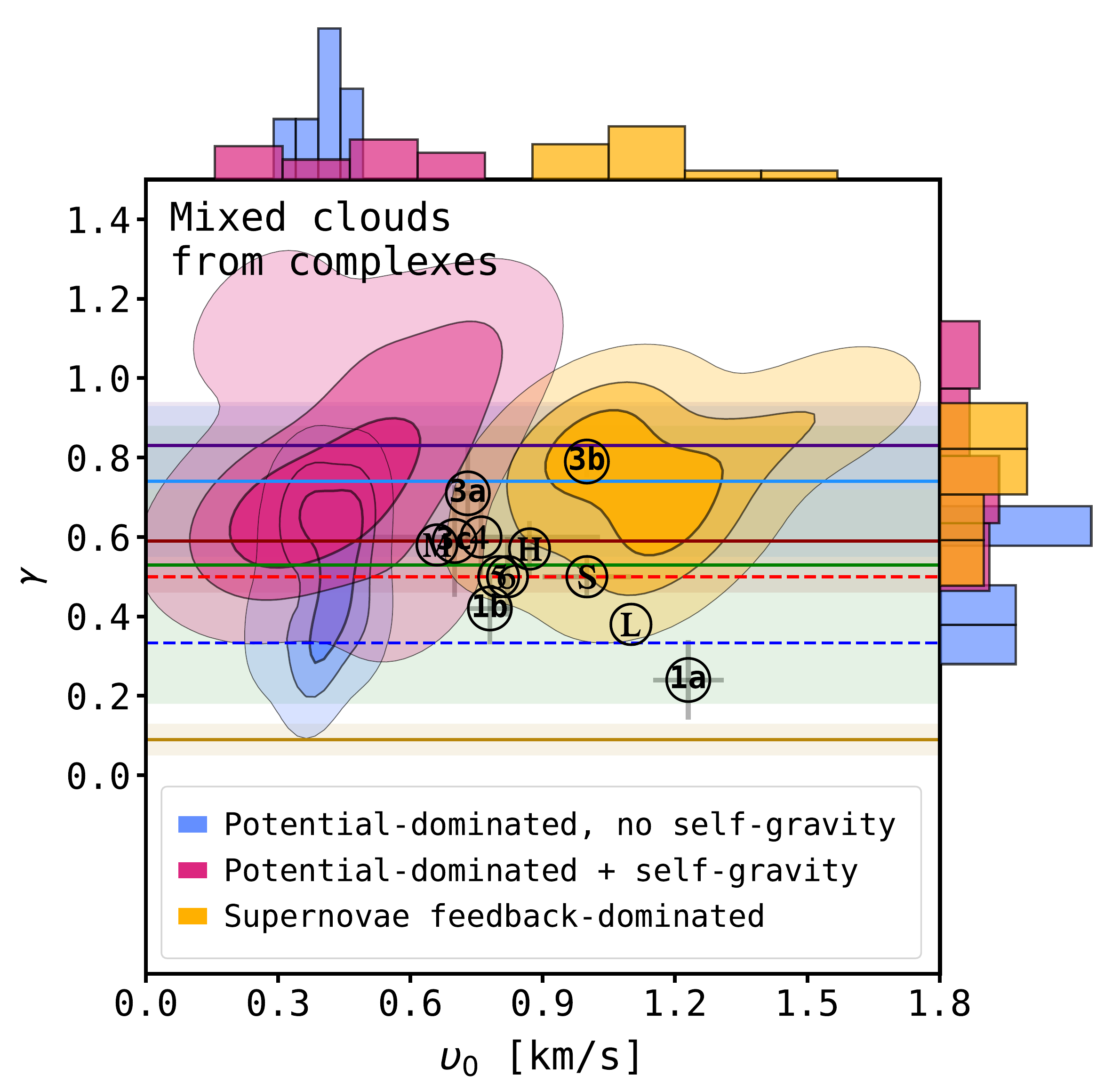}
\end{overpic} 
\begin{overpic}[width=0.33\textwidth]{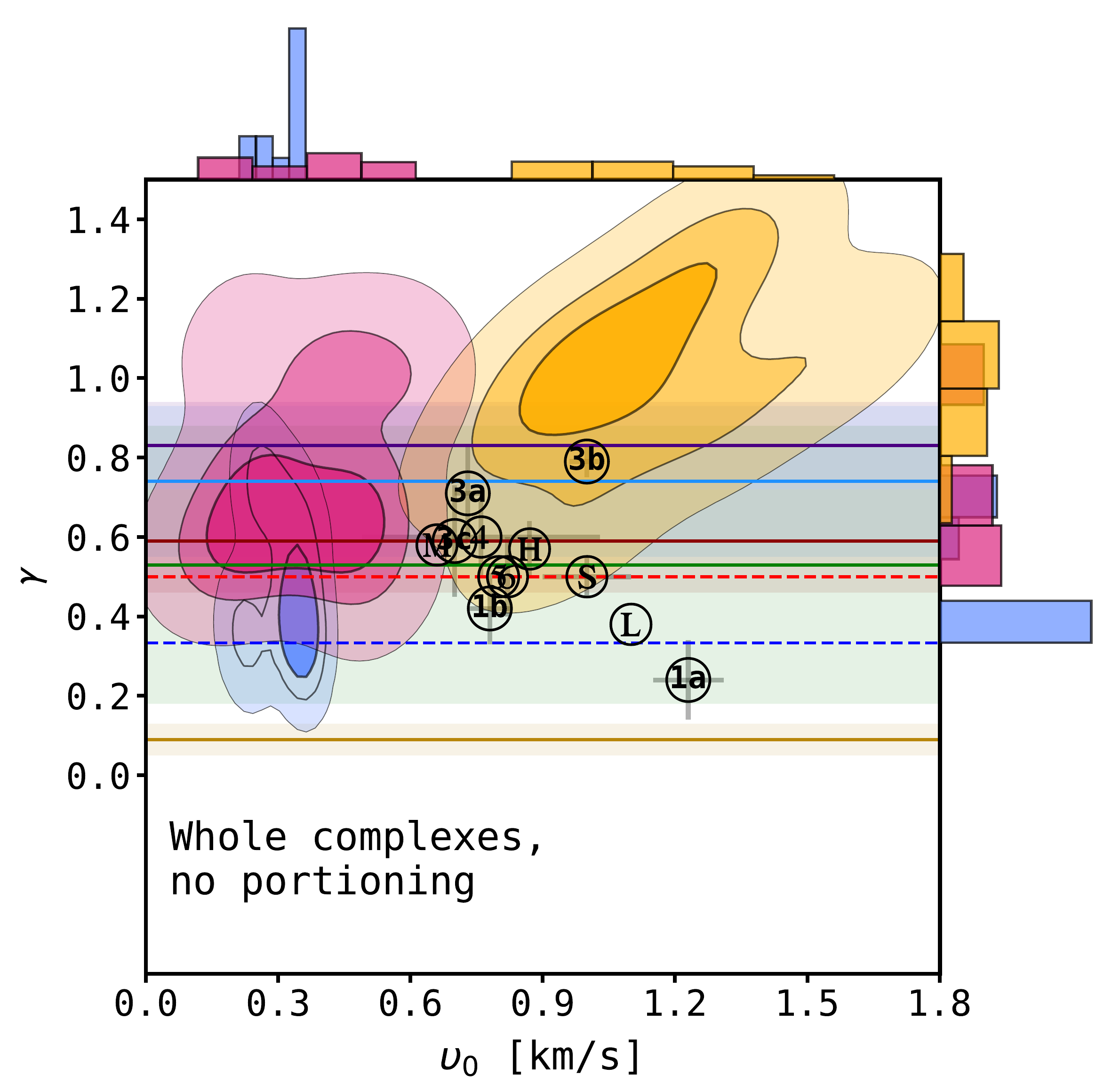}
\end{overpic} 
\begin{overpic}[width=0.33\textwidth]{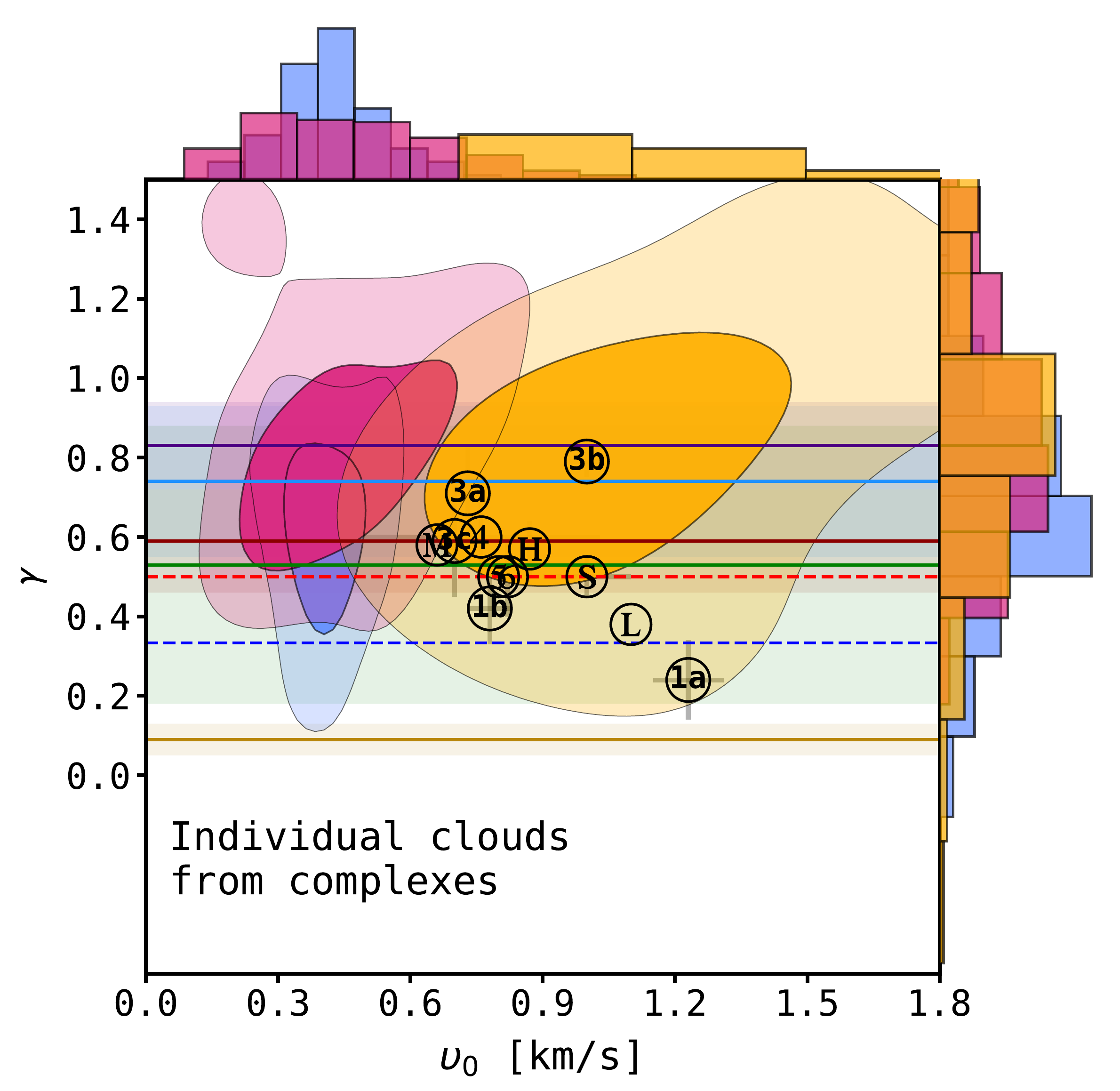}
\end{overpic} 

\caption{Distribution of structure function parameters for different PCA extraction methods: from left to right, Mixed, Complex and Individual Clouds (see Sec. \ref{sec:methodology}). The filled contours correspond to 33, 68 and 95 per cent levels. Mean parameters and standard deviations for each method and scenario are summarised in Table \ref{table:mean_pars_pdf}. Horizontal lines and circles are the literature parameters listed in Table \ref{table:references}.  
}
\label{fig:pca_summary_distribution}
\end{center}
\end{figure*}

Figure \ref{fig:pca_summary_distribution} shows the distribution of structure function scaling parameters derived from each of the PCA extraction methods and physical scenarios, and Table \ref{table:mean_pars_pdf} summarises the mean values and standard deviations. The Mixed and the Complex methods yield very similar parameter dispersion for all the physical scenarios. However, studying cloud complexes as a whole with the Complex method shifts the potential-dominated cases (A$_0$, B$_0$, A and B) to lower scaling coefficients $\vz$ ($\Delta_{\vz} \sim -0.1$ km s$^{-1}$) and the feedback-dominated cases (C, D) to higher scaling exponents $\gamma$ ($\Delta_\gamma \sim 0.22$) compared to the Mixed method. We attribute these variations to intermittency of density and velocity fields that make the PCA scaling exponents not to follow 2$^{\rm nd}$ order velocity fluctuations but lower orders only \citep[see][]{brunt+2003c, roman-duval+2011}. Similarly, the difference in scaling coefficients is due to the PCA-derived scales describing low-order velocity fluctuations within the Complex (see e.g. Fig. \ref{fig:pca_fits_faceon}). This effect is more prominent in the Complex analysis method, in which PCA is computed on entire cloud complexes and hence intermittent fields are more likely to appear; especially in the feedback-dominated cases where extreme fluctuations in the turbulent flow are expected. 

In Figure \ref{fig:pca_summary_distribution} we also present the scaling parameter distribution for the Individual Cloud method, which computes the structure function of individual molecular clouds and treats them as individual objects. Due to the much higher number of objects, the parameter scattering is naturally larger for all the scenarios compared to the Mixed and Complex methods. However, the feedback-dominated scenario produces the widest range of scaling parameters as a consequence of the variation in localised internal and external supernova feedback plus the local gravitational influence. 

\begin{figure}
	\includegraphics[width=\columnwidth]{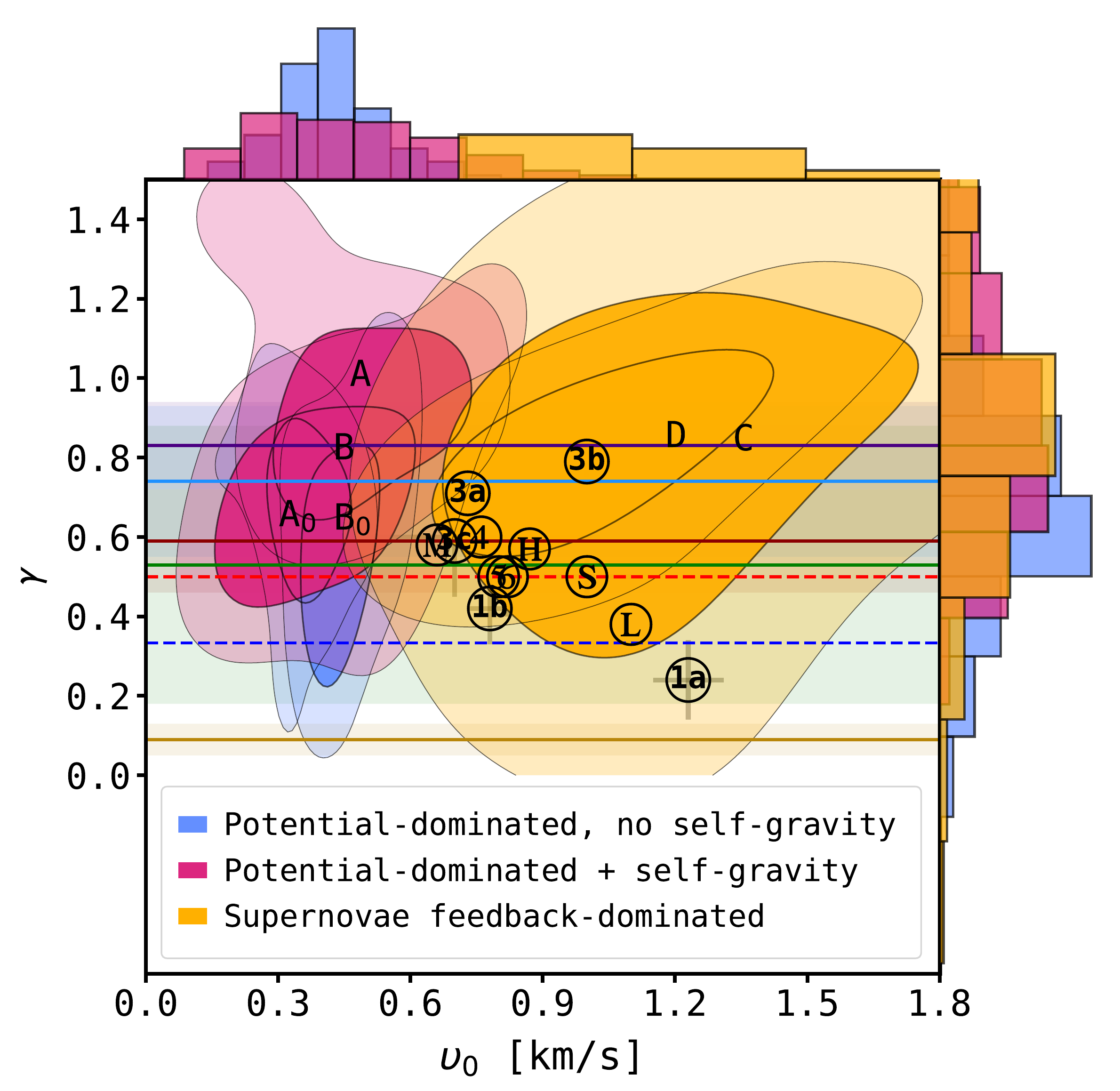}
    \caption{Same as the third panel of Figure \ref{fig:pca_summary_distribution} but discriminating individual clouds according to their hosting cloud complex. Lines and circles refer to the literature values indicated in Table \ref{table:references}.}
    \label{fig:individual_clouds_pars}
\end{figure}

Figure \ref{fig:individual_clouds_pars} splits individual cloud distributions depending on the origin cloud complex to illustrate the influence of the surrounding environment. Parameter variations due to different density contexts are not significant for potential-dominated complexes though.  
On the other hand, both cloud complexes in the feedback-dominated scenario produce similar scaling parameters, however, complex D spans a smaller range of parameters as noticed from its individual cloud values. This is due to the supernova explosions embedded in complex D which destroy cloud structures faster than in the quieter complex C.

As expected, due to the prescribed burst of supernovae in our feedback-dominated scenario, the mean scaling parameters lie close to values resulting from hypersonic-turbulence samples such as the ionised zone of the Rosette cloud reported by \cite{heyer+2006} (marker $\mathcircled{\textrm{\bf \scriptsize 3b}}$ in Figs. \ref{fig:pca_summary_distribution} and \ref{fig:individual_clouds_pars}) and the \cite{federrath+2010} simulations using purely compressive forces (solid blue line \solidline{dodgerblue_mpl}).    

In summary,
\begin{itemize}
    \item Using different analysis scales can reveal density and velocity intermittent fields. 
    \item Feedback-dominated cases are more prone to intermittency. This is evidenced as higher dispersion of individual cloud scaling parameters than potential-dominated complexes.
    \item Environmental conditions of clouds may also split scaling parameters ($\vz$, $\gamma$) into separate clusters of points for each physical scenario.
    \item The feedback-dominated scenario produces the largest range of scaling parameters.
    \item The parameter distribution from clouds with embedded supernovae is more confined in $\gamma$ than those with external feedback only. We attribute this to the faster disruption of coherent structures in the former case.  
\end{itemize}

\section{discussion} \label{sec:discussion}

\subsection{Comparison to Observational and Synthetic Structure Functions}
 
From Figure \ref{fig:pca_summary_distribution}, the potential-dominated complexes without self-gravity are mostly clustered between (incompressible) subsonic \citep{kolmogorov+1941, onsager+1949} and supersonic shock-dominated turbulence \citep{kraichnan+1974, frisch+2001} regimes, or just above the latter. This suggests that large-scale gravitational forces along with random (isolated) supernova feedback are able to reproduce classical scaling exponents ($\gamma$) from theoretical fluids. They also agree with observational exponents found by \cite{larson+1981} and \cite{solomon+1987}, and with simulations of pure solenoidal turbulence forcing performed by \cite{federrath+2010}. The solenoidal turbulence is associated with regions of low star formation activity such as the quiescent zone of the Rosette cloud complex \citep{heyer+2006} or the Musca cloud \citep{hacar+2016}. However, the magnitude of velocity fluctuations ($\vz$) remains too low in this scenario to fully reproduce the observations.

Turning on self-gravity generally produces steeper structure function scaling exponents ($\gamma$). Depending on the evolutionary stage of cloud complexes, gravity-driven (non-thermal) motions such as collapse and/or accretion flows can increase the velocity dispersion. This results in a sustained increase of the magnitude of velocity fluctuations ($\vz$), triggered by the emergence of local potential wells in the region. In pre-core stages this occurs on small scales characterised by the core accretion zone, and in core stages on medium/large scales given by the separation between cores \citep{smith+2016}. Complex B is an excellent sample to study local gravitational effects as it is composed of two quiescent filamentary structures that evolve differently. At early stages, one filament is dense and prominent and the other is diffuse and discontinuous. Figure \ref{fig:time_complexB} shows that the PCA-derived parameters from individual clouds of this complex are clearly different depending on the analysed filament. Both $\vz$ and $\alpha$ are in general higher for the longer, denser filament, which agrees very well with the idea that massive regions yield higher velocity dispersion for larger column densities due to bound clumps and cores undergoing gravitational collapse \citep{ballesteros-paredes+2011}.

It is particularly interesting to compare the results from this physical scenario with real giant molecular clouds that have little stellar feedback. This is the case for the Musca cloud in the Musca-Chamaeleonis molecular complex reported by \cite{hacar+2016} as the first observational evidence of a filament that is largely coherent with negligible internal turbulence. They found two observational relations of the form $\dv=\vz l^\gamma$ that depend on the scale size range of the cloud: (i) a transonic $\dv = 0.55 l^{0.25}$ for scales $<$ 1.0\,pc and (ii) a supersonic $\dv = 0.66 l^{0.58}$ on scales between 1.0$-$3.0\,pc. We multiplied their reported $\dv$ by a factor of $\sqrt{3}$ (assuming isotropic 3D velocity fluctuations) to facilitate comparison with our three-dimensional parameters. Our self-gravitating cloud complexes without clustered feedback are consistent with the scaling parameters derived from this quiescent region in the supersonic regime, especially the most evolved complexes (see middle panel of Fig. \ref{fig:pca_time_evolution}). Additionally, some cloud portions in complex B also manage to reproduce Musca's parameters in the transonic regime. This is the case for clouds in the diffuse filament of complex B where core formation has just commenced at $t=3$\,Myr (see Fig. \ref{fig:time_complexB}). 

Furthermore, \cite{hacar+2016} also found that an independent analysis of individual portions in the Musca cloud leads to a wide range of scaling parameters, suggesting the presence of local fluctuations that can substantially differ from the structure function of the cloud as a whole. This resembles our result that individual cloud parameters exhibit a high degree of scatter when compared to the parent larger-scale cloud complex parameters. We found that this level of scattering is related to several gravitational stages governing different scales of the cloud/complex (see Fig. \ref{fig:time_complexB}), which at the same time, is closely related to the density distribution of the region. Thus, we attribute the broken power-law reported for the Musca cloud to asynchronous evolution of parameters driven by local gravitational effects.

Scaling exponents ($\gamma$) derived from our feedback-dominated cloud complexes are consistent with \cite{federrath+2010} simulations with purely compressive forces or with the value reported by \cite{bertram+2014} for \twCO{} emission of molecular clouds with artificial turbulent fields. When considering both scaling parameters ($\gamma$, $\vz$), our cloud complexes and individual clouds can reproduce a range of observations including those of \cite{larson+1981}, \cite{solomon+1987} and \cite{heyer+2004} (see Figs. \ref{fig:pca_summary_distribution} and \ref{fig:individual_clouds_pars}). The distribution of parameters is centred around the point $\mathcircled{\textrm{\bf \scriptsize 3b}}$ corresponding to the zone II of the Rosette cloud complex, which is dominated by strong stellar feedback from nearby massive stars \citep{heyer+2006}. This is compatible with the burst of supernova explosions that inject strong energy feedback in our simulations. 

\subsection{Are Supernovae Important for Driving Turbulence in Molecular Clouds?}

It is useful to briefly discuss the role that supernovae play in driving cloud-scale turbulence in these simulations, given their overall importance for the energy balance of the ISM \citep[see e.g.][]{maclow+2004}. We find, in common with several previous studies \citep{ibanez-mejia+2017, seifried+2018}, that randomly-distributed supernovae that explode with a rate similar to the supernova rate in the Milky Way are unable to drive turbulence in molecular clouds at a level consistent with observations of Galactic GMCs. As \cite{seifried+2018} explore in some detail, the reason for this is that in the random supernova scenario, the chances of supernovae exploding close to the clouds on a regular basis are small and so any turbulence injected into the clouds by a nearby supernova tends to decay away long before the next nearby supernova occurs. We therefore conclude that supernovae are not important for driving the turbulence observed in quiescent (i.e. non-star-forming) molecular clouds and rather gravity plays a more dominant role \citep[see e.g.,][]{klessen+2010}. On the other hand, in cloud complexes actively forming massive stars, supernova explosions do not occur with a spatially random distribution, but instead are highly correlated with the gas distribution. In this case, the rate of nearby supernova explosions is much higher than with the random distribution, allowing the supernovae to play a dominant role in driving the turbulence in these clouds. Future work will include the analysis of other important driving mechanisms such as stellar winds, jets and photoionising radiation, as well as the influence of magnetic fields. 

\section{conclusions} \label{sec:conclusions}

We have performed principal component analysis (PCA) on full non-LTE radiative transfer simulations of molecular cloud complexes, self-consistently generated using our Cloud Factory galactic-scale ISM simulation suite. We explore PCA-derived velocity structure functions from three different physical scenarios set up in our Cloud Factory: (a) one where the ISM dynamics is dominated by the large-scale Galactic potential, with (isolated) supernovae explosions randomly distributed across the Galaxy, (b) same as the previous case but self-gravity is turned on, and (c) a feedback-dominated scenario where supernova explosions are random but also tied to star formation sites, which results in strong clustered feedback. Large-scale potential and local gravitational effects are both active in this case.

Regardless of the physical scenario, we find that all the cloud complexes analysed from our Cloud Factory zooms agree with distinct types of turbulence reported in the literature.
Clearly, large-scale gravitational forces alone when combined with turbulent decay are enough to reproduce Kolmogorov's and Burgers-like turbulence scaling exponents, but scaling coefficients remain too low compared with observations. Nonetheless, under weak influence of isolated supernova explosions, local gravitational forces can make structure functions evolve over time and reproduce observations of quiescent molecular clouds. 

We report time-dependent trajectories in the structure function parameter space driven by local gravitational effects and supersonic turbulent flows. The magnitude of velocity fluctuations ($\vz$) increases steadily for self-gravitating regions with low stellar feedback. Typically, just-assembled clouds display low magnitudes and then migrate through the ($\vz$, $\gamma$) parameter space as star-forming cores emerge within. The scaling exponents ($\gamma$) are generally less predictable because they depend upon the stage of gravitational collapse, which varies locally as a function of the boundness conditions of sub-structures in the cloud complex. This could explain power-law breaks and variations in structure function parameters observed for different size-scales in quiescent molecular clouds. 

However, gravitational forces alone (when combined with random supernovae feedback) are not enough to reproduce both the scaling coefficient and exponent of molecular clouds with active star formation. We find that clustered feedback from supernovae tied to sites of star formation is key to self-consistently generate clouds that reproduce the scaling parameters reported by observations with similar size-scales and resolutions to those used in our simulations. 

Our results suggest that a PCA-based statistical study is a robust method to diagnose the physical mechanisms driving gravoturbulent fluctuations in molecular clouds by providing a quantitative description of the velocity field. The analysis tools developed in this work are all condensed in our new open source \pcafactory{} package.

\section*{Data Availability}
The data underlying this article are publicly available on \url{yt.Hub}, and can be accessed by following instructions at \url{https://github.com/andizq/andizq.github.io/tree/master/pcafactory-data}.

\section*{Acknowledgements}

The authors would like to thank C. Brunt, C. Dickinson, and the anonymous referee for their inspiring discussions and comments which improved the results, conclusions, and key theoretical aspects of this work. AFI acknowledges the studentship funded by the UK's Science and Technology Facilities Council (STFC) through the Radio Astronomy for Development in the Americas (RADA) project, grant number ST/R001944/1, as well as support from the Deutsche Forschungsgemeinschaft (DFG, German Research Foundation) - Ref no. FOR 2634/1 TE 1024/1-1. AFI also acknowledges support by the Agencia de Educaci\'on Superior de Medell\'in (Sapiencia), attached to the Alcald\'ia de Medell\'in, and the Centro de Ciencia y Tecnolog\'ia de Antioquia (CTA), through the grant Premio Medell\'in Investiga 2019, which encourages high-impact research and innovation in the city of Medellin, Colombia. RJS gratefully acknowledges support from an STFC Ernest Rutherford Fellowship (grant ST/N00485X/1), without which this work would not have been possible. SCOG and RSK acknowledge support from the Deutsche Forschungsgemeinschaft (DFG) via the Collaborative Research Center (SFB 881, Project-ID 138713538) ``The Milky Way System'' (sub-projects A1, B1, B2 and B8) and from the Heidelberg cluster of excellence (EXC 2181 - 390900948) ``{STRUCTURES}: A unifying approach to emergent phenomena in the physical world, mathematics, and complex data'', funded by the German Excellence Strategy. The authors gratefully acknowledge support by the state of Baden-W\"utemberg through bwHPC and the German Research Foundation (DFG) through grant INST 35/1134-1 FUGG. They furthermore acknowledge the data storage service SDS@hd supported by the Ministry of Science, Research and the Arts Baden-W\"utemberg (MWK) and the German Research Foundation (DFG) through grant INST 35/1314-1 FUGG. This work used the COSMA Data Centric system at Durham University, operated by the Institute for Computational Cosmology on behalf of the STFC DiRAC HPC Facility (\url{www.dirac.ac.uk}). This equipment was funded by a BIS National E-infrastructure capital grant ST/K00042X/1, DiRAC Operations grant ST/K003267/1 and Durham University.

%%%%%%%%%%%%%%%%%%%%%%%%%%%%%%%%%%%%%%%%%%%%%%%%%%

%%%%%%%%%%%%%%%%%%%% REFERENCES %%%%%%%%%%%%%%%%%%

% The best way to enter references is to use BibTeX:

\bibliographystyle{mnras}
\bibliography{references} % if your bibtex file is called example.bib

%%%%%%%%%%%%%%%%%%%%%%%%%%%%%%%%%%%%%%%%%%%%%%%%%%

%%%%%%%%%%%%%%%%% APPENDICES %%%%%%%%%%%%%%%%%%%%%

\appendix

\section{Supporting Figures} \label{sec:appendixfigures}

\begin{figure}
	\includegraphics[width=0.95\columnwidth]{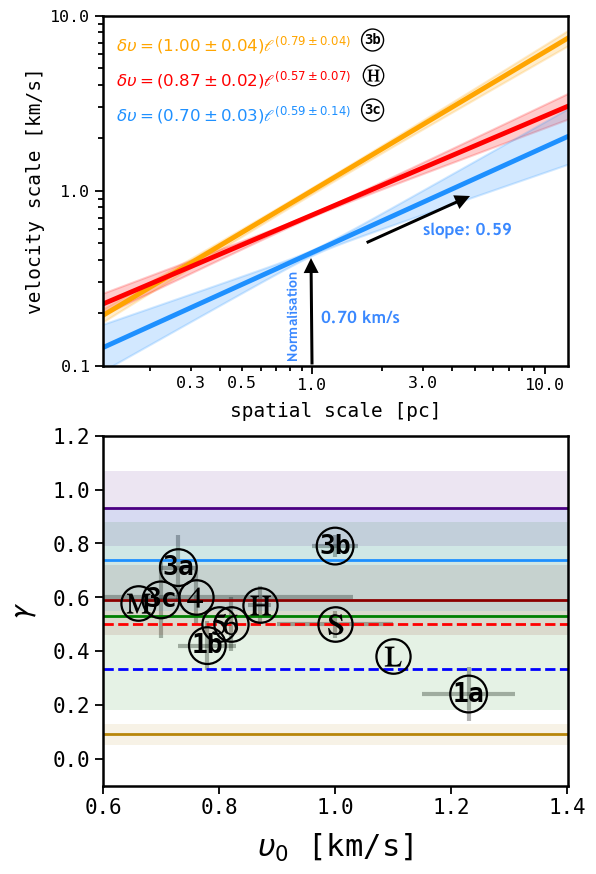}
    \caption{Comparison of selected structure function scaling parameters reported in previous literature. Line and marker codes are listed in Table \ref{table:references}.}
    \label{fig:literature_summary}
\end{figure}

\begin{figure}
	\includegraphics[width=0.95\columnwidth]{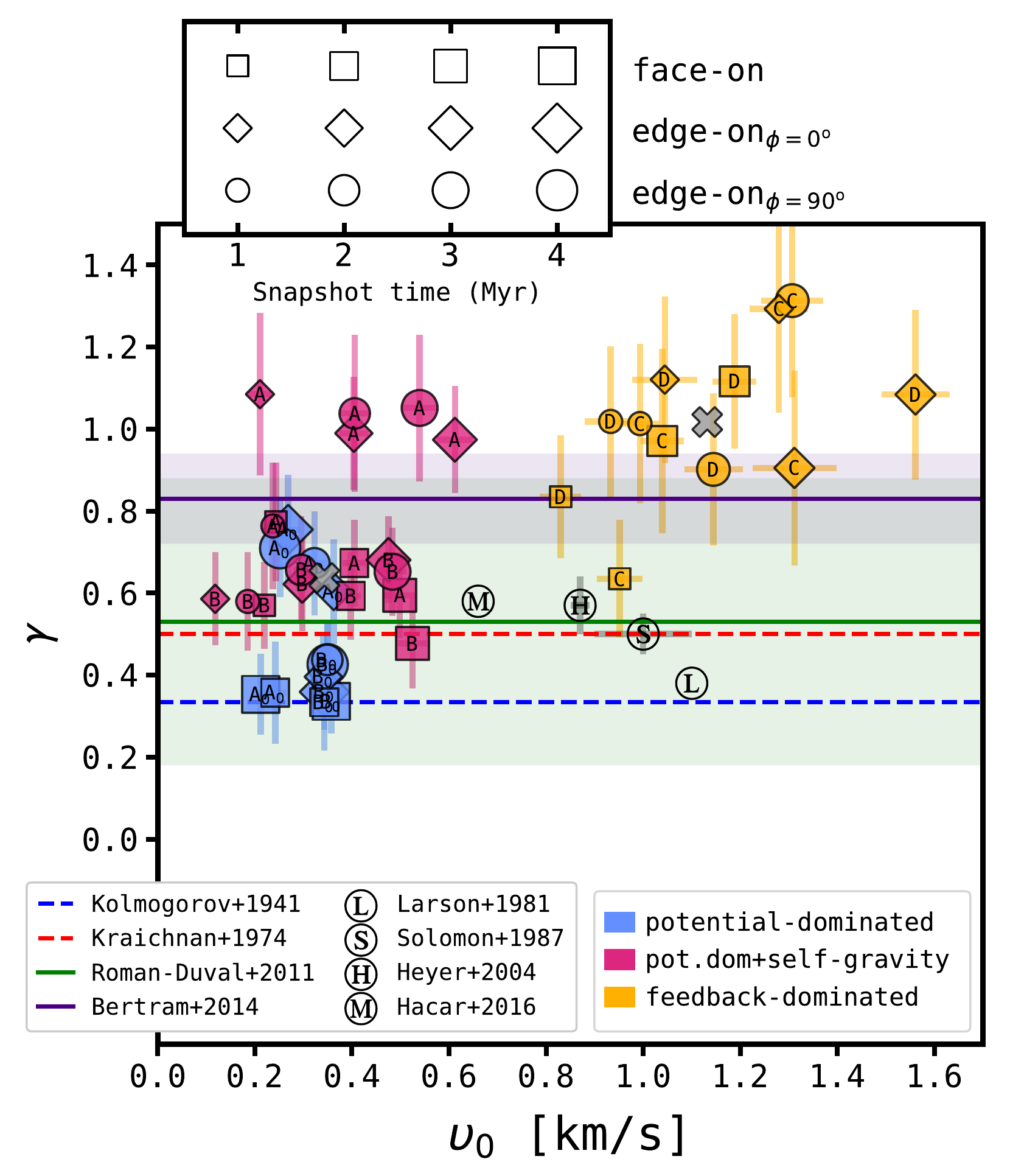}
    \caption{Same as Fig. \ref{fig:pca_summary} but using the Complex method (see Sec. \ref{sec:methodology}).}
    \label{fig:pca_summary_cloud}
\end{figure}

\begin{figure*}
\begin{center}
\begin{tabular}{cc}

\begin{overpic}[width=0.77\textwidth]{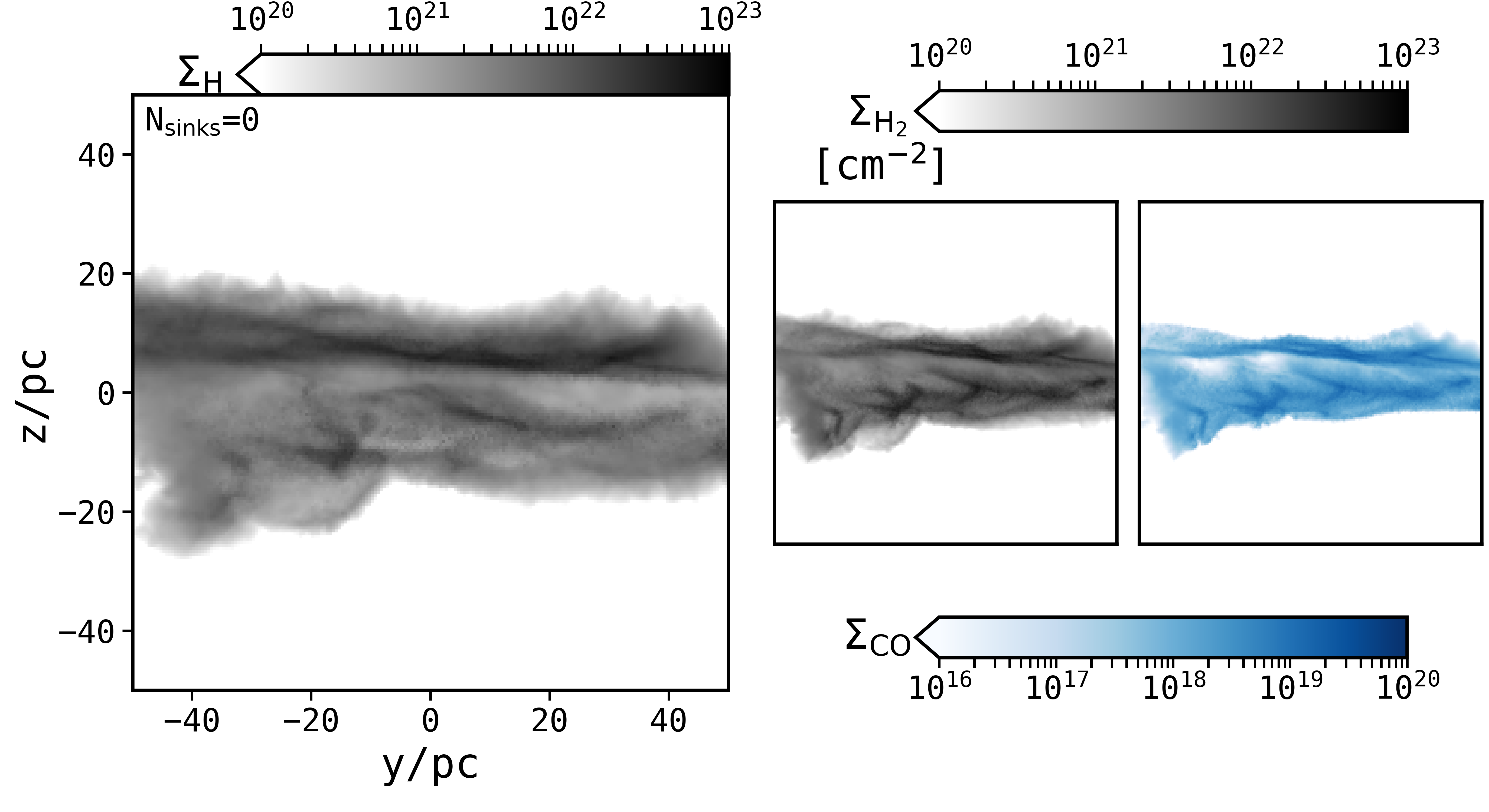}
\end{overpic} \hspace{0.5cm}
\put (40,120) {\makebox(0,0){{\huge A$_0$}}} \hspace{0.5cm}
\\

\begin{overpic}[width=0.77\textwidth]{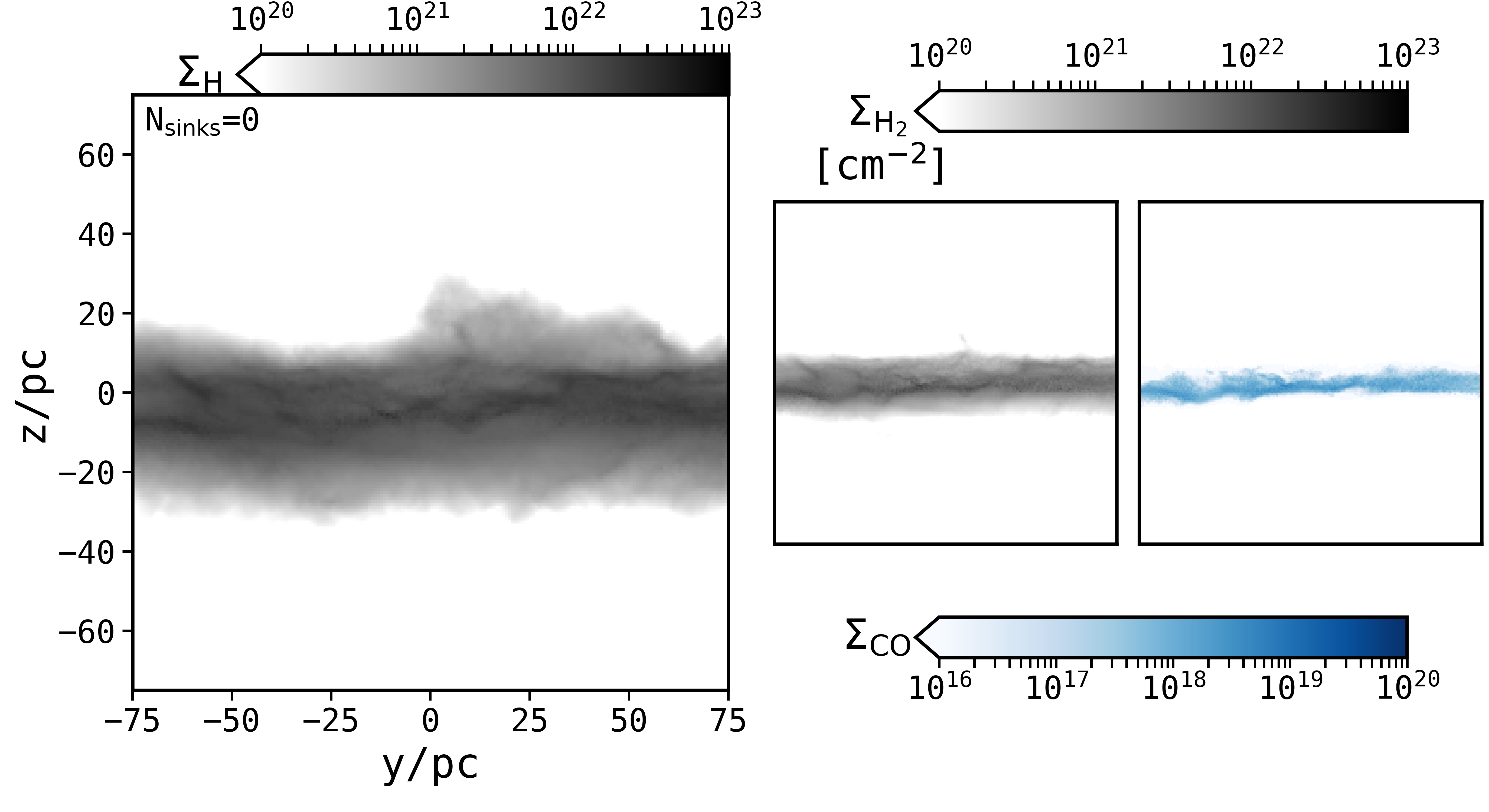}
\end{overpic} \hspace{0.5cm}
\put (40,120) {\makebox(0,0){{\huge B$_0$}}} \hspace{0.5cm}
\\

\begin{overpic}[width=0.77\textwidth]{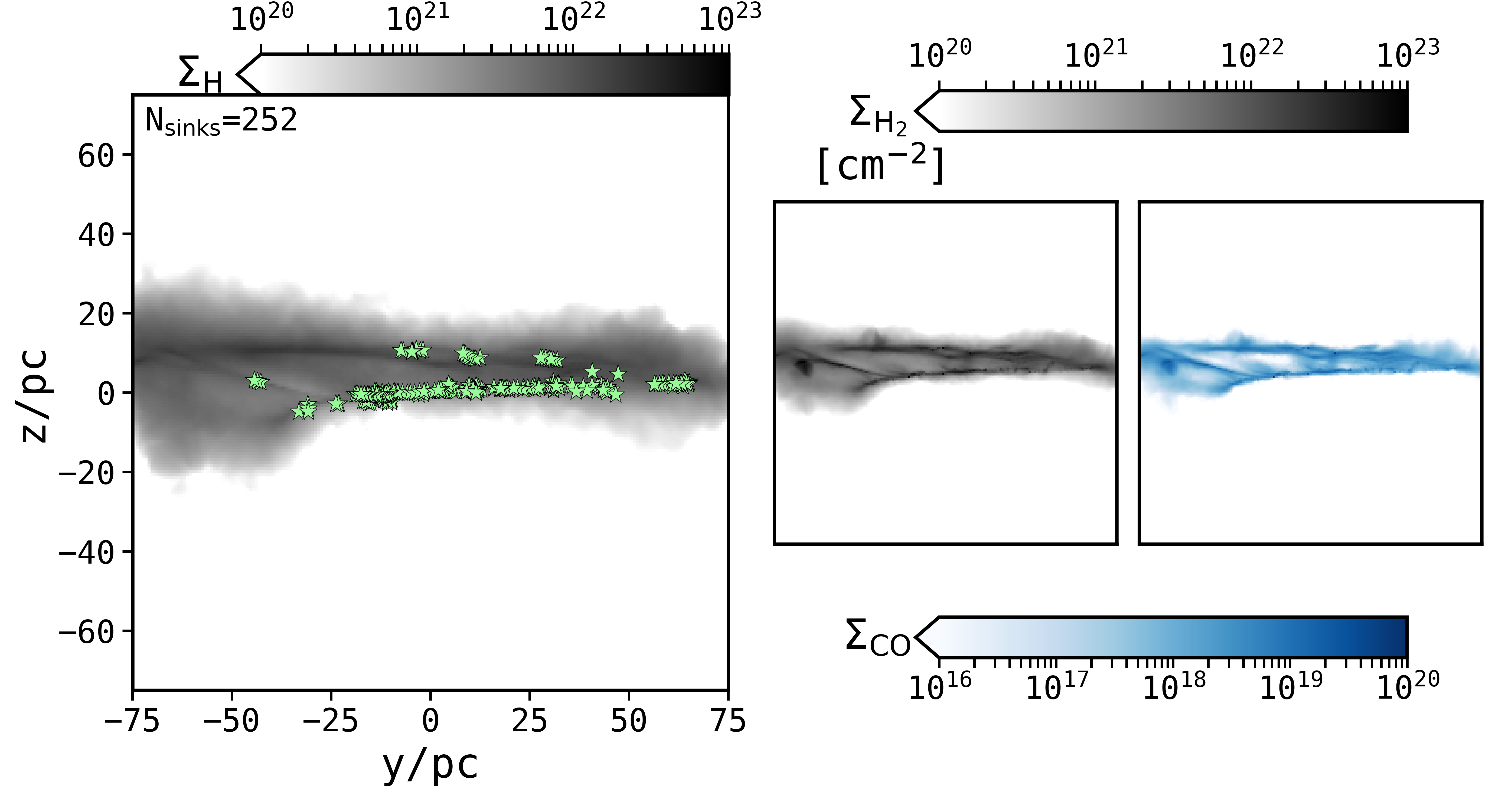}
\end{overpic} \hspace{0.5cm}
\put (40,120) {\makebox(0,0){{\huge A}}} \hspace{0.5cm}

\end{tabular}
\caption[]{Edge-on$_{\phi=90^{\circ}}$ projections of H, H$_2$ and $^{12}$CO column densities ($\Sigma$) from cloud complexes (labeled on the right) extracted 2\,Myr after injecting tracer particles in the simulations. If any, sink particles are overlaid on H maps as star markers.}
\label{fig:appendix_column_densities}
\end{center}
\end{figure*}

\begin{figure*}
\ContinuedFloat
\begin{center}
\begin{tabular}{cc}

\begin{overpic}[width=0.77\textwidth]{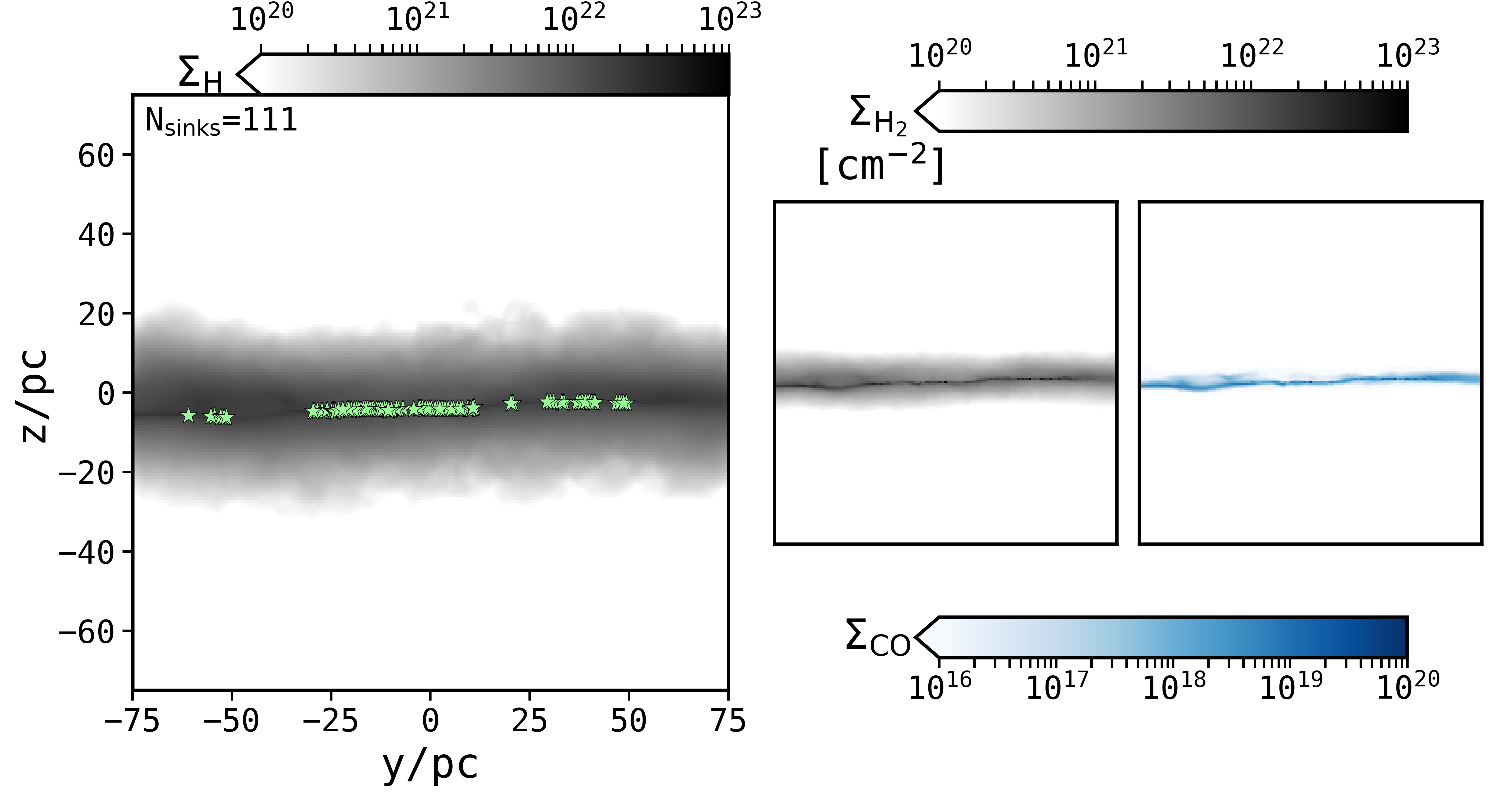}
\end{overpic} \hspace{0.5cm}
\put (40,120) {\makebox(0,0){{\huge B}}} \hspace{0.5cm}
\\

\begin{overpic}[width=0.77\textwidth]{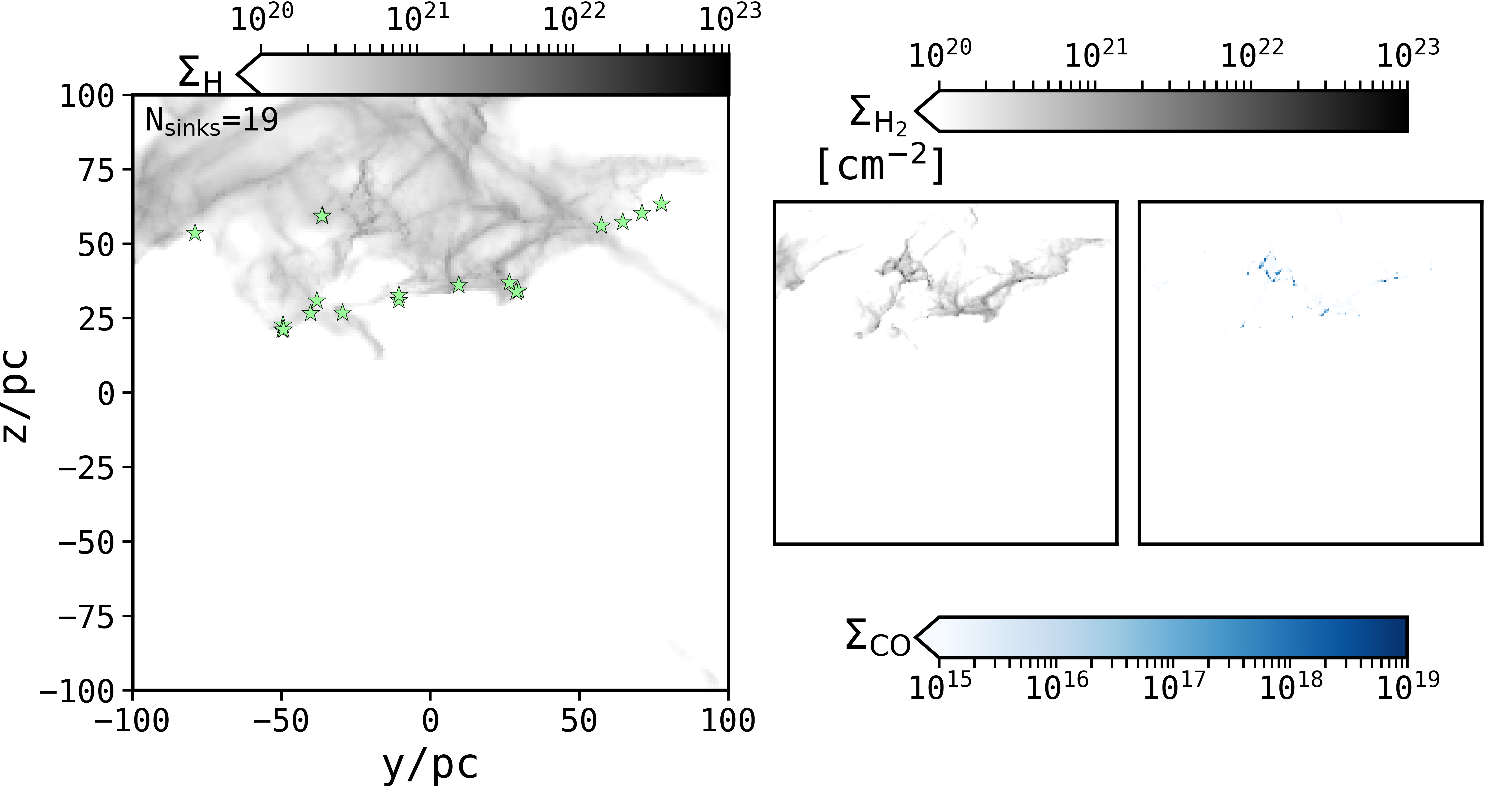}
\end{overpic} \hspace{0.5cm}
\put (40,120) {\makebox(0,0){{\huge C}}} \hspace{0.5cm}
\\

\begin{overpic}[width=0.77\textwidth]{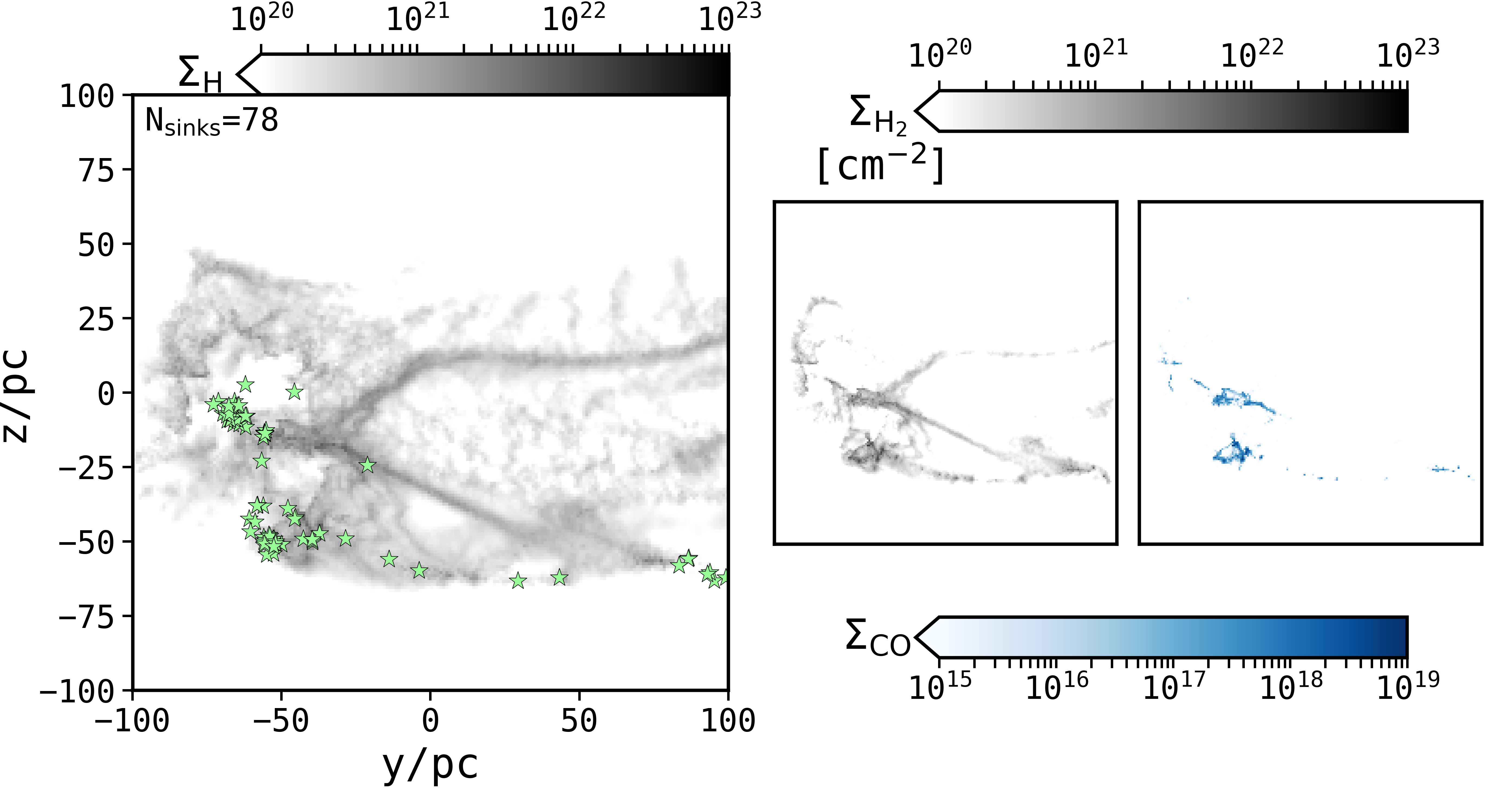}
\end{overpic} \hspace{0.5cm}
\put (40,120) {\makebox(0,0){{\huge D}}} \hspace{0.5cm}

\end{tabular}
\caption[]{(continued)}
\end{center}
\end{figure*}

\begin{figure*}
%\begin{tabular}{l}
\includegraphics[width=1.0\textwidth]{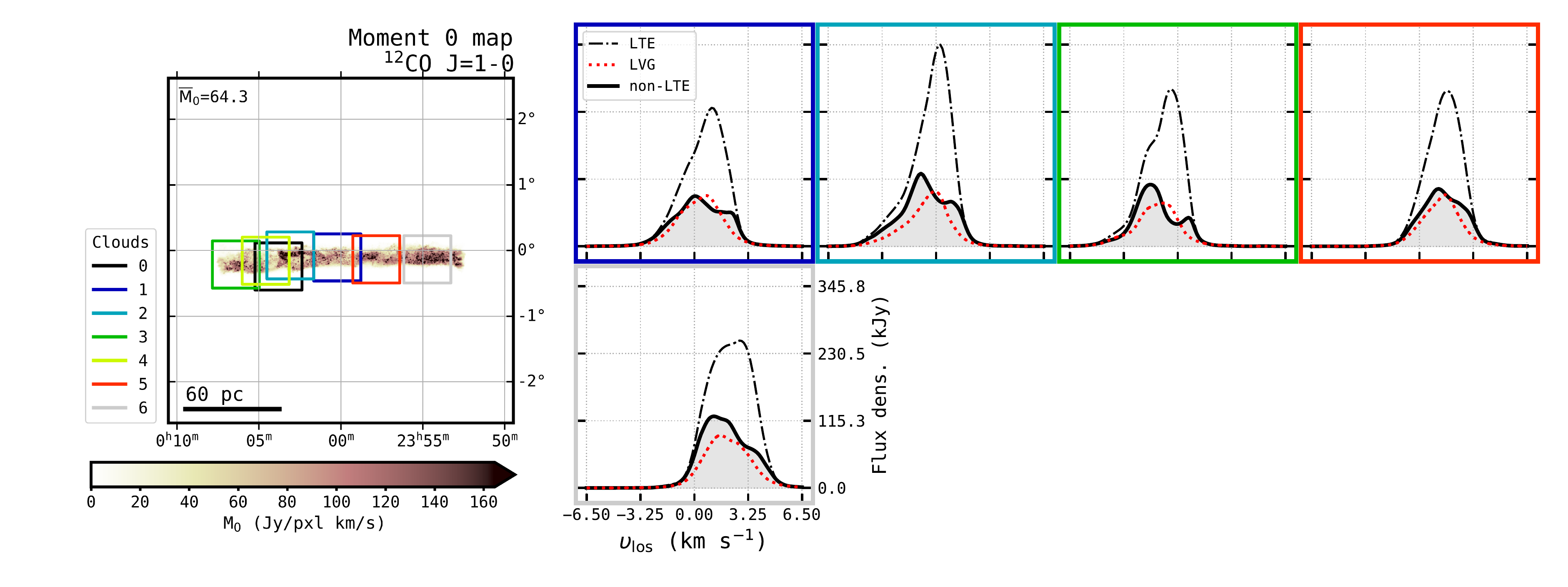}\\
\includegraphics[width=1.0\textwidth]{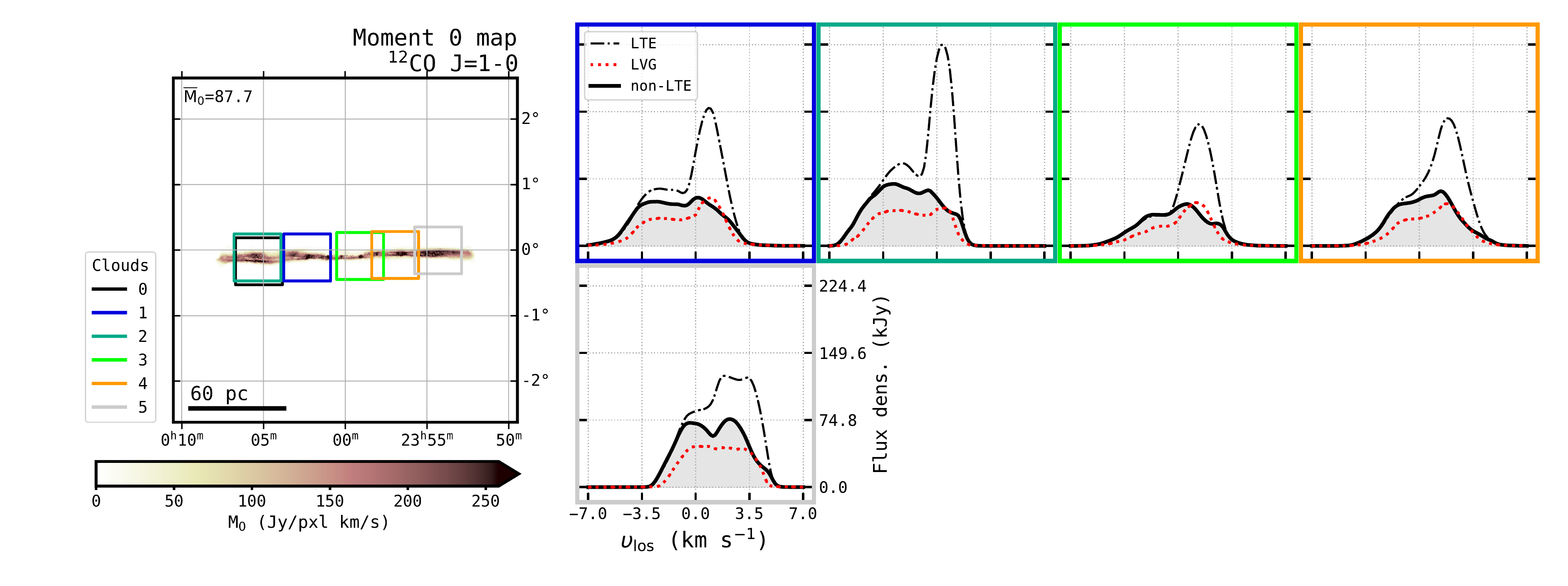}\\
\includegraphics[width=1.0\textwidth]{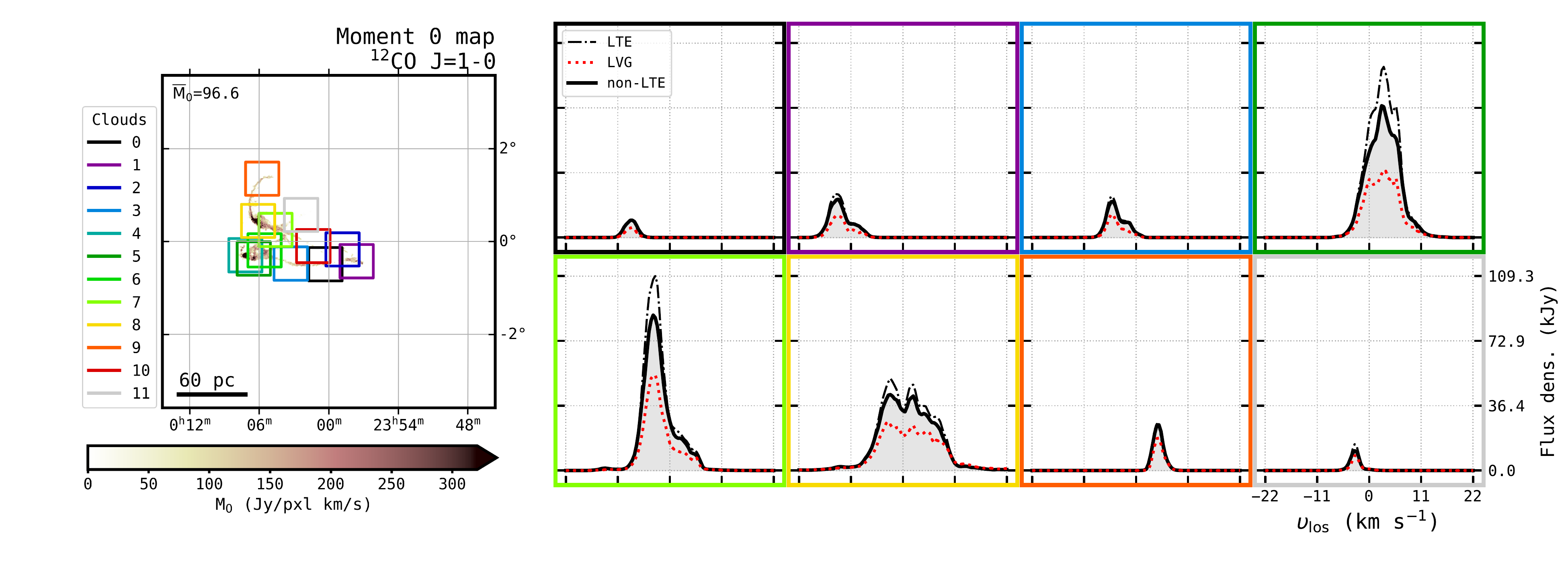}

%\end{tabular}
\caption{LTE (dashed lines) LVG (dotted red lines) and non-LTE (solid lines and shades) \twCOfull{} emission profiles of individual clouds in cloud complexes B$_0$ (top), B (middle) and D (bottom). These are \edgeonphi{} views of the complexes analysed in Fig. \ref{fig:pca_fits_faceon}. The colours of the axes spines in the right panels correspond to the colour code in the left panels and indicate from which cloud the line profile is extracted. 
}
\label{fig:line_profiles}
\end{figure*}

\begin{figure*}
%\begin{tabular}{l}

\includegraphics[width=1.0\textwidth]{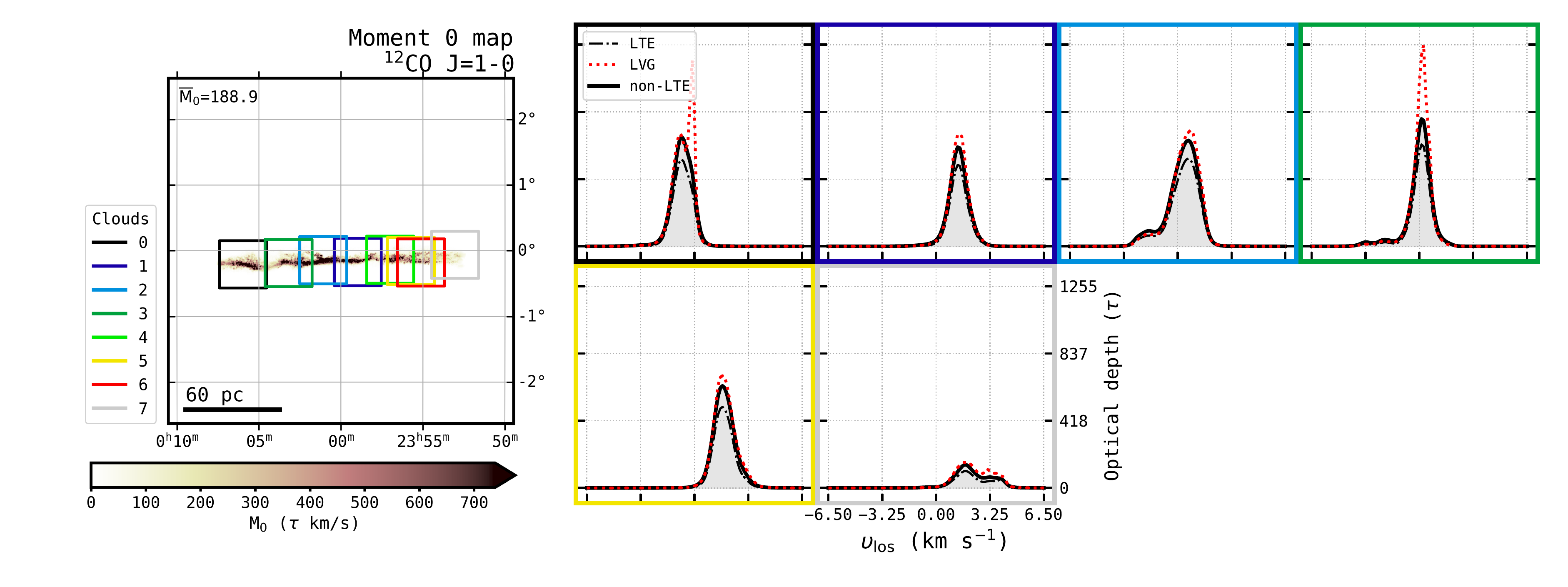}\\
\includegraphics[width=1.0\textwidth]{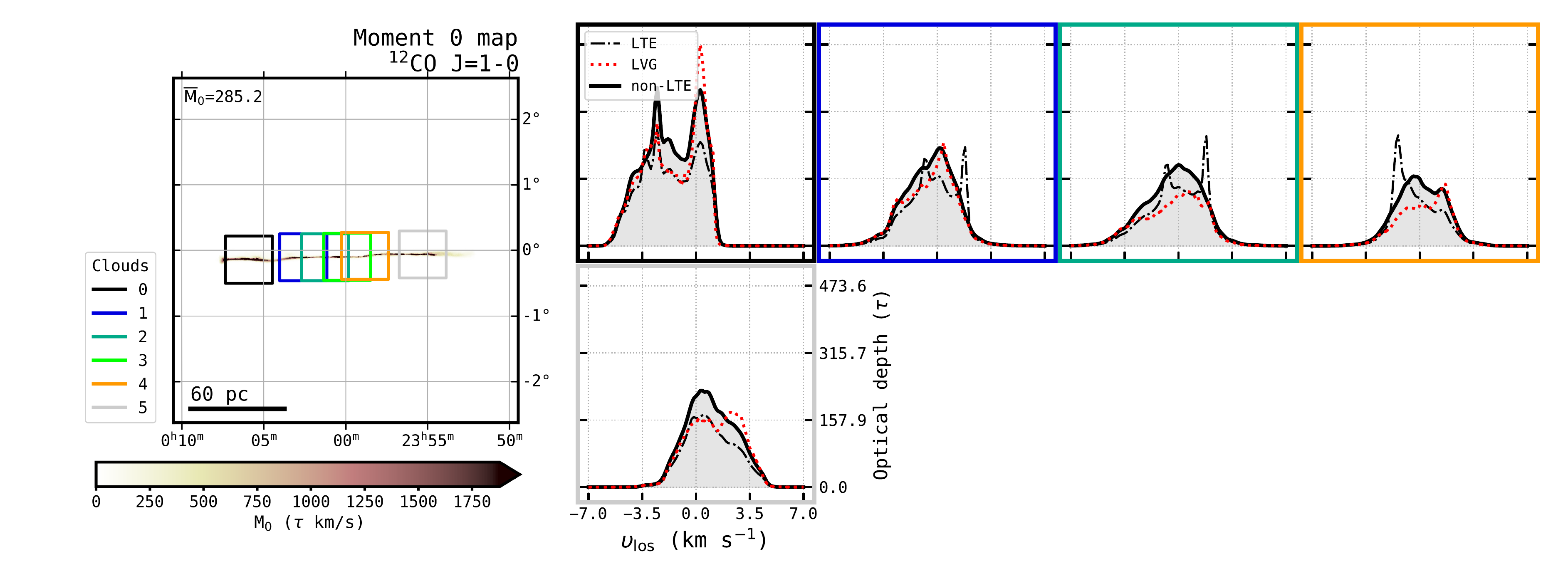}\\
\includegraphics[width=1.0\textwidth]{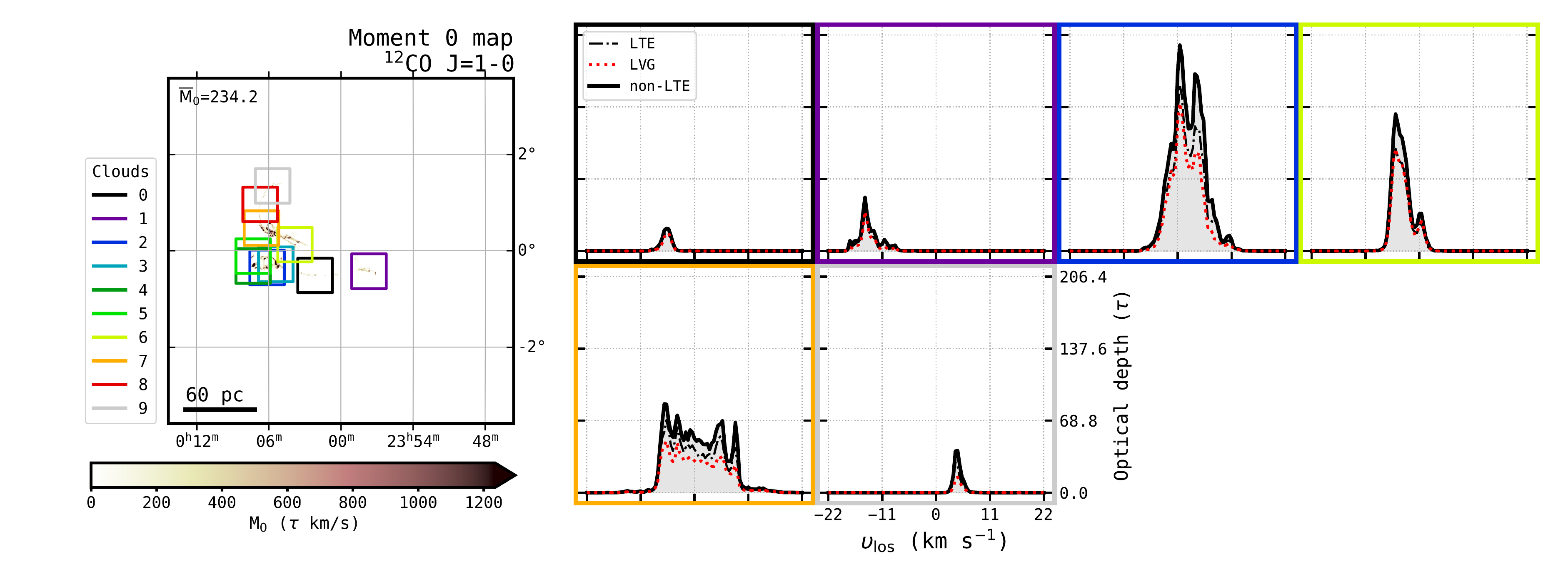}

%\end{tabular}
\caption{Same as Fig. \ref{fig:line_profiles} but using optical depth maps.
}
\label{fig:line_profiles_tau}
\end{figure*}

\begin{figure*}

\includegraphics[width=0.495\textwidth]{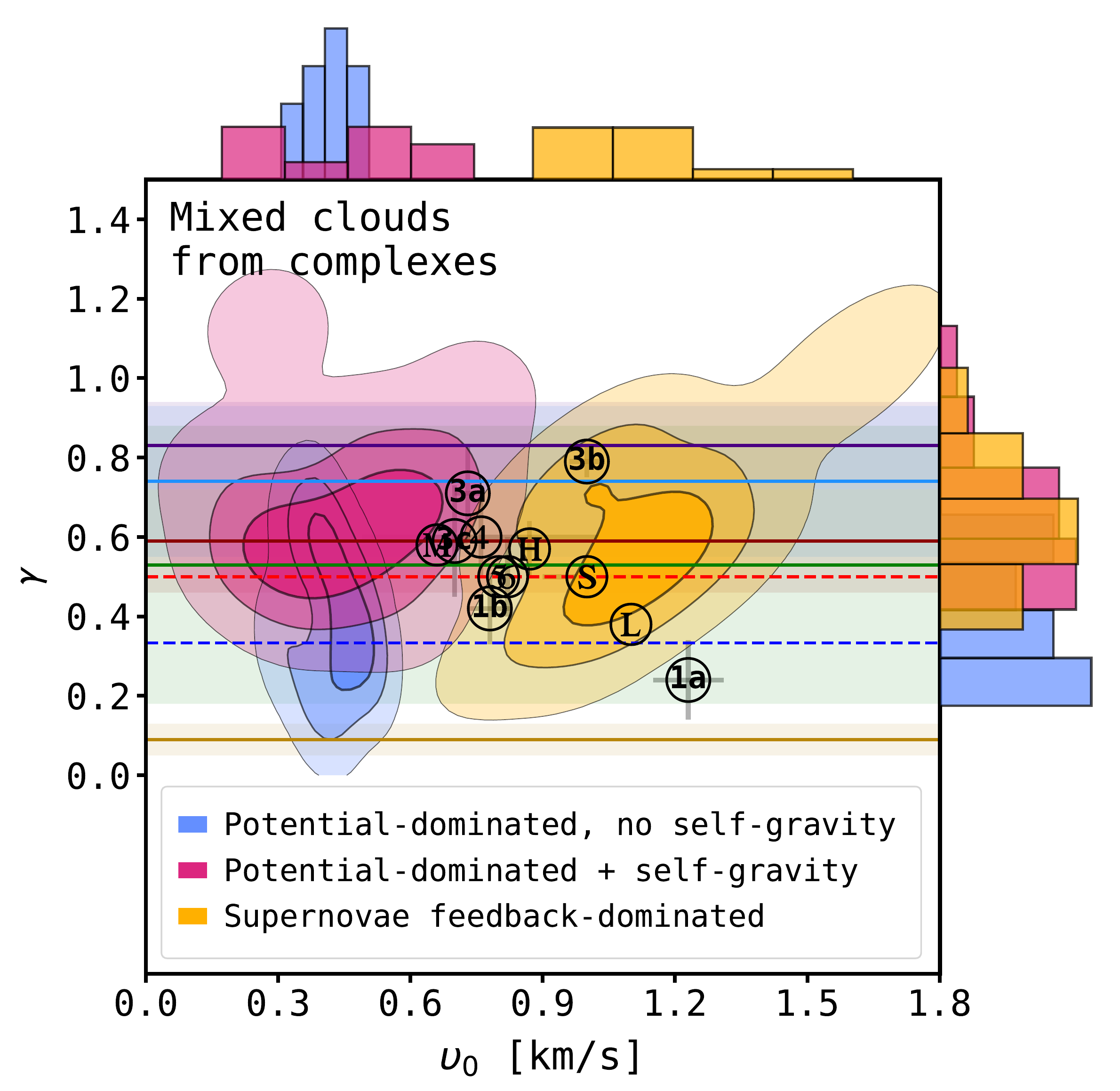}
\includegraphics[width=0.495\textwidth]{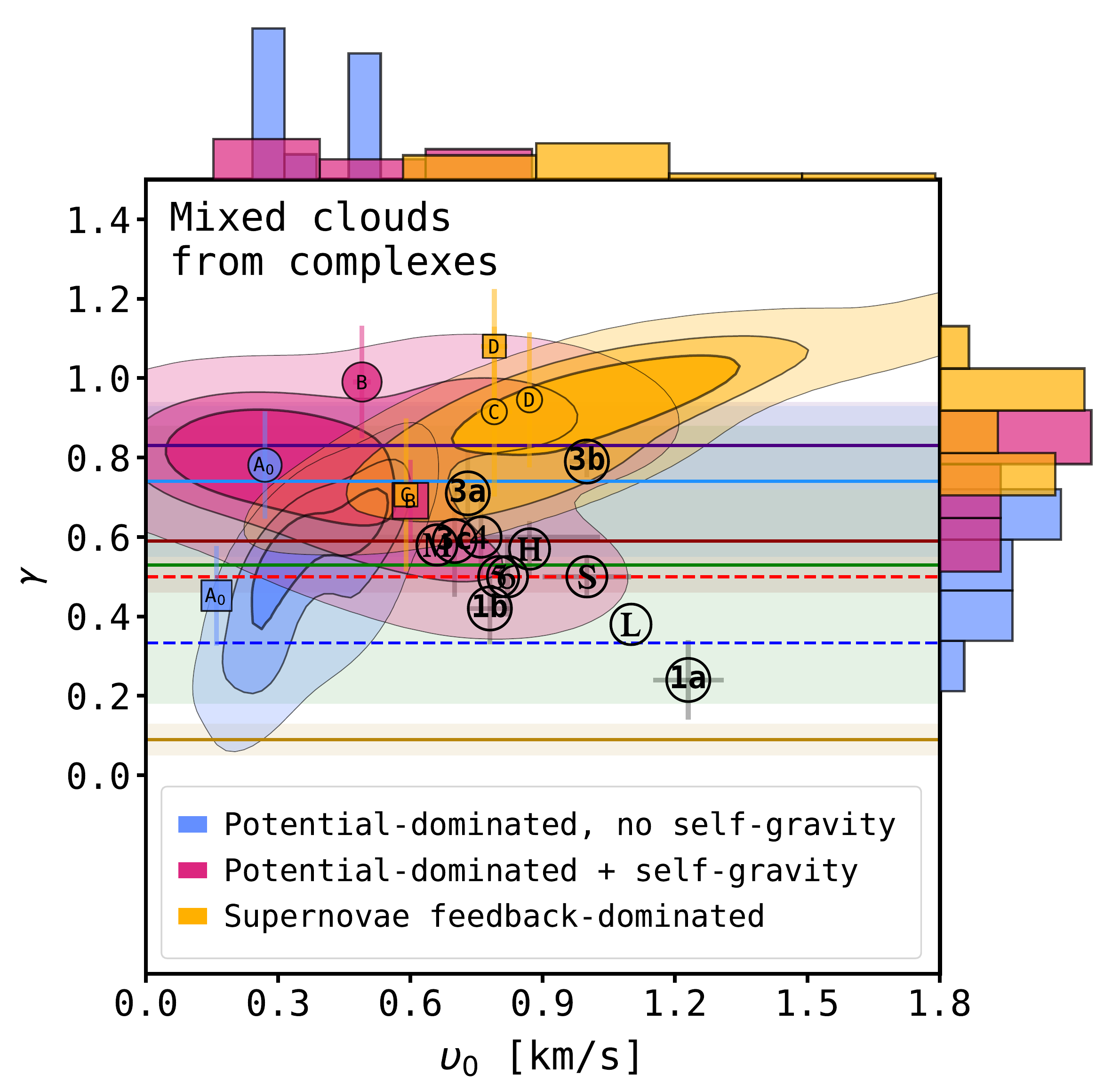} \hfill

\caption{Same as left panel of Fig. \ref{fig:pca_summary_distribution} but using LTE (left) and LVG (right) level populations for the radiative transfer of \twCO{} from our cloud complexes. Coloured markers in the right panel are the $^{13}$CO parameters presented in Table \ref{table:12co_vs_13co}.
}
\label{fig:pca_summary_distribution_LTE}

\end{figure*}
%%%%%%%%%%%%%%%%%%%%%%%%%%%%%%%%%%%%%%%%%%%%%%%%%%

% Don't change these lines
\bsp	% typesetting comment
\label{lastpage}
\end{document}